\documentclass[]{aa} % 
\usepackage{natbib}
\bibpunct{(}{)}{;}{a}{}{,}
\usepackage{graphicx}
\usepackage{txfonts}
\begin{document}
   \title{Tidal and rotational effects in the perturbations of hierarchical triple stellar systems}
   \subtitle{II. Eccentric systems -- the case of \object{AS Camelopardalis}}

   \author{T. Borkovits\inst{1} \and E. Forg\'acs-Dajka\inst{2} \and Zs. Reg\'aly\inst{3}}

   \offprints{T. Borkovits}

   \institute{\inst{1}Baja Astronomical Observatory of B\'acs-Kiskun County,
              H-6500 Baja, Szegedi \'ut, Pf. 766, Hungary, \\
              \email{borko@alcyone.bajaobs.hu} \\	  
              \inst{2}E\"otv\"os University, Department of Astronomy, H-1518 Budapest, Pf. 32, Hungary, \\
              \email{E.Forgacs-Dajka@astro.elte.hu} \\        
              \inst{3}Konkoly Observatory of HAS, H-1525 Budapest, Pf. 67, Hungary \\
              \email{regaly@konkoly.hu}}         

   \date{Received <date> / Accepted <date>}

   \titlerunning{}

\abstract{}
{We study the perturbations of a relatively close third star on a tidally distorted 
eccentric eclipsing binary. We consider both the observational consequences of the
variations of the orbital elements and the interactions of the stellar rotation with
the orbital revolution in the presence of dissipation. We concentrate mainly on the
effect of a hypothetical third companion on both the real, and the observed apsidal
motion period. We investigate how the observed period derived mainly from some 
variants of the O--C relates to the real apsidal motion period.}
{We carried out both analytical and numerical investigations and give the time variations of
the orbital elements of the binary both in the dynamical and the observational reference frames.
We give the direct analytical form of an eclipsing O--C affected simultaneously by
the mutual tidal forces and the gravitational interactions with a tertiary. We also integrated
numerically simultaneously the orbital and rotational equations for the possible hierarchical 
triple stellar system \object{AS~Camelopardalis}.}
{We find that there is a significant domain of the possible hierarchical triple system configurations, where
both the dynamical and the observational effects tend to measure longer apsidal advance rate 
than is expected theoretically. This happens when the mutual inclination
of the close and the wide orbits is large, and the orbital plane of the tertiary 
almost coincides with the plane of the sky. We also obtain new numerical results on 
the interaction of the orbital evolution and stellar rotation in such triplets. 
The most important fact is that resonances might occur as the stellar rotational rate varies 
during the dissipation-driven synchronization process, for example in the case when 
the rotational rate of one of the stars reaches the average Keplerian angular velocity of the orbital revolution.}
{}

\maketitle

%
%________________________________________________________________

\section{Introduction}
In a previous paper \citep{borkoetal04}, we introduced a new numerical code
which integrated simultaneously the orbital equations of hierarchical triple systems,
including the tidal and dissipative terms, and the Eulerian equations of stellar rotation.
First we applied the method for the \object{Algol} system itself. 
In this triple the inner binary had an almost circular orbit
with approximately synchronized stellar rotation (another application of an
earlier version of the code, not including tidal dissipation was also presented
for the ternary system \object{IM Aur} in \citealp{borkoetal02}).

In the present study we concentrate on a dynamically less-relaxed scenario, namely when the
inner binary has a significant eccentricity, i.e. the system is far from its synchronized and
circularized state. Perhaps the most important feature of such systems is the apsidal motion effect (AME),
i.e. the revolution of the orbital axis with a constant period which is determined mainly by the orbital
separation, eccentricity, masses and the inner mass-distribution of the binary members.
Nevertheless, several other physical processes also force AME. The two most significant ones are 
the perturbation of a third body, and the relativistic apsidal motion.
There is a small subgroup amongst these eccentric eclipsing binaries which have an additional importance,
as their apsidal advance period is significantly (by more than 10-20\%) affected by the relativistic
apsidal motion contribution. It is well-known, that the period of AME in these systems can be used as
further confirmation or even as challenge for the General Relativity Theory. Unfortunately, these binaries necessarily
have larger separation, so in such systems the apsidal motion period falls into the order of centuries or even of millenia.
Consequently, in these systems first we have to solve the problem of the accurate determination of the apsidal period from
a small portion of one revolution of the apsidal line, before we can label them as a challenge for the General Relativity Theory.

For our study we chose the eclipsing binary \object{AS Camelopardalis}, 
which is a member of an even smaller subgroup of the previously mentioned small group
of the eccentric eclipsing binaries, as this system, together with approximately six others,
shows a significantly lower apsidal motion rate than what is calculated from theory. 
Since this disrepancy was found for the first time at \object{DI Herculis} \citep{semeniuk68}, 
several authors have investigated this phenomenon. A summary of their results can be found 
in \citet{claret98}. One of the possible explanations is the perturbations by a third component.
The effect of the perturbations of a tertiary for the apsidal motion period has already been investigated 
only in a few previous studies \citep[][and references therein]{khodykinvedeneyev97,khodykinetal04}. 
These papers mainly focused on the above-mentioned two systems. Furthermore, in
our opinion, these earlier studies have two fundamental disadvantages. First, the third body effect
and the tidal effect were considered independently, and the resultant net apsidal motion period was
calculated simply as an algebraic sum, which is very far from the reality, as it will be shown in the present paper. 
Second, the relation between the observed parameters and the physical quantities were not included in the scope of these papers.
However, as we show, how the observed quantities, which are mainly deduced from some variants of 
the eclipsing O--C curves, relate to the real apsidal motion period, is a sophisticated problem. Note, as we know, 
\citet{claret98} was the first to mention this problem, nevertheless, in his paper this was not
examined in the light of the perturbations by a third body.

In this paper we mainly focus on the short term observational consequences
of the perturbations of a third body for an eccentric binary. We carry out both analytical and numerical
studies. We give an analytical form of the time-dependence of the orbital elements of the close binary both
in the dynamical and the observational reference frames up to fifth order in the eccentricity and related quantities.
We also present the analytical form of the O--C diagram of such a binary. We show that the complex variations of the orbital elements on a time-scale similar to the tidally 
forced apsidal motion period may result in significant discrepancies in the shape of the O--C curve from the pure eccentric two body case,
even without the remarkable real variation of the apsidal advance rate. Finally, we carry out some longer-time numerical integration
to investigate the variation of the orbital as well as the stellar rotational parameters with and without dissipation.

It is important to note, that we restrict ourselves only for the simultaneous investigation of the third body and the
tidally forced perturbations in the orbital elements in the frame of the classical, Newtonian mechanics. It may seem contradictory that despite the fact 
that our purpose is to give some acceptable explanations for the anomalously slow apsidal motion for those systems, where the relativistic contribution is
remarkable, we do not consider the relativistic apsidal motion contribution at all. However, this contradiction can be resolved easily, as follows.
The previous papers considered the three effects (tidal, third body, relativistic) as being independent. If we accept this,
then our results should only be modified by some additive constants, which does not influence our qualitative results.
Nevertheless, in the following we show that in the sense of the tidal and third body terms this is not the case. Similarly,
we can assume, that neither the third body nor the relativistic terms can be considered independently from the tidal contribution.
Consequently, to correctly consider the varying relativistic apsidal motion rate we would have to apply the
relativistic formalism, which is far beyond the scope of this paper. However, in contrast to the tidal term, the relativistic apsidal motion 
has a notably smaller dependence on the eccentricity. This suggests that the linear, additive approximation is
more realistic in this latter case. Consequently, we believe that our calculations give significant and well-applicable results.

We also omit the investigation of the effect of the non-synchronized and even non-aligned stellar rotational axes
on the perturbations of the orbital elements. Theoretically, it can be expected that such relatively young early-type eclipsing systems such as
e.g. \object{DI Her} or \object{AS Cam} could have non-aligned rotational axes \citep[cf.][]{zahn77}, 
which can produce even a reversed net apsidal revolution. Analytical formulae are given e.g. in \citet{shakura85}, and \citet{companyetal88}. 
However, for the case of \object{AS Cam} in the thorough discussion \citet{maloneyetal89} showed that this solution might be excluded, 
as the $v_{\mathrm{rot}}\sin{i}$ values derived from radial velocity measurements of \citet{hilditch72b} strongly suggest nearly synchronized rotation.
\citet{claret98} also refutes this solution in the case of \object{DI Herculis}. We also note that
although in the presence of a third companion, stellar precession could be forced by the misalignment of the orbital planes
even in the case when perfect synchronization is expected. A small amount of amplitude precession of the rotational axes actually occured in our
numerical integrations presented in Sect.~\ref{sec:numstudies}. Nevertheless, their amplitudes are so small
that they could not affect the apsidal motion significantly

In the next section we give the general mathematical form of the orbital elements and the O--C curve of an eccentric
eclipsing binary when the revolution of the stars are affected by both tidal
interactions and third-body perturbations. Then in Sect.~\ref{sec:numstudies} we present several short-time numerical integrations
with different initial configurations of the \object{AS Camelopardalis} system for supporting the analytical results of Sect.~\ref{sec:O-Canal}, 
and, furthermore, we also study the dynamical evolution of the system on a longer time-scale, including also dissipative forces.
In Sect.~\ref{sec:discussion} we further discuss our results and conclude. Finally, in Appendix~\ref{appendix} we describe our mathematical
calculations in details.

\section{Mathematical form of the O--C in a tidally and third-body perturbed eccenteric eclipsing binary \label{sec:O-Canal}}

In a previous paper \citep{borkoetal03} we calculated the effect of the third-body
perturbations on the moments of the eclipsing minima of such eclipsing binaries which are
members of close hierarchical triple stellar systems. Here we mainly follow the same
method described that paper, so we give here only a brief summary, except the
steps where we substantially modified the earlier methods. 

\subsection{General considerations and equations of the problem}
As is well-known, at the moment of the mid-eclipse 
\begin{equation}
u\approx\pm\frac{\pi}{2}+2k\pi,
\label{eq:fundamental}
\end{equation}
where $u$ is the true longitude measured from the intersection of the orbital
plane and the plane of sky, and $k$ is an integer. An exact equality stands only
if the binary has a circular orbit, or if the orbit is seen edge-on exactly 
\citep[for the correct inclination dependence of the occurrence of the mid-eclipses see][]{gimenezgarcia-pelayo83}.
This latter condition is almost satisfied in those binaries which are of interest to us now.
It is known from the textbooks of celestial mechanics, that
\begin{eqnarray}
\dot{u}&=&\frac{c}{\rho_1^2}-\dot\Omega\cos{i}, \nonumber \\
&=&\mu^{1/2}a^{-3/2}(1-e^2)^{-3/2}(1+e\cos{v})^2-\dot\Omega\cos{i},
\label{eq:upont}
\end{eqnarray}
consequently, the moment of the $N$-th primary minimum after an epoch $t_0$ can be
calculated as
\begin{eqnarray}
\int^{t_N}_{t_0}\mathrm{d}t&=&\int_{\pi/2}^{2N\pi+\pi/2}\frac{a^{3/2}}{\mu^{1/2}}\frac{(1-e^2)^{3/2}}{[1+e\cos(u-\omega)]^2}\frac{\mathrm{d}u}{1-\frac{\rho_1^2}{c_1}\dot\Omega\cos{i}}, \nonumber\\
&\approx&\int\frac{a^{3/2}}{\mu^{1/2}}\frac{(1-e^2)^{3/2}}{[1+e\cos(u-\omega)]^2}\left(1+\frac{\rho_1^2}{c_1}\dot\Omega\cos{i}\right)\mathrm{d}u.
\label{eq:TN}
\end{eqnarray}
In the equations above $c_1$ denotes the specific angular momentum of the inner binary, $\rho_1$ is the radius vector
of the secondary with respect to the primary, while the orbital elements have their usual meanings. Furthermore, in Eq.~(\ref{eq:TN}) we 
applied that the true anomaly can be written as $v=u-\omega$. Nevertheless, to 
avoid any confusion we emphasize that the angular elements (i.e. $u$, $\omega$, $\Omega$, $i$) are expressed in the ``observational'' frame of
reference, that is, its fundamental plane is the plane of the sky, and $u$, as well as $\omega$ is measured from the intersection
of the binary's orbital plane with that plane, while $\Omega$ is measured along the plane of the sky from an arbitrary origin.
In order to evaluate Eq.~(\ref{eq:TN}) first we have to express the perturbations in the orbital elements with respect to $u$. 

It is well known from the basic works of the three-body problem that in the present problem
the perturbations in the orbital elements are effective on three different time-scales.
Nevertheless, the so-called ``short-term'', as well as the ``long-term'' perturbations can be omitted due to their small amplitude.
Strictly speaking, the second kind of the above perturbations might reach the
limit of detectability in some systems \citep[see][]{borkoetal03}, but from our point of view the
``apse-node'' terms have an exclusive importance.
So, in what follows we concentrate on the so-called ``apse-node'' time-scale perturbative terms. 
They can be divided into two groups according to their different origin in Eq.~(\ref{eq:TN}). First, the ``apse-node'' time-scale
perturbations in the orbital elements $e$ and $\omega$ arises also naturally in the formula above (as is well-known, there 
are neither ``apse-node''-type, nor secular perturbations in the semi-major axis $a$). We will refer to this group in the following as indirect perturbations. 
Furthermore, some other terms which represent low-amplitude, short-period perturbations in $a$, $e$, $\omega$ give large-amplitude ``apse-node'' terms in $\dot{u}$ due
to the multiplication with some of the $\cos{nv}$ terms. These are the direct perturbations in the orbital motion. Although our calculation
of these latter direct perturbations would give back the first group too, we found that it is more convenient to calculate the two groups
in two different ways.

First, we consider the indirect perturbations.
As one can see later (e.g. Eqs.~[\ref{eq:excentricitas3}], [\ref{eq:gharmad}]), the variation in both the eccentricity, and the argument of periastron during a few revolutions
can be expressed as
\begin{eqnarray}
\Delta{e}&\sim&e\left(\frac{P}{P'}\right)^2\Delta u, \\
\Delta\omega&\sim&\left(\frac{P}{P'}\right)^2\Delta u,
\end{eqnarray}
%and, furthermore, as it is well known, that in the present level of approximation there is no secular change in
%the semi-major axis $a$, 
so it is a quite good approximation to carry out the integration Eq.~(\ref{eq:TN}) first for one revolution 
treating $a$, $e$ and $\omega$ formally as constant. In this case taking into account only 
the first term on the right hand side (rhs) we arrive at an analogue of the well-known Keplerian equation, which has the following closed solution
\begin{eqnarray}
\overline{P}_I&\!=\!&\!\frac{P}{2\pi}\!\left[2\arctan\left(\!\sqrt{\frac{1-e}{1+e}}\frac{\cos\omega}{1+\sin\omega}\right)\!-(1-e^2)^{1/2}\frac{e\cos\omega}{1+e\sin\omega}\right], \nonumber \\
\overline{P}_{II}&\!=\!&\!\frac{P}{2\pi}\!\left[2\arctan\left(\!\sqrt{\frac{1-e}{1+e}}\frac{-\cos\omega}{1-\sin\omega}\right)\!+(1-e^2)^{1/2}\frac{e\cos\omega}{1-e\sin\omega}\right], \nonumber \\
\label{eq:apsisclosed}
\end{eqnarray}
for the two types of minima, respectively. (Here $P$ denotes the anomalistic or Keplerian period which is considered to be constant.)
Note, that instead of the exact forms above, naturally its expansion is used widely (as in this paper),
which is as follows, up to the fifth order in $e$:
\begin{eqnarray}
\overline{P}_{I,II}&=&P_\mathrm{s}E+\frac{P}{2\pi}\left[\pm\frac{1}{2}\pi\mp2e\cos\omega+\left(\frac{3}{4}e^2+\frac{1}{8}e^4\right)\sin2\omega\right. \nonumber \\
&&\left.\pm\left(\frac{1}{3}e^3+\frac{1}{8}e^5\right)\cos3\omega-\frac{5}{32}e^4\sin4\omega\mp\frac{3}{40}e^5\cos5\omega\right], \nonumber \\
\label{eq:apsisexp}
\end{eqnarray}
where $P_\mathrm{s}$ is the sidereal (or eclipsing) period of, for example, the first cycle, and $E$ is the cycle-number.
Nevertheless, the difference of the
two quantities (usually denoted by $D$) is often quoted in the literature in its closed form. 

We now formulate the direct perturbations. To do this we write $e\cos{v}$ as
\begin{eqnarray}
e\cos{v}&=&e\cos\omega\cos{u}+e\sin\omega\sin{u}, \\
&=&(e\cos\omega)_0\cos{u}+(e\sin\omega)_0\sin{u}+\nonumber \\
&&+\int_{u_0}^u\left(\frac{\mathrm{d}e}{\mathrm{d}u'}\cos\omega-e\frac{\mathrm{d}\omega}{\mathrm{d}u'}\sin\omega\right)\mathrm{d}u'\cos{u}+\nonumber \\
&&+\int_{u_0}^u\left(\frac{\mathrm{d}e}{\mathrm{d}u'}\sin\omega+e\frac{\mathrm{d}\omega}{\mathrm{d}u'}\cos\omega\right)\mathrm{d}u'\sin{u}.
\end{eqnarray}
When we concentrate only on the short-period terms in the derivatives (i.e. those which are functions of $\cos{nu}$ or $\sin{nu}$),
we can take the $\cos\omega$ and $\sin\omega$ terms out of the integrand, and carry out the integrations 
only for the derivatives, so the ``apse-node'' time-scale direct perturbations in $e\cos{v}$ could be derived from
\begin{equation}
e\cos{v}_\mathrm{dir}=\cos{v}\int_{u_0}^u\left(\frac{\mathrm{d}e}{\mathrm{d}u'}\right)_\mathrm{u}\mathrm{d}u'-\sin{v}\int_{u_0}^u\left(e\frac{\mathrm{d}\omega}{\mathrm{d}u'}\right)_\mathrm{u}\mathrm{d}u',
\end{equation}
where the subscript $_\mathrm{u}$ refers to those terms which contain $\pm{u}$ in their arguments, and, generally
\begin{eqnarray}
e^m\cos{nv}_\mathrm{dir}&=&me^{m-1}\cos{nv}\int_{u_0}^u\left(\frac{\mathrm{d}e}{\mathrm{d}u'}\right)_\mathrm{nu}\mathrm{d}u'-\nonumber \\
&&-ne^{m-1}\sin{nv}\int_{u_0}^u\left(e\frac{\mathrm{d}\omega}{\mathrm{d}u'}\right)_\mathrm{nu}\mathrm{d}u'.
\end{eqnarray}
Furthermore, the direct perturbations coming from the semi-major axis can be calculated as
\begin{eqnarray}
\left(\dot{u}\right)^{-1}_\mathrm{a-dir}&=&\frac{3}{2}\frac{1}{a}\frac{\mu^{-1/2}a^{3/2}(1-e^2)^{3/2}}{(1+e\cos{v})^2}\int\frac{\mathrm{d}a}{\mathrm{d}u'}\mathrm{d}u'.
\end{eqnarray}
The derivatives are as follows:
\begin{eqnarray}
\frac{\mathrm{d}a}{\mathrm{d}u}&=&\frac{2a^{3/2}}{\sqrt{\mu(1-e^2)}}\left[f_\mathrm{r}e\sin{v}+f_\mathrm{t}(1+e\cos{v})\right]\frac{\mathrm{d}t}{\mathrm{d}u} \nonumber \\
&\approx&\frac{2a^3}{\mu}\frac{1-e^2}{(1+e\cos{v})^2}\left[f_\mathrm{r}e\sin{v}+f_\mathrm{t}(1+e\cos{v})\right], \label{eq:dadu}\\
\frac{\mathrm{d}e}{\mathrm{d}u}&\approx&\frac{a^2}{\mu}\frac{(1-e^2)^2}{(1+e\cos{v}^2}\left[f_\mathrm{r}\sin{v}+f_\mathrm{t}\left(\cos{v}+\frac{\cos{v}+e}{1+e\cos{v}}\right)\right], \label{eq:dedu}\\
e\frac{\mathrm{d}\omega}{\mathrm{d}u}&\approx&\frac{a^2}{\mu}\frac{(1-e^2)^2}{(1+e\cos{v}^2}\left[-f_\mathrm{r}\cos{v}+f_\mathrm{t}\left(\sin{v}+\frac{\sin{v}}{1+e\cos{v}}\right)-\right. \nonumber \\
&&\left.-f_\mathrm{n}\frac{e\cot{i}\sin{u}}{1+e\cos{v}}\right], \label{eq:edomdu}\\
\frac{\mathrm{d}\Omega}{\mathrm{d}u}&\approx&\frac{a^2}{\mu}\frac{(1-e^2)^2}{(1+e\cos{v})^2}f_\mathrm{n}\frac{e\sin{u}}{\sin{i}(1+e\cos{v})}, \label{eq:dOmdu}
\end{eqnarray}
where $f_\mathrm{r,t,n}$ represent the radial, transversal and normal components of the perturbing force (see later).
We applied the following approximation:
\begin{equation}
\frac{\mathrm{d}t}{\mathrm{d}u}\approx\frac{\rho_1^2}{c_1}.
\end{equation}
Finally, considering the last term on the rhs of Eq.~(\ref{eq:TN}), by the use of Eqs.~(\ref{eq:edomdu}) and (\ref{eq:dOmdu}) it
can be seen that all the direct ``apse-node'' terms in $e^m\cos{nv}$-s which would occur from the $f_\mathrm{n}$ force-component are cancelled
by the opposite-sign term in $\frac{\rho_1^2}{c_1}\frac{\mathrm{d}\Omega}{\mathrm{d}u}\cos{i}$, 
and only the indirect term from $\mu^{-1/2}a^{3/2}\frac{\mathrm{d}\Omega}{\mathrm{d}u}\cos{i}$ gives a further contribution.

\begin{figure}
\centering
\includegraphics[width=\hsize]{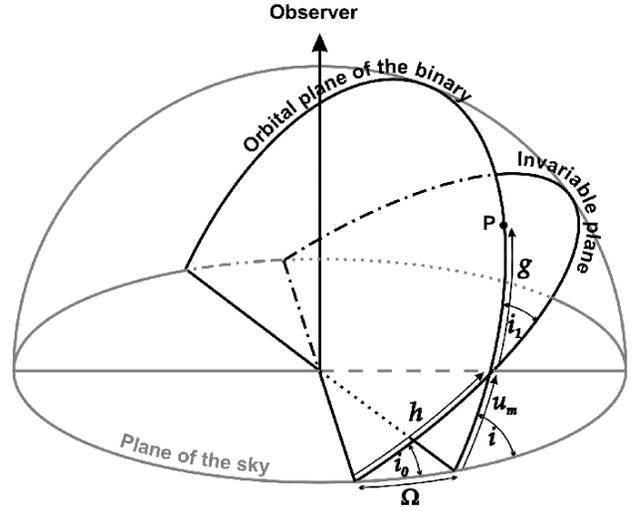}
\caption{The spatial configuration of the system.}
\label{fig:krsz1}
\end{figure}

As a next step we calculate the secular perturbations in the above-listed orbital elements. For the calculations we
truncated the perturbing force at the second order term. In this case the force components effective on the close binary are as follows:
\begin{eqnarray}
f_\mathrm{r1}&=&\frac{3}{8}\frac{Gm_3}{\rho_2^2}\frac{\rho_1}{\rho_2}\left[(1+I)^2\cos(2u'-2u-\alpha)\right. \nonumber \\
&&\left.\!\!\!+(1-I)^2\!\cos(2u'+2u-\beta)\!+\!2(1-I^2)\cos(2u-2u_\mathrm{m})\right], \\
f_\mathrm{r2}&=&\frac{3}{4}\frac{Gm_3}{\rho_2^2}\frac{\rho_1}{\rho_2}\left[(1-I^2)\cos(2u'-2u'_\mathrm{m})+I^2-\frac{1}{3}\right], \\
f_\mathrm{r3}&=&-\frac{\mu}{\rho_1^4}\left(\frac{{\cal{T}}_2}{\rho_1^3}+{\cal{R}}\right), \label{fr3}\\
f_\mathrm{t}&=&\frac{3}{8}\frac{Gm_3}{\rho_2^2}\frac{\rho_1}{\rho_2}\left[(1+I)^2\sin(2u'-2u-\alpha)\right. \nonumber \\
&&\left.\!\!\!-(1-I)^2\!\sin(2u'+2u-\beta)\!-\!2(1-I^2)\sin(2u-2u_\mathrm{m})\right], \label{ft}\\
f_\mathrm{n}&=&-\frac{3}{4}\frac{Gm_3}{\rho_2^2}\frac{\rho_1}{\rho_2}\{2\cos(u-u_\mathrm{m})\sin2(u'-u'_\mathrm{m})\sin{i}_\mathrm{m}+ \nonumber \\
&&[1-\cos2(u'-u'_\mathrm{m})]\sin(u-u_\mathrm{m})\sin2i_\mathrm{m}\}. \label{fn}
\end{eqnarray}
We divided the radial component into three parts, relative to their different significances. 
Namely, $f_\mathrm{r3}$ is the tidal term, while amongst the two three-body terms, $f_\mathrm{r2}$ is formally analogous
to $f_\mathrm{r3}$, i.e. gives a similar secular contribution to the considered perturbations (see later). Furthermore, $I$ stands for the cosine of the mutual inclination ($i_\mathrm{m}$)
of the two orbits, $\rho_2$ is the radius vector of the tertiary, $u$, $u'$ are the true longitude of the close and the wide orbits measured from
the plane of the sky, while $u_\mathrm{m}$, $u'_\mathrm{m}$ are the true longitudes of the intersection of the orbital planes measured from the
plane of the sky along the close and the wide orbital planes, respectively (see Fig.~\ref{fig:krsz1}).
Finally, the contributions of the mutual tidal, and the rotational oblateness are as follows:
\begin{eqnarray}
{\cal{T}}_2&=&6\left(\frac{m_2}{m_1}k_2^{(1)}R_1^5+\frac{m_1}{m_2}k_2^{(2)}R_2^5\right), \\
{\cal{R}}&=&\frac{k_2^{(1)}R_1^5\omega_{z'_1}^2}{Gm_1}+\frac{k_2^{(2)}R_2^5\omega_{z'_2}^2}{Gm_2},
\end{eqnarray}
where $k_2^{(1,2)}$ are the usual apsidal motion constants, $R_{1,2}$ are the
average radii of the stars, while $\omega_{z'_{1,2}}$ are the rotational angular velocities
of the star (which are treated as constant here). 

As in the present approximation there are no ``apse-node'' or secular changes in the orbital
elements of the tertiary, we considered its orbital elements as constant. 
Although the equations of perturbations can be written directly for the above-mentioned orbital
elements, it is more convenient to use the equations for the orbital elements expressed
in the so-called dynamical frame of reference, in which the fundamental plane is the
invariable plane of the triple system. The angular orbital elements in the
observational frame can then be expressed from these by the formulae of spherical geometry. 
To avoid any confusion, the argument of
periastron, and the longitude of the corresponding ascending node in this system are denoted by $g$ and $h$, 
respectively (which are their usual notations in the perturbation theories). 
Furthermore, the inclination of the orbital plane of the binary
with respect to the invariable plane is denoted by $i_1$. It can be clearly seen,
that the relation between the two periastron elements are $\omega=g+u_\mathrm{m}$ (see
Fig.~\ref{fig:krsz1}). \footnote{Note, that
in some studies on apsidal motion for the equivalents of Eq.~\ref{eq:apsisclosed} the sign $\varpi$
is used, which is meant as $g+h$. Naturally, it is correct if the perturbative force lies perfectly
in the orbital plane of the binary, which is fulfilled as far as only the tidal effect of the
distorted binary members with perpendicular rotation axes are considered, nevertheless, in
the present situation this is no longer the case.}
In the frame of our approximation the studied problem is reduced to one degree of freedom. Consequently, first we express the
variations of all the interesting orbital elements in the function of the argument of periastron $g$ of the binary in the dynamical reference frame, and then,
we give the $g(u)$ function.
So, the secular parts of the perturbational equations are as follows: 
\begin{eqnarray}
\frac{\mathrm{d}u}{\mathrm{d}g}&=&\frac{1-\frac{\rho_1^2}{c_1}\dot\Omega\cos{i}}{A+B\cos2g}, \nonumber \\
&\approx&\frac{1}{A+B\cos2g}-\frac{\mathrm{d}\Omega}{\mathrm{d}g}\cos{i}, \label{eq:dudg}\\
\frac{\mathrm{d}e}{\mathrm{d}g}&=&\frac{eA_\mathrm{t}\sin2g}{A+B\cos2g}, \label{eq:dedg}\\
\frac{\mathrm{d}h}{\mathrm{d}g}&=&-\frac{1}{\cos{i_1}}\frac{A_\mathrm{n1}+A_\mathrm{n2}-A_\mathrm{n2}\cos2g}{A+B\cos2g}, \label{eq:dhdg} \\
%&=&\frac{A_\mathrm{h1}B-A_\mathrm{h2}A}{B}\frac{1}{A+B\cos2g}+\frac{A_\mathrm{h2}}{B}, \label{dhdg}\\
\frac{\mathrm{d}i_1}{\mathrm{d}g}&=&\tan{i_1}\frac{A_\mathrm{n2}\sin2g}{A+B\cos2g}, \label{eq:di1dg} \\
&=&-\frac{e}{1-e^2}\cot{i_\mathrm{m}}\frac{\mathrm{d}e}{\mathrm{d}{g}}, \label{eq:di1dgdedgvel} \\
\frac{\mathrm{d}\delta}{\mathrm{d}g}&=&\frac{A_\mathrm{d}+B_\mathrm{d}\cos2g}{A+B\cos2g}, \label{eq:dupontkozvdg}
\end{eqnarray}
where $\delta$ denotes the direct perturbations in $(\dot{u})^{-1}$,
while for the angular elements in the observational frame of reference we obtain that
\begin{eqnarray}
\frac{\mathrm{d}\Omega}{\mathrm{d}g}&=&\frac{\mathrm{d}h}{\mathrm{d}g}\frac{\cos{I_0}-\cos{i_1}\cos{i}}{1-\cos^2i}+\frac{\mathrm{d}i_1}{\mathrm{d}g}\frac{\sin{I_0}\sin{h}}{1-\cos^2i}, \label{eq:dOmegadg} \\
%\frac{\mathrm{d}\Omega}{\mathrm{d}g}&=&\left[\frac{A_\mathrm{h1}B-A_\mathrm{h2}A}{B}\frac{1}{A+B\cos2g}+\frac{A_\mathrm{h2}}{B}\right]\frac{\sin{i_1}\sin{i}\cos{u_\mathrm{m}}}{1-\cos^2{i}}\nonumber \\
%&&+A_\mathrm{h1}\frac{\sin2g}{A+B\cos2g}\frac{\sin{i_1}\sin{i}\sin{u_\mathrm{m}}}{1-\cos^2{i}}. \label{dOmegadg}
\frac{\mathrm{d}\omega}{\mathrm{d}g}&=&1+\frac{\mathrm{d}u_\mathrm{m}}{\mathrm{d}g} \nonumber \\
&=&1+\frac{\mathrm{d}h}{\mathrm{d}g}\cos{i_1}-\frac{\mathrm{d}\Omega}{\mathrm{d}g}\cos{i} \nonumber \\
&=&1+\frac{\mathrm{d}h}{\mathrm{d}g}\frac{\cos{i_1}-\cos{I_0}\cos{i}}{1-\cos^2i}-\frac{\mathrm{d}i_1}{\mathrm{d}g}\frac{\sin{I_0}\cos{i}\sin{h}}{1-\cos^2i} \nonumber \\
&=&\frac{A-A_\mathrm{n1}-A_\mathrm{n2}+A_\mathrm{t}\cos2g}{A+B\cos2g}-\frac{\mathrm{d}\Omega}{\mathrm{d}g}\cos{i}, \label{eq:domegadg}\\
%&=&\frac{(A_\mathrm{r2}+A_\mathrm{r3})B-A_\mathrm{t}A}{B}\frac{1}{A+B\cos2g}+\frac{A_\mathrm{t}}{B}-\frac{\mathrm{d}\Omega}{\mathrm{d}g}\cos{i}, \label{domegadg} \\
\frac{\mathrm{d}i}{\mathrm{d}g}&=&-\frac{\mathrm{d}h}{\mathrm{d}g}\frac{\sin{I_0}\sin{i_1}\sin{h}}{\sin{i}}\nonumber \\
&&+\frac{\mathrm{d}i_1}{\mathrm{d}g}\frac{\cos{I_0}\sin{i_1}+\cos{i_1}\sin{I_0}\cos{h}}{\sin{i}}. \label{eq:didg}
\end{eqnarray}
In the equations above the tidal contributions are:
\begin{eqnarray}
A_\mathrm{r3}&=&\frac{1}{2a^5}\left[5{\cal{T}}_2\frac{1+\frac{3}{2}e^2+\frac{1}{8}e^4}{(1-e^2)^5}+2{\cal{R}}\frac{1}{(1-e^2)^2}\right], \label{eq:Ar3}\\
A_\mathrm{dr3}&=&-\frac{2}{a^5}\left[{\cal{T}}_2\frac{1+\frac{37}{8}e^2+\frac{59}{16}e^4+\frac{113}{32}e^6}{(1-e^2)^5}+\right. \nonumber \\
&&\left.+{\cal{R}}\frac{1+\frac{5}{4}e^2+\frac{5}{4}e^4+\frac{5}{4}e^6}{(1-e^2)^2}\right],
\label{eq:Adr3}
\end{eqnarray}
while the third body terms are as follows:
\begin{eqnarray}
A_\mathrm{G}&=&\frac{15}{8}\frac{m_3}{m_{123}}\left(\frac{P}{P'}\right)^2(1-e'^2)^{-3/2}, \label{eq:AG}\\
A_\mathrm{r2}&=&\frac{3}{5}A_\mathrm{G}(1-e^2)^{1/2}\left(I^2-\frac{1}{3}\right), \label{eq:Ar2}\\
A_\mathrm{dr2}&=&\frac{4}{3}\left(1+\frac{25}{8}e^2+\frac{15}{8}e^4+\frac{95}{64}e^6\right)A_\mathrm{r2}, \label{eq:Adr2}\\
A_\mathrm{t}&=&A_\mathrm{G}(1-e^2)^{1/2}(1-I^2), \label{eq:At}\\
A_\mathrm{dt}&=&\frac{51}{20}e^2\left(1+\frac{31}{51}e^2+\frac{23}{48}e^4\right)A_\mathrm{t}, \label{eq:Adt}\\
A_\mathrm{n1}&=&\frac{2}{5}A_\mathrm{G}(1-e^2)^{1/2}\left[I^2+\frac{C_1}{C_2}I\right], \label{eq:An1}\\
A_\mathrm{n2}&=&A_\mathrm{G}\frac{e^2}{(1-e^2)^{1/2}}\left[I^2+\frac{C_1}{C_2}I\right], \label{eq:An2}
\end{eqnarray}
and, finally,
\begin{eqnarray}
A&=&A_\mathrm{r3}+A_\mathrm{r2}+A_\mathrm{n1}+A_\mathrm{n_2} \nonumber \\
&=&\frac{1}{2a^5}\left[5{\cal{T}}_2\frac{1+\frac{3}{2}e^2+\frac{1}{8}e^4}{(1-e^2)^5}+2{\cal{R}}\frac{1}{(1-e^2)^2}\right] \nonumber \\
&&+A_\mathrm{G}\left[\frac{1}{(1-e^2)^{1/2}}I^2-\frac{1}{5}(1-e^2)^{1/2}+\frac{2}{5}\frac{1+\frac{3}{2}e^2}{(1-e^2)^{1/2}}\frac{C_1}{C_2}I\right], \nonumber \\
\label{eq:A}\\
B&=&A_\mathrm{t}-A_\mathrm{n2} \nonumber \\
&=&A_\mathrm{G}\left[(1-e^2)^{1/2}-\frac{1}{(1-e^2)^{1/2}}I^2-\frac{e^2}{(1-e^2)^{1/2}}\frac{C_1}{C_2}I\right],  \label{eq:B}\\
A_\mathrm{d}&=&A_\mathrm{dr3}+A_\mathrm{dr2}, \label{eq:Ad}\\
B_\mathrm{d}&=&A_\mathrm{dt}, \label{eq:Bd}
\end{eqnarray}
At the calculation of the formulae above it was also used that
\begin{eqnarray}
\cos{i_1}&=&\frac{\vec{C}\vec{C}_1}{CC_1}, \\
\sin{i_1}&=&\frac{|\vec{C}\times\vec{C}_1|}{CC_1},
\end{eqnarray}
where $\vec{C}_1$ means the orbital angular momentum of the close binary, while $\vec{C}$ is the same
for the whole system. Supposing that the rotational angular momenta of the three stars are negligible
we obtain that
\begin{eqnarray}
\cos{i_1}&=&\frac{C_1}{C}+\frac{C_2}{C}I, \\
\sin{i_1}&=&-\frac{C_2}{C}\sin{i_\mathrm{m}},
\end{eqnarray}
and it is well known, that
\begin{eqnarray}
C_1&=&\frac{m_1m_2}{m_{12}}\sqrt{Gm_{12}a(1-e^2)}, \\
C_2&=&\frac{m_{12}m_2}{m_{123}}\sqrt{Gm_{123}a'(1-e'^2)}.
\end{eqnarray}

\subsection{Solution for edge-on binary orbits with small eccentricity variations}
\subsubsection{Closed form solutions for the binary's orbital elements}
For the first time we assume an edge-on binary orbit in the observational frame, which is a plausible expectation
in the case of the relatively wider eccentric eclipsing binaries. Consequently, at this stage we omit terms multiplied by $\cos{i}$.
We consider Eqs.~(\ref{eq:dudg})--(\ref{eq:domegadg}). One can see that if $B\geq A$ these equations become singular 
at certain directions of the axis. This is exactly the case which defines the so-called Kozai resonance \citep{kozai62}. 
We are interested in such binaries where $A>B$, i.e. this resonance does not occur. In this case as far as 
the coefficients at the rhs of the equations can be treated as constant, or at least their variations are small, 
all the equations have closed solution, which, for $B\neq0$ are as follows:
\begin{eqnarray}
u&=&u_0+\frac{1}{A}\frac{1}{\sqrt{1-E^2}}\arctan\left(\sqrt{\frac{1-E}{1+E}}\tan{g}\right)_{g_0}^g, \label{eq:ueconst}\\
e&=&e_0-\frac{1}{2}e\frac{A_\mathrm{t}}{B}\ln\left(\frac{1+E\cos2g}{1+E\cos2g_0}\right), \label{eq:eeconst}\\
h&=&h_0+\frac{1}{\cos{i_1}}\frac{A_\mathrm{n2}}{B}(g-g_0)-\frac{1}{\cos{i_1}}\left(\frac{A_\mathrm{n1}+A_\mathrm{n2}}{A}+\frac{A_\mathrm{n2}}{B}\right) \nonumber \\
&&\times\frac{1}{\sqrt{1-E^2}}\arctan\left(\sqrt{\frac{1-E}{1+E}}\tan{g}\right)_{g_0}^g, \label{eq:heconst}\\
i_1&=&(i_1)_0-\frac{1}{2}\tan{i_1}\frac{A_\mathrm{n2}}{B}\ln\left(\frac{1+E\cos2g}{1+E\cos2g_0}\right), \label{eq:i1econst}\\
\omega&=&\omega_0+\frac{A_\mathrm{t}}{B}(g-g_0)-\left(\frac{A_\mathrm{n1}+A_\mathrm{n2}}{A}+\frac{A_\mathrm{n2}}{B}\right)\nonumber \\
&&\times\frac{1}{\sqrt{1-E^2}}\arctan\left(\sqrt{\frac{1-E}{1+E}}\tan{g}\right)_{g}^g, \label{eq:omegaeconst} \\
\delta&=&\delta_0+\frac{B_\mathrm{d}}{B}(g-g_0) \nonumber \\
&&+\left(\frac{A_\mathrm{d}}{A}-\frac{B_\mathrm{d}}{B}\right)\frac{1}{\sqrt{1-E^2}}\arctan\left(\sqrt{\frac{1-E}{1+E}}\tan{g}\right)_{g_0}^g,
\label{eq:upontkozveconst}
\end{eqnarray}
where 
\begin{equation}
E=\frac{B}{A}.
\end{equation}
Eqs.~(\ref{eq:ueconst}) and (\ref{eq:heconst}) are equivalent to the results of \citet{soderhjelm84}. Nevertheless, in his paper the eccentricity
equation was not calculated, as the eccentricity was considered as strictly constant. Furthermore, we stress again, that the last equation for $\omega$ is calculated
in the observational and not in the dynamical frame of reference.
The first equation reveals that in this case we get the following constant angular velocity (in $P_\mathrm{s}$ units) for the apsidal motion
in the dynamical system:
\begin{equation}
\Pi=A\sqrt{1-E^2}.
\end{equation}
Similarly, the secular terms, i.e. the mean angular velocities of the invariant node ($h$), and the observable argument of periastron ($\omega$)
are as follows:
\begin{eqnarray}
{\cal{H}}_0&=&\frac{1}{\cos{i_1}}\left[\frac{A_\mathrm{n2}}{B}-\left(\frac{A_\mathrm{n1}+A_\mathrm{n2}}{A}+\frac{A_\mathrm{n2}}{B}\right)\frac{1}{\sqrt{1-E^2}}\right]\Pi, \\
{\cal{O}}_0&=&\Pi+\cos{i_1}{\cal{H}}_0 \nonumber \\
&=&\left[\frac{A_\mathrm{t}}{B}-\left(\frac{A_\mathrm{n1}+A_\mathrm{n2}}{A}+\frac{A_\mathrm{n2}}{B}\right)\frac{1}{\sqrt{1-E^2}}\right]\Pi,
\end{eqnarray}
while the secular part of the direct perturbations gives
\begin{equation}
{\cal{D}}_0=\left[\frac{B_\mathrm{d}}{B}+\left(\frac{A_\mathrm{d}}{A}+\frac{B_\mathrm{d}}{B}\right)\frac{1}{\sqrt{1-E^2}}\right]\Pi. \label{eq:kozvszekularis}
\end{equation}
In the following we introduce the new variable ${\cal{G}}$ as
\begin{equation}
{\cal{G}}=\arctan\left(\sqrt{\frac{1-E_0}{1+E_0}}\tan{g}\right).
\label{Wdef}
\end{equation}
It can be seen easily that 
\begin{equation}
\frac{\mathrm{d}u}{\mathrm{d}g}\frac{\mathrm{d}g}{\mathrm{d}{\cal{G}}}=\frac{1}{A}\frac{1-E_0\cos2{\cal{G}}}{1-EE_0+(E-E_0)\cos2{\cal{G}}}\frac{\sqrt{1-E_0^2}}{1-E_0\cos2{\cal{G}}}, 
\end{equation}
and, as in the current approximation $E=E_0=const.$,
\begin{equation}
u-u_0=\Pi^{-1}({\cal{G}}-{\cal{G}}_0).
\end{equation}
Similarly,
\begin{eqnarray}
e&=&e_0(1+\epsilon_0)+\frac{1}{2}e_0\left(\frac{A_\mathrm{t}}{B}\right)_0\ln(1-E_0\cos2{\cal{G}}), \label{eq:e(W)0}\\
g&=&\arctan\left(\sqrt{\frac{1+E_0}{1-E_0}}\tan{{\cal{G}}}\right), \label{eq:g(W)0}\\
h&=&h_0^*-\frac{1}{\cos{i_1}}\left(A_\mathrm{n1}+A_\mathrm{n2}+\frac{A_\mathrm{n2}}{E}\right)_0\frac{1}{\Pi_0}{\cal{G}} \nonumber \\
&&+\frac{1}{\cos{i_1}}\left(\frac{A_\mathrm{n2}}{B}\right)_0\arctan\left(\sqrt{\frac{1+E_0}{1-E_0}}\tan{{\cal{G}}}\right), \label{eq:h(W)0}\\
i_1&=&(i_1)_0^*+\frac{1}{2}\tan{i_1}\left(\frac{A_\mathrm{n}}{B}\right)_0\ln(1-E_0\cos2{\cal{G}}), \label{eq:i1(W)0}\\
\omega&=&\omega_0^*-\left(A_\mathrm{n1}+A_\mathrm{n2}+\frac{A_\mathrm{n2}}{E}\right)_0\frac{1}{\Pi_0}{\cal{G}} \nonumber \\
&&+\left(\frac{A_\mathrm{t}}{B}\right)_0\arctan\left(\sqrt{\frac{1+E_0}{1-E_0}}\tan{{\cal{G}}}\right), \label{eq:omega(W)0} \\
\delta&=&\delta_0^*+\left(A_\mathrm{d}-\frac{B_\mathrm{d}}{E}\right)_0\frac{1}{\Pi_0}{\cal{G}} \nonumber \\
&&+\left(\frac{B_\mathrm{d}}{B}\right)_0\arctan\left(\sqrt{\frac{1+E_0}{1-E_0}}\tan{{\cal{G}}}\right), \label{eq:upontkozv(W)0} 
\end{eqnarray}
where
\begin{eqnarray}
\epsilon_0&=&-\frac{1}{2}\left(\frac{A_\mathrm{t}}{B}\right)_0\ln(1-E_0\cos2{\cal{G}}_0), \\
h_0^*&=&h_0-\frac{1}{\cos{i_1}}\left(\frac{A_\mathrm{n2}}{B}\right)_0g_0\nonumber \\
&&+\left(A_\mathrm{n1}+A_\mathrm{n2}+\frac{A_\mathrm{n2}}{E}\right)_0\frac{1}{\Pi_0}{\cal{G}}_0, \\
(i_1)_0^*&=&(i_1)_0-\frac{1}{2}\left(\tan{i_1}\frac{A_\mathrm{n}}{B}\right)_0\ln(1-E_0\cos2{\cal{G}}_0), \\
\omega_0^*&=&\omega_0-\left(\frac{A_\mathrm{t}}{B}\right)_0g_0+\left(A_\mathrm{n1}+A_\mathrm{n2}+\frac{A_\mathrm{n2}}{E}\right)_0\frac{1}{\Pi_0}{\cal{G}}_0 \nonumber \\
&=&\left(u_\mathrm{m}\right)_0+\left(h_0^*-h_0\right)\cos{i_1}, \\
\delta_0^*&=&\delta_0-\left(\frac{B_\mathrm{d}}{B}\right)_0g_0-\left(A_\mathrm{d}-\frac{B_\mathrm{d}}{E}\right)_0\frac{1}{\Pi_0}{\cal{G}}_0,
\end{eqnarray}
respectively. In what follows, we omit the subscript $_0$ from the parameters $E$, ${\cal{E}}$, 
because these parameters will always be used as constants, with the value calculated at $e=e_0$.
\subsubsection{The analytical form of the apsidal part of the O--C \label{subsubsect:analO-C0}}
By the use of the Taylorian expansion of Eqs.~(\ref{eq:e(W)0}), (\ref{eq:omega(W)0}) and (\ref{eq:upontkozv(W)0}) we obtain for the analytical form of the apsidal
part of the O--C as follows:
\begin{eqnarray}
\frac{2\pi}{P}\overline{P}&=&j\frac{1}{2}\pi-2je_0(1+\epsilon_0)\cos[\omega_0^*+(1+{\cal{U}}){\cal{G}}] \nonumber \\
&&+\frac{3}{4}e_0^2(1+\epsilon_0)^2\sin[2\omega_0^*+(2+2{\cal{U}}){\cal{G}}]\nonumber \\
&&+j\frac{1}{3}e_0^3(1+\epsilon_0)^3\cos[3\omega_0^*+(3+3{\cal{U}}){\cal{G}}] \nonumber \\
&&+je_0\left[\frac{1}{8}(1+\epsilon_0){\cal{E}}^2+\frac{1}{4}{\cal{E}}E\right]\cos[\omega_0^*+(1+{\cal{U}}){\cal{G}}] \nonumber \\
&&+je_0\frac{1}{2}\epsilon_0{\cal{E}}\cos[\omega_0^*+(3+{\cal{U}}){\cal{G}}] \nonumber \\
&&+je_0\left[{\cal{E}}+\frac{1}{2}\epsilon_0{\cal{E}}\right]\cos[\omega_0^*-(1-{\cal{U}}){\cal{G}}] \nonumber \\
&&+je_0\left[\frac{1}{16}(1-\epsilon_0){\cal{E}}^2-\frac{1}{8}\epsilon_0{\cal{E}}E\right]\cos[\omega_0^*+(5+{\cal{U}}){\cal{G}}]\nonumber \\
&&+je_0\left[-\frac{1}{16}(3+\epsilon_0){\cal{E}}^2\right. \nonumber \\
&&\left.+\frac{1}{8}(2+\epsilon_0){\cal{E}}E\right]\cos[\omega_0^*-(3-{\cal{U}}){\cal{G}}]\nonumber \\
&&+e_0^2\left[\frac{3}{8}\epsilon_0(1+\epsilon_0){\cal{E}}\right]\sin[2\omega_0^*+(4+2{\cal{U}}){\cal{G}})] \nonumber \\
&&+e_0^2\left[-\frac{3}{8}(2+3\epsilon_0+\epsilon_0^2){\cal{E}}\right]\sin(2\omega_0^*+2{\cal{U}}{\cal{G}})\nonumber \\
&&+\frac{51}{40}e_0^2{\cal{E}}\sin{2{\cal{G}}}+{\cal{O}}[(e,E)^4].
\label{eq:O-C(W)0}
\end{eqnarray}
From now on we suppose, that ${\cal{O}}(e)={\cal{O}}(E)={\cal{O}}({\cal{E}})$, or more generally,
${\cal{O}}(e)={\cal{O}}\left(\frac{A_\mathrm{G}}{A}\right)$. Furthermore,
\begin{eqnarray}
{\cal{E}}&=&\frac{A_\mathrm{t}}{A}, \\
{\cal{U}}&=&\frac{A_\mathrm{n2}}{B}-\left(\frac{A_\mathrm{n1}+A_\mathrm{n2}}{A}+\frac{A_\mathrm{n2}}{B}\right)\frac{1}{\sqrt{1-E^2}},
\end{eqnarray}
the latter is the period ratio of $h\cos{i_1}$ (or in the present approximation, $u_\mathrm{m}$) and $g$.
This term gives the relative difference of the speed of the apsidal advance in the observational and the dynamical
frames of reference. Furthermore, note, that in this approximation the direct terms give only a small contribution to the O--C 
(the last term of Eq.~[\ref{eq:O-C(W)0}]). Nevertheless, the secular part of such perturbations are added to the observed eclipsing
period. Finally, $j\pm1$ for the two different types of minima.

We consider the two extreme cases. First, if the two orbits are coplanar (i.e. $I^2=1$), then the angular velocity of the apsidal motion is
\begin{equation}
(1+{\cal{U}})\Pi=A_\mathrm{r2}+A_\mathrm{r3}.
\label{eq:Picoplanar}
\end{equation}
Furthermore, as ${\cal{E}}=0$,  (i.e. $\Delta{e}=0$, which is true as far as we do not consider the octuple term
in the perturbing force), Eq.~(\ref{eq:O-C(W)0}) reduces for its usual form, apart from the different period given by Eq.~(\ref{eq:Picoplanar}).
Second, in the case of two perpendicular orbits (i.e. $I=0$), ${\cal{U}}$ diminishes, as well as ${\cal{E}}=E$ occurs,
and, consequently, Eq.~(\ref{eq:O-C(W)0}) also becomes somewhat simpler. Moreover, the more important feature is that
these are generally the only two cases, when the O--C curve has only one fundamental period. Finally, in these
two extreme cases
\begin{equation}
\dot\Omega=0,
\end{equation}
so the results above are rigorously correct for not only edge-on visible orbits, as far as the orbital eccentricity can be considered as constant on the
rhs of the perturbation equations (\ref{eq:dudg})--(\ref{eq:didg}).

\subsection{Solution for the general case}
\subsubsection{Formulae for $e$, $g$ and $h$ -- results and discussion on the dynamical apsidal motion period and on the nodal regression \label{subsubsect:egh}}
One can see that the assumption of the constant eccentricity remains plausible only if $B<<A$, or
if $A_\mathrm{t}\approx 0$, which may happen in two different ways. The trivial case, when no
third body exists in the system, or at least its influence can be disregarded with respect to the tidal forces, 
or the other possibility is, that the two orbits are nearly coplanar. (Note, that this latter case
usually also could satisfy the $B<<A$ condition, as in this case $B$ is in the order of $e^2A_\mathrm{G}$.)
Nevertheless, in the really interesting systems neither of these conditions are fulfilled, so we have to solve
the equations above in several iteration steps. 
In order to do this we used the Taylorian expansions of Eqs.~(\ref{eq:dudg}), (\ref{eq:dedg}) with respect to $e$. Our calculation
is listed in Appendix~\ref{appendix}. Here we give only the final forms up to the third order in the inner eccentricity, together
with such assumption that the other quantities given above are also in the first order of eccentricity.
So, in this approximation the modified angular velocity of the apsidal advance, the inner eccentricity, and the argument
of periastron in the dynamical frame ($g$) become
\begin{eqnarray}
(\Pi^*)^{-1}&=&\Pi^{-1}\left(1-\frac{1}{16}A_2{\cal{E}}^2-\frac{1}{2}A_1E{\cal{E}}-A_1\epsilon_0-\frac{1}{2}A_2\epsilon_0^2\right),
\label{eq:Picsillag-13}
\end{eqnarray}
\begin{eqnarray}
e&=&e_0\left(1+\epsilon_0+\frac{1}{4}E{\cal{E}}-\frac{1}{8}A_1{\cal{E}}^2\right)\nonumber \\
&&\!-e_0\left\{\frac{1}{2}\left[1-\!\frac{1}{32}{\cal{E}}^2+\!\frac{1}{16}E{\cal{E}}+\!\frac{1}{4}E^2+(1-A_1)\epsilon_0\right]{\cal{E}}\cos2{\cal{G}}\right. \nonumber \\
&&-\left[\frac{1}{16}(1-A_1){\cal{E}}+\frac{1}{8}E+\left(\frac{1}{16}{\cal{E}}+\frac{1}{8}E\right)\epsilon_0\right]{\cal{E}}\cos4{\cal{G}} \nonumber \\
&&\left.+\left[\frac{1}{192}{\cal{E}}^2+\frac{1}{32}E{\cal{E}}+\frac{1}{24}E^2\right]{\cal{E}}\cos6{\cal{G}}\right\},
\label{eq:excentricitas3}
\end{eqnarray}
\begin{eqnarray}
g&=&{\cal{G}}+\frac{1}{2}\left(1+\frac{1}{4}E^2\right)E\sin2{\cal{G}}+\frac{1}{8}E^2\sin4{\cal{G}}\nonumber \\
&&+\frac{1}{24}E^3\sin6{\cal{G}}\nonumber \\
&&-\left[\frac{1}{4}A_1{\cal{E}}+\left[\frac{1}{4}(A_1+A_2){\cal{E}}+\frac{1}{2}A_1E\right]\epsilon_0\right\}\sin2{\cal{G}} \nonumber \\
&&+\left[\frac{1}{64}(A_1+A_2){\cal{E}}^2-\frac{3}{32}A_1E{\cal{E}}\right]\sin4{\cal{G}},
\label{eq:gharmad}
\end{eqnarray}
where
\begin{equation}
{\cal{G}}=\Pi^*(u-u_0^*),
\end{equation}
while
\begin{eqnarray}
\epsilon_0&=&\frac{1}{8}(1-A_1){\cal{E}}^2+\left[\frac{1}{2}+\frac{3}{64}{\cal{E}}^2-\frac{1}{32}E{\cal{E}}+\frac{1}{8}E^2\right]{\cal{E}}\cos2g_0 \nonumber \\
&&+\left[\frac{1}{16}(1-A_1){\cal{E}}-\frac{1}{8}E\right]{\cal{E}}\cos4g_0\nonumber \\
&&+\left[\frac{1}{192}{\cal{E}}^2-\frac{1}{32}E{\cal{E}}+\frac{1}{24}E^2\right]{\cal{E}}\cos6g_0, \label{eq:eps0harmad}\\
A_1&=&e_0\frac{1}{A}\frac{\mathrm{d}A}{\mathrm{d}e}, \\
A_2&=&e_0^2\frac{1}{A}\frac{\mathrm{d}^2A}{\mathrm{d}e^2}. 
\end{eqnarray}

Before continuing with the effects of those quantities which relate to the observational system, we discuss
our results for the dynamical apsidal advance rate. First, we consider the instantaneous angular velocity of the
apsidal line, i.e $\Pi=\sqrt{A^2-B^2}$. As we are concentrating on hierarchical triple systems, where the total angular
momentum is highly concentrated in the wider orbit, we omit the terms in $A$ and $B$ which are multiplied by $C_1/C_2$.
In this case we can easily have that up to $i_\mathrm{m}=63\fdg43$ (i.e. $I^2>0.2$) $A>A_\mathrm{r3}$.
So, when 
\begin{equation}
A_\mathrm{r2}+A_\mathrm{n1}+A_\mathrm{n2}>A_\mathrm{t}-A_\mathrm{n2}
\label{eq:Kozaifeltetel}
\end{equation}
then the third body affected instantaneous apsidal angular velocity is surely larger, and, consequently, 
the apsidal motion period is shorter than in the only tidally perturbed case. Note, that Eq.~(\ref{eq:Kozaifeltetel})
is also the zero order condition for the occurrence of the so-called Kozai resonance in a mass-point three-body
model. This gives $i_\mathrm{m}<39\fdg23$ for $e=0$. We now concentrate on the average angular velocity, $\Pi^*$.
There is only one quantity in the $\Pi/\Pi^*$ ratio, Eq.~(\ref{eq:Picsillag-13}) which can be negative, namely $\epsilon_0$. 
Consequently, all of the other terms would produce a shorter average period than the instantaneous one. 
According to Eq.~(\ref{eq:eps0harmad}), there is a first order term in $\epsilon_0$, $\frac{1}{2}{\cal{E}}\cos2g_0$,
which may give the largest contribution in the whole $\Pi/\Pi^*$ ratio. One can see that we could expect a longer
average apsidal motion period than the instantaneous one, when $\cos2g_0\approx-1$. This happens when the argument of periastron
is $g_0\approx\pm90\degr$ in the dynamical system in the moment of the calculation of the orbital element (i.e. at the time of observation). 
Note that as one can see from e.g. Eq~(\ref{eq:eeconst}), for $g_=\pm90\degr$ the eccentricity
takes its maximum value, and as is well-known, the larger the eccentricity the faster the tidally-forced apsidal 
advance speed, this result is not an unexpected one.

For the sake of completeness, we give our result for the dynamical nodal regression, although up to
third order in $e,E$ it is identical with the Taylorian of Eq.~(\ref{eq:h(W)0}):
\begin{equation}
h=h_0^*+H_0{\cal{G}}-\frac{1}{2}\frac{1}{\cos{i_1}}\frac{A_\mathrm{n2}}{\Pi^*}\sin2{\cal{G}},
\end{equation}
where
\begin{equation}
H_0=-\frac{1}{\Pi^*}\frac{1}{\cos{i_1}}\left\{A_\mathrm{n1}+A_\mathrm{n2}\left(1+\frac{1}{2}E\right)+\frac{1}{4}e_0\frac{\mathrm{d}A_\mathrm{n2}}{\mathrm{d}e}{\cal{E}}\right\}.
\end{equation}
We emphasize again, that this term gives further, negative contribution to the observable apsidal motion period,
through the $h\cos{i_1}$ expression, which is independent of the observable inclination of the system.

At this point we refer to the paper of \citet{hegenuspl86}. In their work they investigate the possible
observational effects of nodal motion forced by inclined stellar rotational axes. Although their
treatment is entirely correct, they unfortunately denoted the dynamical node (which is $h$ in the present paper) 
by $\Omega$, which usually means the longitude of the node in the sky (as used in the present work) 
in the observational astrophysics. From this notation some misinterpretations occured later. So, when \citet{khodykin89}
reacts to the previously mentioned paper, and states that the orbital plane precession of an eclipsing binary is unable 
to significantly distort the observed apsidal motion rate \citep[see also][]{khaliullinetal91},
he is right in the sense of the observed node (see the $\mathrm{d}\Omega\cos{i}$ term in $\mathrm{d}\omega$), but
\citet{hegenuspl86} consider the dynamical node (the contribution of $\mathrm{d}h\cos{i_1}$ in $\mathrm{d}\omega$). 
We note also, that in the inclined rotation case $i_1$ is small, which justifies the omission of the $\cos{i_1}$ multiplicator in
the work of \citet{hegenuspl86}.

\subsubsection{Non edge-on orbits: further observational effects from nodal motion ($\omega$, $u_\mathrm{m}$)}
We know continue with allowing non-exactly edge-on orbits (i.e.~$\cos{i}\neq0$). We then have to take into
account the $\mathrm{d}\Omega\cos{i}$ terms, too. According to Eq.~(\ref{eq:dOmegadg}), as far as we omit the
small amplitude term $\mathrm{d}i_1$ this can be written as a function of $\mathrm{d}h\cos{i_1}$. The detailed
calculations together with the results up to fifth order in $e$ and $E,{\cal{E}}$ are given in Appendix~\ref{appendix} (see Eqs.~[\ref{eq:omegaapp}]--[\ref{eq:Onmp6}]).
Here we list the form of the result up to the fourth order in the secular term, and to the third order in the trigonometric ones. Namely,
\begin{equation}
\omega=g+u_\mathrm{m},
\end{equation}
where $g$ is given by (\ref{eq:gharmad}), while
\begin{eqnarray}
u_\mathrm{m}&=&(u_\mathrm{m})_0^*+U_0{\cal{G}}-\frac{1}{2}\frac{A_\mathrm{n2}}{\Pi^*}(1+{\cal{C}}_0)\sin2{\cal{G}}+{\cal{C}}_1\sin(h_0^*+H_0{\cal{G}}) \nonumber \\
&&+\frac{1}{2}{\cal{C}}_2\sin(2h_0^*+2H_0{\cal{G}})+\frac{1}{3}{\cal{C}}_3\sin(3h_0^*+3H_0{\cal{G}}),
\end{eqnarray}
where
\begin{equation}
U_0=-\frac{1}{\Pi^*}\left\{A_\mathrm{n1}+A_\mathrm{n2}\left(1+\frac{1}{2}E\right)+\frac{1}{4}e_0\frac{\mathrm{d}A_\mathrm{n2}}{\mathrm{d}e}{\cal{E}}\right\}(1+{\cal{C}}_0).
\end{equation} 
The ${\cal{C}}_{0-3}$ terms are coming from the Taylorian expansion of $(1-\cos^2{i})^{-1}$ which arises in
the equation for $\mathrm{d}\Omega\cos{i}$. These are trigonometric functions of $I_0$ and $i_1$. They
are listed only in Appendix~\ref{appendix}, Eqs.~(\ref{eq:dOmegacosi})--(\ref{eq:CSnndef}), (\ref{eq:C0def}), (\ref{eq:C1def}).
In our calculations we consider ${\cal{C}}_n$ as it would be $n$-th order in $e$. 

\begin{figure*}
\centering
\includegraphics[width=8.5cm]{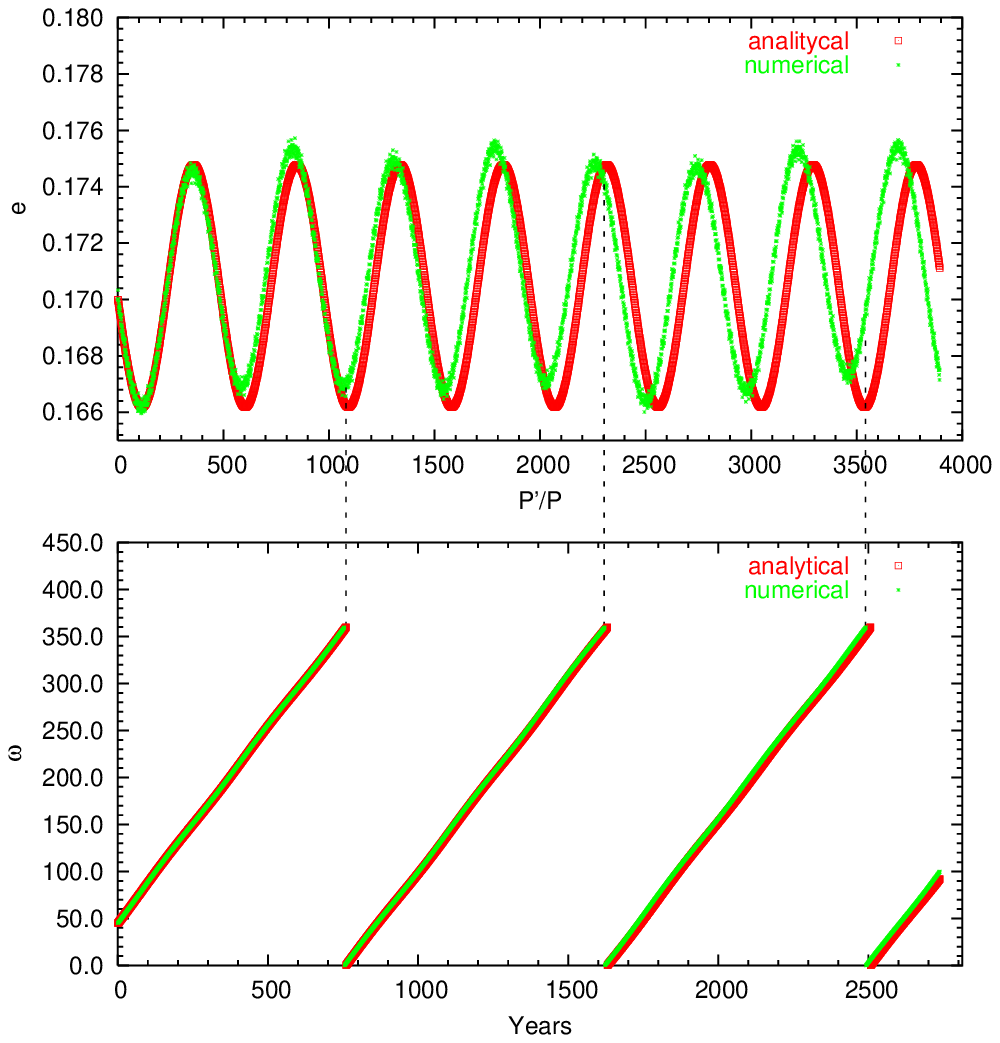}\includegraphics[width=8.5cm]{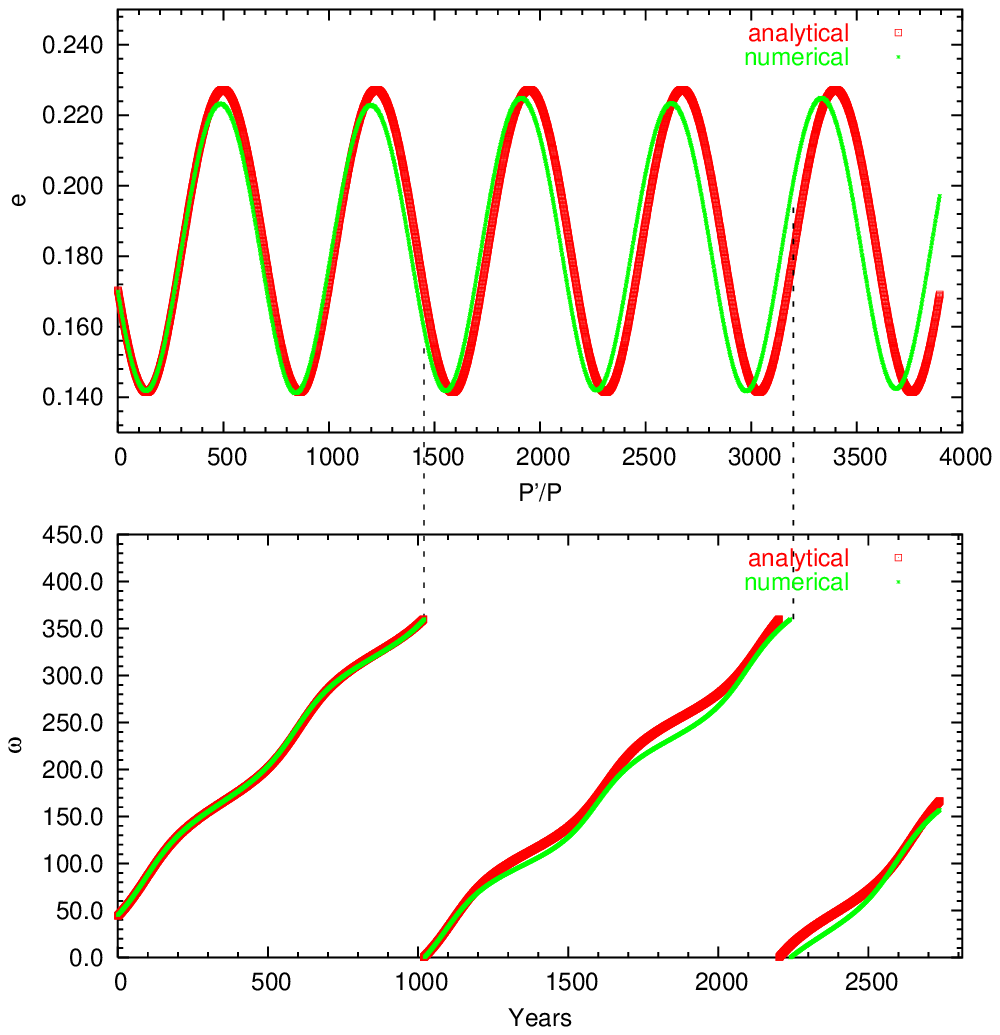}
\caption{The variation of the eccentricity and the observable argument of periastron in the case of the \object{AS Camelopardalis}
due to the perturbations of a third body. The initial mutual inclinations are $i_\mathrm{m}=20\degr$ (left panel), 
and $i_\mathrm{m}=60\degr$ (right panel). The dashed lines connect the $\omega=2\pi$ values with the corresponding eccentricity,
illustrating that the period of the $e$-cycles and the observable apsidal motion is different. (See text for an explanation of 
the discrepancy between the numerical and analytical results.)}
\label{fig:numanal2060}
\end{figure*}

Here we would like to stress that in this case a further important difference to those mentioned earlier is 
that there will no longer be one true period for the apsidal advance in the observational system, now we can speak only about only quasi-periodicity.
(This latter statement applies even better for higher orders where the different linear combinations of the angular
velocities of $g$ and $h$ appear.)
One can see this clearly in Fig.~\ref{fig:numanal2060}, where we plotted the variations of the eccentricity and
the observable argument of periastron ($\omega$) in a hypothetical eccentric triple system for two different initial mutual
inclinations ($i_\mathrm{m}=20\degr$, $60\degr$) of the close and the wide orbital planes. (The initial parameters are listed in Tables~\ref{tab:fixparams} and \ref{tab:inputparams}.) 
In these figures we connected the $\omega=2\pi$ values with the corresponding eccentricity values by dashed lines. Here it is clearly
visible, that the periods are not equal to the corresponding $e$ and $\omega$. It can be seen also that in the
case of $i_\mathrm{m}=20\degr$ (left panel) the eccentricity obtained from numerical integration (see next section) significantly
departs from the analytical value, while the argument of periastron ($\omega$) shows a good fit, at least as far as
one omits the small cumulative error in the period. On the contrary in the case of $i_\mathrm{m}=60\degr$ (right
panel) eccentricity curves produce very satisfactory fits, while the analytical $\omega$ departs from the
numerical curve suddenly after approx. 1000 years. In the first case, the explanation can be found in the
small mutual inclination, in which case, as was mentioned earlier, the higher order terms in the perturbative
forces are also important, while in the second case the departure arises from the fact, that due to the large
amplitude orbital precession the $\cos{i}^{-1}$ term becomes so large that our approximation will no longer be satisfactory.

\subsubsection{Direct perturbations in the orbital motion}
We also calculate the direct terms for higher accuracy. This is slightly problematic,
as due to the very strong $e$-dependence, the derivatives of $A_\mathrm{d}$ could increase to very large values already for
medium eccentricities. Fortunately, if the time differences of the two types of minima are used instead of the usual
O--C function, these direct terms will fall out, as at the same time they have the same value for primary and secondary minima.
Nevertheless, we use them in the following, considering the quantities derived from the derivatives of $A_\mathrm{d}$ as
first order, and those from $B_\mathrm{d}$ as second order (in our present sample configuration of \object{AS~Cam}
this can be done approx. up to $e'=0.3$ for perpendicular orbits). The results up to fifth order can be found in
Eqs.~(\ref{eq:vkozvapp})--(\ref{eq:D8}), while up to third order are listed below:
\begin{eqnarray}
\delta&=&\delta_0^*+D_0{\cal{G}} \nonumber \\
&&\!+\left\{-\frac{1}{4}\left[V_1(1+\epsilon_0)+V_2\epsilon_0\right]{\cal{E}}+\frac{1}{2}W(1+\epsilon_0)\right\}\sin2{\cal{G}} \nonumber \\
&&\!+\!\left[\frac{1}{64}\left(V_1+V_2\right){\cal{E}}^2-\frac{1}{32}V_1E{\cal{E}}-\frac{1}{16}W_1{\cal{E}}+\frac{1}{8}WE\right]\!\sin4{\cal{G}}, \nonumber \\
\end{eqnarray}
where
\begin{eqnarray}
D_0&=&\!\!\frac{1}{\Pi^*}\left[A_\mathrm{d}+e_0\frac{\mathrm{d}A_\mathrm{d}}{\mathrm{d}e}\left(\epsilon_0+\frac{1}{4}E{\cal{E}}\right)\!+\!e_0^2\frac{\mathrm{d}^2A_\mathrm{d}}{\mathrm{d}e^2}\left(\frac{1}{16}{\cal{E}}^2+\frac{1}{2}V_2\epsilon_0^2\right)\right.\nonumber \\
&&\left.-\frac{1}{2}B_\mathrm{d}E-\frac{1}{4}e_0\frac{\mathrm{d}B_\mathrm{d}}{\mathrm{d}e}(\epsilon_0+{\cal{E}})-\frac{1}{4}e_0^2\frac{\mathrm{d}^2B_\mathrm{d}}{\mathrm{d}e^2}\epsilon_0\right],
\end{eqnarray} 
while
\begin{eqnarray}
V&=&\frac{1}{\Pi^*}A_\mathrm{d}, \\
V_1&=&e_0\frac{1}{\Pi^*}\frac{\mathrm{d}A_\mathrm{d}}{\mathrm{d}e}, \\
V_2&=&e_0^2\frac{1}{\Pi^*}\frac{\mathrm{d^2}A_\mathrm{d}}{\mathrm{d}e^2}, \\
W&=&\frac{1}{\Pi^*}B_\mathrm{d}, \\
W_1&=&e_0\frac{1}{\Pi^*}\frac{\mathrm{d}B_\mathrm{d}}{\mathrm{d}e}, \\
W_2&=&e_0^2\frac{1}{\Pi^*}\frac{\mathrm{d^2}B_\mathrm{d}}{\mathrm{d}e^2}.
\end{eqnarray}

\subsubsection{Generalized form of the O--C}
Finally, now we report the generalized form of the O--C curve, which is as follows:
\begin{eqnarray}
\frac{2\pi}{P}O-C&\approx&V_{100}\cos[\omega_0^*+(1+U_0){\cal{G}}] \nonumber \\
&&+V_{200}\sin[2\omega_0^*+(2+2U_0){\cal{G}}] \nonumber \\
&&+V_{300}\cos[3\omega_0^*+(3+3U_0){\cal{G}}] \nonumber \\
&&+V_{1-20}\cos[\omega_0^*-(1-U_0){\cal{G}}]\nonumber \\
&&+V_{120}\cos[\omega_0^*+(3+U_0){\cal{G}}] \nonumber \\
&&+V_{1-40}\cos[\omega_0^*-(3-U_0){\cal{G}}] \nonumber \\
&&+V_{140}\cos[\omega_0^*+(5+U_0){\cal{G}}]\nonumber \\
&&+V_{2-20}\sin[2\omega_0^*+2U_0{\cal{G}}]\nonumber \\
&&+V_{220}\sin[2\omega_0^*+(4+2U_0){\cal{G}}] \nonumber \\      
&&\nonumber\\
&&+V_{10-1}\cos[\omega_0^*-h_0^*+(1+U_0-H_0){\cal{G}}] \nonumber \\           
&&+V_{101}\cos[\omega_0^*+h_0^*+(1+U_0+H_0){\cal{G}}] \nonumber \\
&&+V_{10-2}\cos[\omega_0^*-2h_0^*+(1+U_0-2H_0){\cal{G}}] \nonumber \\
&&+V_{102}\cos[\omega_0^*+2h_0^*+(1+U_0+2H_0){\cal{G}}] \nonumber \\
&&+V_{1-2-1}\cos[\omega_0^*-h_0^*-(1-U_0+H_0){\cal{G}}] \nonumber \\
&&+V_{1-21}\cos[\omega_0^*+h_0^*-(1-U_0-H_0){\cal{G}}] \nonumber \\
&&+V_{12-1}\cos[\omega_0^*-h_0^*+(3+U_0-H_0){\cal{G}}] \nonumber \\
&&+V_{121}\cos[\omega_0^*+h_0^*+(3+U_0+H_0){\cal{G}}]\nonumber \\
&&+V_{20-1}\sin[2\omega_0^*-h_0^*+(2+2U_0-H_0){\cal{G}}] \nonumber \\
&&+V_{201}\sin[2\omega_0^*+h_0^*+(2+2U_0+H_0){\cal{G}}]\nonumber\\
&&\nonumber \\
&&+V_{020}\sin2{\cal{G}} \nonumber \\
&&+V_{040}\sin4{\cal{G}} \nonumber \\
\nonumber\\
&&+V_{001}\sin[h_0^*+H_0{\cal{G}}]\nonumber \\
&&+V_{002}\sin[2h_0^*+2H_0{\cal{G}}]\nonumber \\
&&+V_{003}\sin[3h_0^*+3H_0{\cal{G}}]+{\cal{O}}(e^4,E^4), \nonumber \\
\label{eq:OminCanalvegleg}
\end{eqnarray}
where
\begin{eqnarray}
V_{100}&=&-2je_0\left(1-\frac{1}{4}{\cal{C}}_1^2\right)(1+\epsilon_0), \\
V_{1-20}&=&je_0\left[\frac{1}{2}({\cal{E}}+E)(1+\epsilon_0)-\frac{1}{4}A_1{\cal{E}}\right], \\
V_{120}&=&je_0\left[\frac{1}{2}({\cal{E}}-E)(1+\epsilon_0)+\frac{1}{4}A_1{\cal{E}}\right], \\
V_{1-40}&=&\frac{1}{16}je_0(-{\cal{E}}^2+E^2), \\
V_{140}&=&je_0\left(-\frac{1}{16}{\cal{E}}^2+\frac{1}{4}E{\cal{E}}-\frac{3}{16}{\cal{E}}^2\right), \\
V_{200}&=&\frac{3}{4}e_0^2(1+2\epsilon_0), \\
V_{2-20}&=&-\frac{3}{8}e^2({\cal{E}}+E), \\
V_{220}&=&\frac{3}{8}e^2(-{\cal{E}}+E), \\
V_{300}&=&\frac{1}{3}je_0^3, 
\end{eqnarray}
\begin{eqnarray}
V_{10\pm1}&=&\mp je_0{\cal{C}}_1(1+\epsilon_0), \\
V_{10-2}&=&-\frac{1}{4}je_0({\cal{C}}_1^2-2{\cal{C}}_2), \\
V_{102}&=&\frac{1}{4}je_0({\cal{C}}_1^2+2{\cal{C}}_2), \\
V_{1-2\pm1}&=&\mp\frac{1}{4}je_0{\cal{C}}_1({\cal{E}}+E), \\
V_{12\pm1}&=&\pm\frac{1}{4}je_0{\cal{C}}_1({\cal{E}}-E), \\
V_{20\pm1}&=&\pm\frac{3}{4}e_0^2{\cal{C}}_1, 
\end{eqnarray}
\begin{eqnarray}
V_{020}&=&-\frac{1}{4}\left[V_1(1+\epsilon_0)+V_2\epsilon_0\right]{\cal{E}}+\frac{1}{2}W(1+\epsilon_0)+\frac{1}{2}M{\cal{C}}_0, \\
V_{040}&=&\frac{1}{64}(A_1+A_2){\cal{E}}^2-\frac{3}{32}A_1E{\cal{E}}, \\
V_{00n}&=&-\frac{1}{n}{\cal{C}}_n.
\end{eqnarray}
Furthermore,
\begin{equation}
M=\frac{1}{\Pi^*}A_\mathrm{n2}.
\end{equation}
A more detailed result up to fifth order containing 102 trigonometric terms is also listed in Appendix~\ref{appendix}, Eqs.~(\ref{eq:O-Cfinalaotig})--(\ref{eq:V_005}). 
Here we note, that the indices refer to the multiplicators of $\omega$, $g$, and $h$ in the trigonometric terms, respectively.
We separated the different terms by blank lines. All four groups besides the perturbations in $e$ and $g$, give the additional contribution 
of the precession of the orbital plane with respect to the observer.\footnote{We do not use intentionally the usual phrase 
`nodal regression', as we concentrate on the observational frame of reference, where
we can observe even nodal progression (see Sect.~\ref{sec:numstudies}).} The first, and second groups give the nodal contribution to
the observable argument of periastron ($\omega$), while the third one comes both directly from the $\Omega\cos{i}$, and
the direct orbiatl motion terms in $\dot{u}$. Finally, the fourth one also contains the contribution of $\Omega\cos{i}$. These latter two terms,
naturally have $e$-independent parts, consequently they do not disappear even in the case of a
circular inner orbit. Nevertheless, as was mentioned above, due to the $\cos{i}$ multiplicator,
their significance is very limited in the case of the relatively wider eccentric eclipsing binaries. Similarly, due to
this latter condition, the contribution of the second group is also very minimal, as these terms arise exactly from the same
reason as the terms of the fourth groups, manifesting the same precession indirectly, through the variation of $\omega$.
On the contrary, the first group contains such terms of the nodal motion which remain effective even in the edge-on $\cos{i}=0$
case, too. These are coming from the $h\cos{i_1}$ term as was discussed in Sects.~\ref{subsubsect:analO-C0} and \ref{subsubsect:egh}.

\begin{figure*}
\centering
\includegraphics[width=8.5cm]{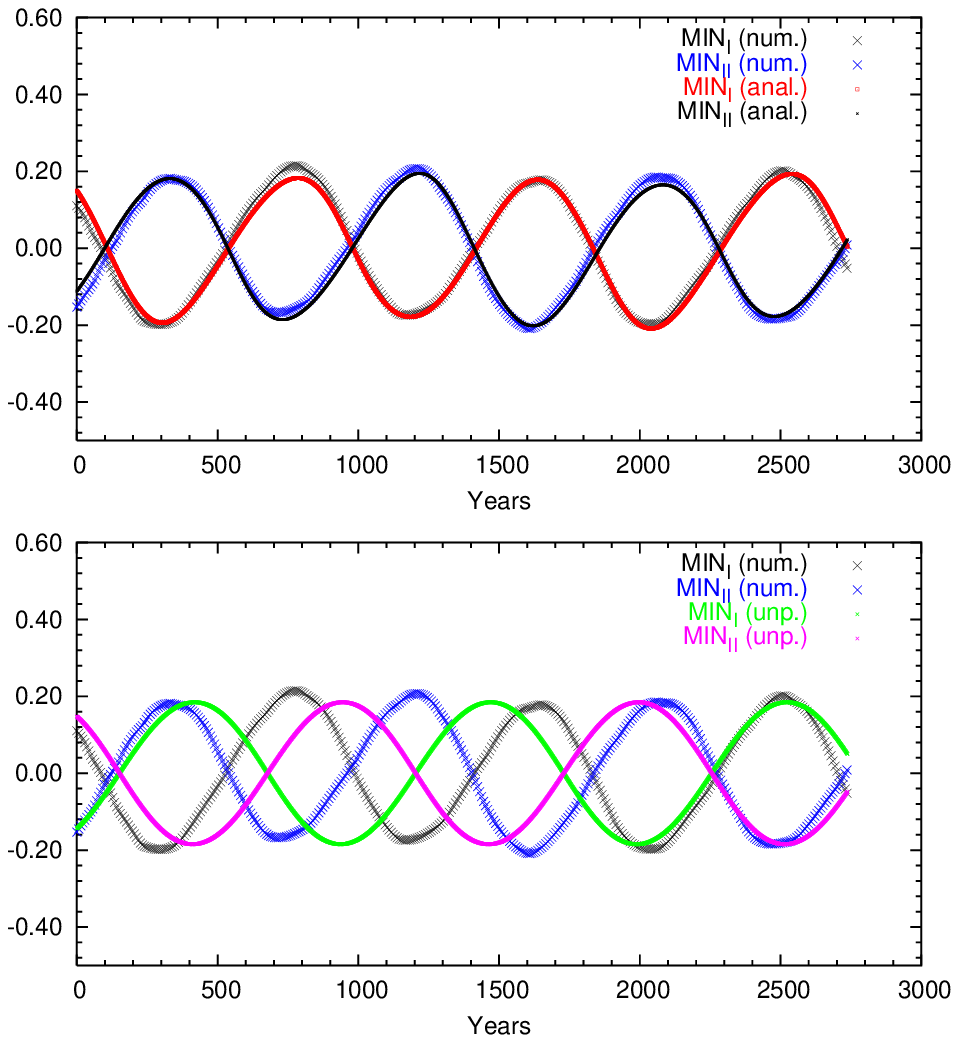}\includegraphics[width=8.5cm]{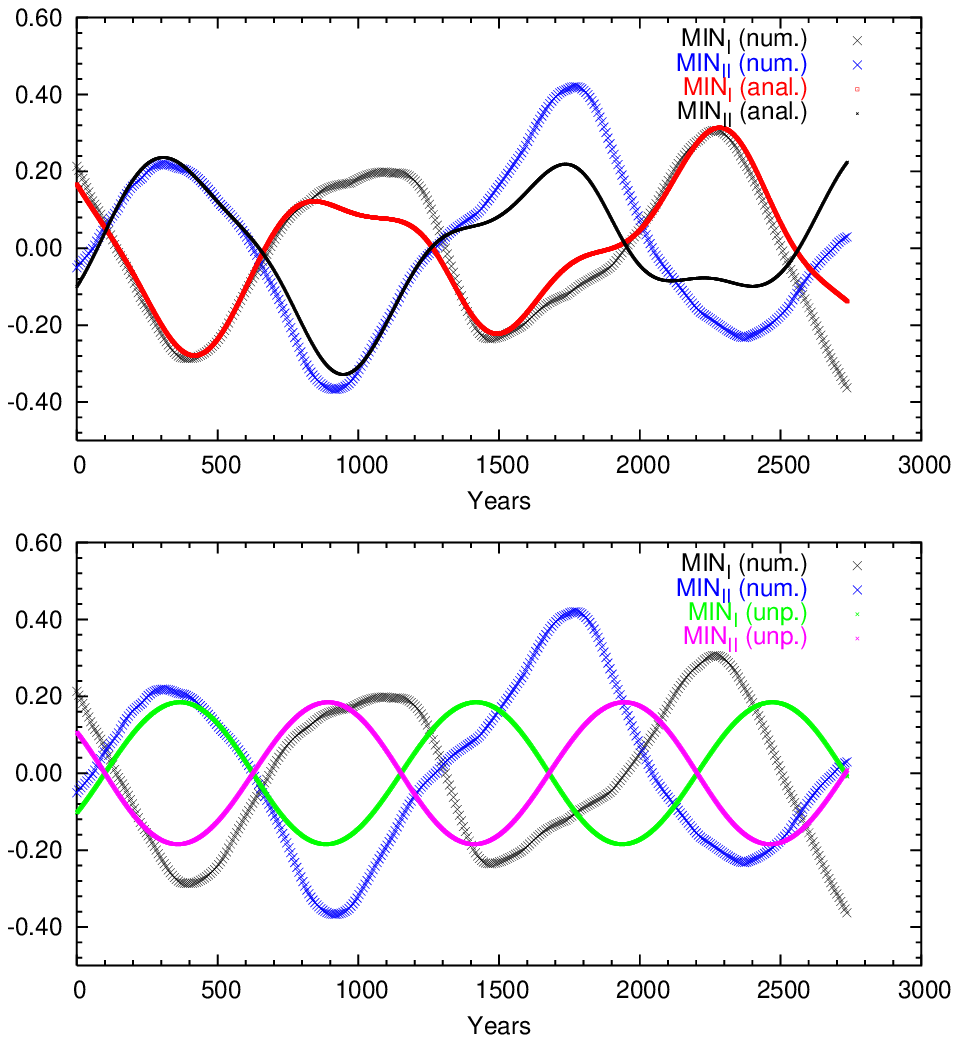}
\caption{The numerically (AS2 and AS3 runs -- see next section) and analytically generated O--C curve in the case of the \object{AS Camelopardalis}
with the perturbations of a third body. The initial mutual inclinations are $i_\mathrm{m}=20\degr$ (left panel), 
and $i_\mathrm{m}=60\degr$ (right panel). Upper panels show the numerically generated O--Cs, as well as the analytical ones
calculated according to the formulae of the present paper. Lower panels demonstrate the difference between the
numerical (i.e. ``observed'') curves, and the only tidally forced apsidal motion produced O--C.}
\label{fig:omincnumanal2060}
\end{figure*}

To illustrate the net effect to the O--C we plotted in Fig.~\ref{fig:omincnumanal2060} the O--C diagram of \object{AS~Camelopardalis}
for the same two configurations for which the variations of $e$ and $\omega$ were shown in Fig.~\ref{fig:numanal2060}.
The upper panels represent the numerical results derived directly from the numerical integrations as well as analytical results
calculated according to the higher order formula listed in Appendix~\ref{appendix}, Eq.~(\ref{eq:O-Cfinalaotig}). 
The lower panels represent the numerical curve and the ``unperturbed'' (i.e. only two-body tidal distortion is present, as usual)
theoretical curve where both the constant eccentricity and apsidal motion period were deduced from the ``measured''
$e_0=e(t_0)$ value. We note, that the significant departure of the numerical and the analytical curves which occurs in the $i_\mathrm{m}=60\degr$
case after approximately 700 years (upper right panel) should arise from the fact that due to the variation of the observable
inclination ($i$) for this time our fundamental principle, i.e. Eq.~(\ref{eq:fundamental}) ceases to be correct.
Nevertheless, at this value of $i$, our system is already no longer an eclipsing one, consequently, this situation is
out of our scope.

\begin{figure*}
\centering
\includegraphics[width=8.5cm]{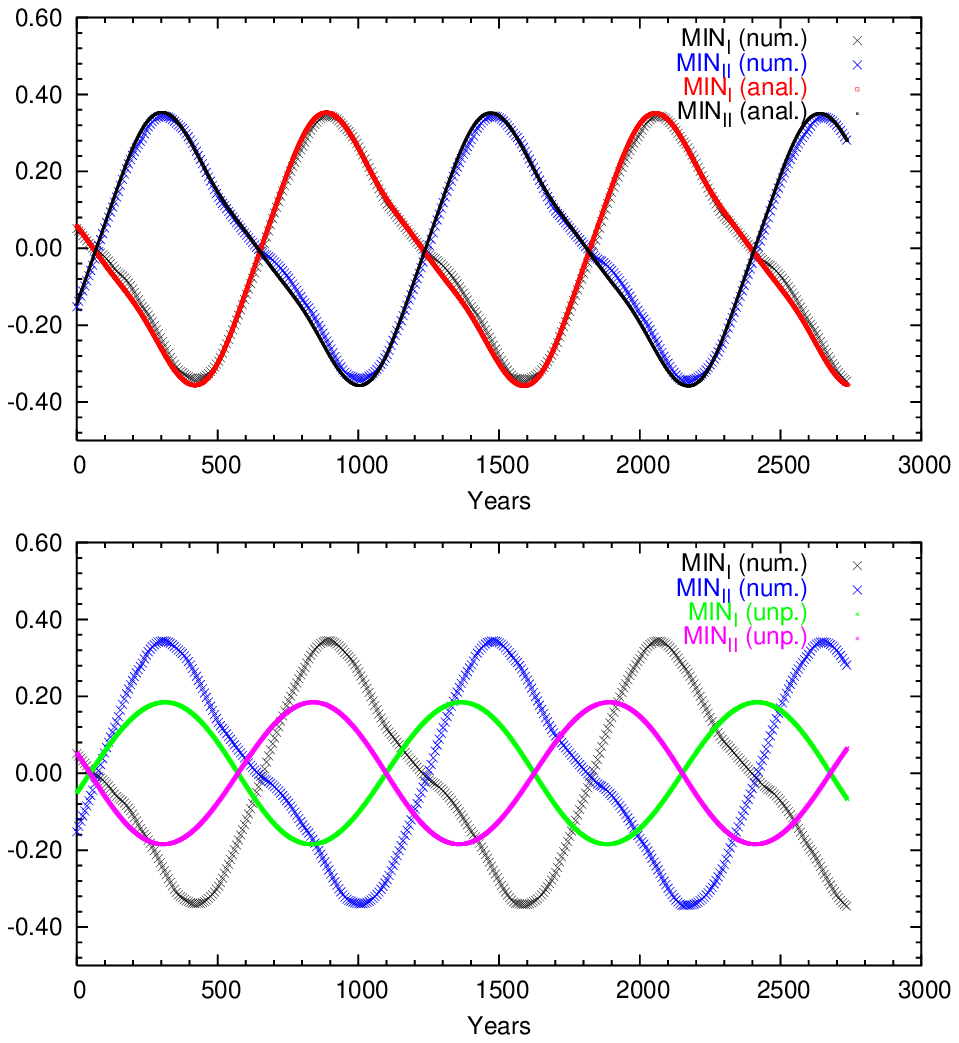}\includegraphics[width=8.5cm]{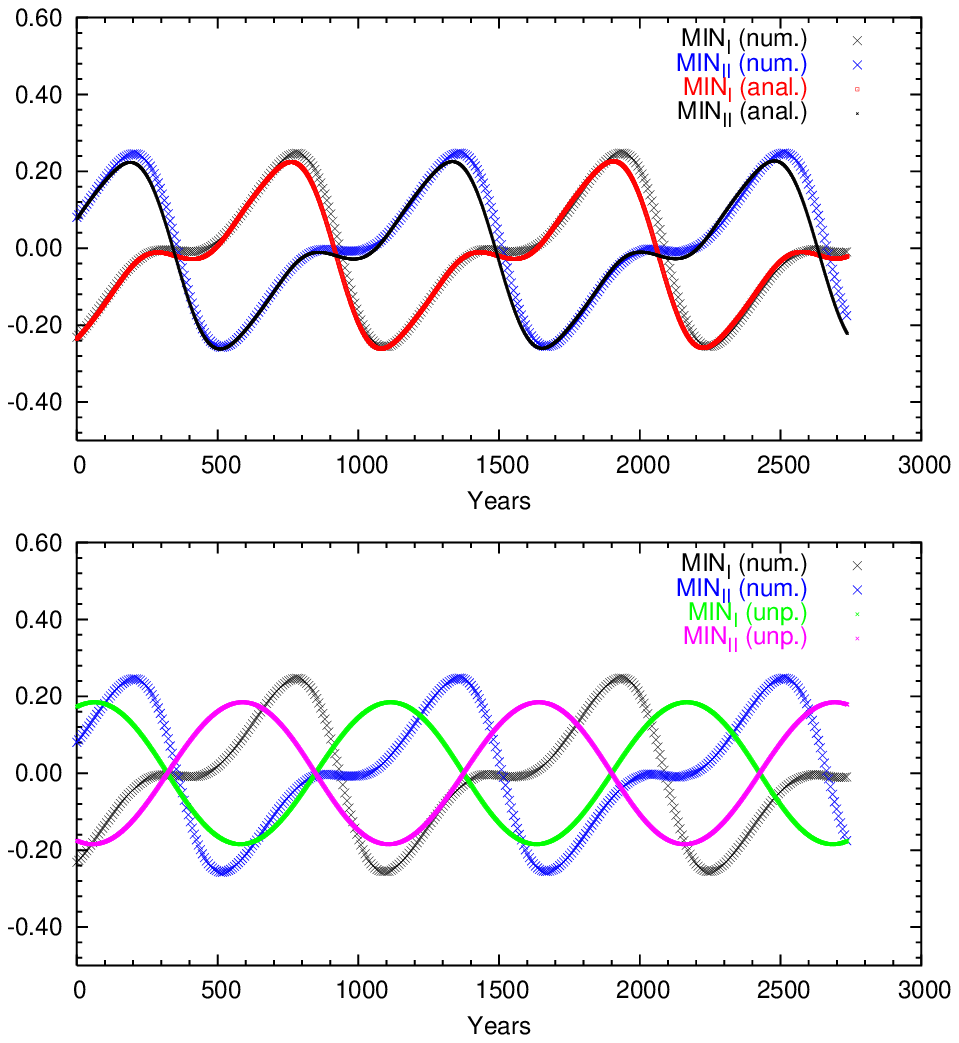}
\caption{As in Fig.~\ref{fig:omincnumanal2060}, but the initial mutual inclinations are $i_\mathrm{m}\approx90\degr$ for both panels
(AS4 and AS4b runs -- see next section).
The only difference is that in the left panel the orbital plane of the tertiary is almost perpendicular to the plane of the
sky, while in the right panel this nearly coincides with it. Note that at this second configuration the amplitude
of the O--C is definitely smaller than in the first case.}
\label{fig:omincnumanal9090+}
\end{figure*}

\subsection{Discussion on observational effects}
\begin{figure}
\centering
\includegraphics[width=8.5cm]{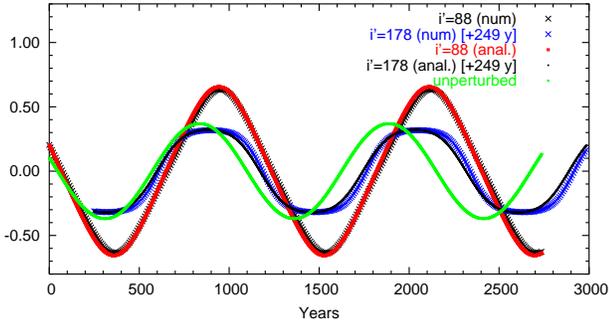}
\caption{The difference of the O--C of the primary and the secondary minima for the O--Cs plotted in Fig.~\ref{fig:omincnumanal9090+}.
We shifted the $i'=178\degr$ curve with approx. 249 years in order to coincide with the ``unperturbed'' curve of the two cases.}
\label{fig:ketO-C}
\end{figure}

When only a small fraction of the O--C curve is observed, several serious problems occur in the sense of the determination
of the apsidal motion period with respectable accuracy. To demonstrate this we consider
only the $V_{100}$ and $V_{1-20}$ terms (we omit $V_{120}$ because we concentrate mainly on such configurations where
the mutual inclination of the inner and the outer orbit is large, and consequently, we suppose that ${\cal{E}}\approx E$).
Furthermore, when the period of the nodal motion is significantly longer than the tidally induced apsidal motion 
(as in the case of \object{AS~Cam}, but not necessarily in e.g. \object{DI~Her}), then $u_\mathrm{m}$,
as well as $h$ can be considered as constant. Then instead of the usual O--C we apply the difference function
of the primary and secondary minima, as
\begin{eqnarray}
\Delta&\approx&\frac{P}{\pi}e_0\left\{-2\cos[\omega_0^*+{\cal{G}}_0+\Pi^*(u-u_0)]\right. \nonumber \\
&&\left.+E\cos[\omega_0^*-{\cal{G}}_0-\Pi^*(u-u_0)]\right\}.
\end{eqnarray}
As far as the nodal motion is neglected it can easily be seen that
\begin{equation}
\omega_0^*=u_\mathrm{m},
\end{equation}
furthermore, according to its definition
\begin{equation}
{\cal{G}}_0=g_0-\frac{1}{2}E\sin2g_0+... ,
\end{equation}
so,
\begin{eqnarray}
\omega_0^*+{\cal{G}}_0&\approx&\omega_0-\frac{1}{2}E\sin2g_0, \\
\omega_0^*-{\cal{G}}_0&\approx&\omega_0-2g_0+\frac{1}{2}E\sin2g_0.
\end{eqnarray}
We concentrate continuously on perpendicular orbital planes configuration. In this case, taking into account that the binary is an eclipsing one,
i.e. its observable inclination is close to 90$\degr$, the plane of the outer orbit should lie either close to the
plane of the sky, or perpendicular to both the planes of the inner orbit and the sky. In the previous case (which is
the more interesting one, because it makes it possible not to observe light-time effect) it is easy to see that
$\omega\approx g+[0,1]\times\pi$, while in the latter one $\omega\approx g\pm\pi/2$ (we also used the fact that
in these hierarchical systems the outer plane is close to the invariable plane). In the two different situations we get that
\begin{eqnarray}
\Delta_{0,\pi}&\approx&-2\frac{P}{\pi}e_0\left(1+\frac{1}{2}{\cal{E}}\cos2\omega_0-\frac{1}{2}{\cal{E}}\right)\cos\left[\omega_0+\Pi^*(u-u_0)\right], \nonumber \\
\label{eq:Delta0pi}\\
\Delta_{\pm\pi/2}&=&-2\frac{P}{\pi}e_0\left(1-\frac{1}{2}{\cal{E}}\cos2\omega_0+\frac{1}{2}{\cal{E}}\right)\cos\left[\omega_0+\Pi^*(u-u_0)\right] \nonumber \\
\label{eq:Deltapmpifel}
\end{eqnarray}
The subscript of $\Delta$ refers to the approximate value of $u_\mathrm{m}$ and we omitted the $\sin2g_0$ terms inside the arguments.
Eq.~(\ref{eq:Delta0pi}) shows a very important result. In the situation when a third body revolves around the eclipsing binary
in a nearly perpendicular plane which lies close to the plane of the sky, the amplitude of the $\Delta$ curve -- which
is used several times for the calculation of the period of the apsidal motion -- could be lower than it is expected
(by the multiplicator $0\geq\delta\geq1-{\cal{E}}$), consequently, the numerical fitting will result in a smaller angular velocity, i.e. longer period.
This is demonstrated in Figs.~\ref{fig:omincnumanal9090+} and \ref{fig:ketO-C}, where two numerically integrated O--C curves are shown (together with the corresponding
analytically calculated ones). There is no difference in the physical configuration of the entire triple system, 
only its orientation rotated with 90$\degr$ along the inner orbital plane, i.e. the outer orbital plane 
from an almost perpendicular position became nearly coincident with the plane of the sky.
Moreover, besides the smaller amplitude, Fig.~\ref{fig:ketO-C} reveals a more important result. As one can see, when the
orbital plane of the tertiary is close to the plane of the sky, large almost horizontal regions can be found in
the difference curve. (This means that in these intervals the primary and the secondary minima vary almost in the same manner.)
The mathematical cause can be understood from the third order approximation. In this case the $V_{100}$, $V_{1\pm20}$ and
the $V_{300}$ terms should be counted. Limiting ourselves only for the third-body-in-the-sky case, then
\begin{eqnarray}
\Delta_{0,\pi}&=&-2\frac{P}{\pi}e_0\left[\left(1-\frac{1}{2}{\cal{E}}+\frac{1}{2}{\cal{E}}\cos2\omega_0+\frac{1}{8}{\cal{E}}^2\right.\right.\nonumber \\
&&\left.-\frac{1}{4}{\cal{E}}^2\cos2\omega_0+\frac{1}{16}{\cal{E}}^2\cos4\omega_0+\frac{1}{8}A_1{\cal{E}}\right]\cos{\cal{G}} \nonumber \\
&&\left.-\left(\frac{1}{8}A_1{\cal{E}}+\frac{1}{6}e_0^2\right)\cos3{\cal{G}}\right].
\end{eqnarray}
The flat extrema occur when ${\cal{G}}=0+k\pi$. Around these intervals the two cosine terms have opposite signs.
This also happens in the ``unperturbed'' case of course. However, a significant difference is that, whilst we considered
both $A_1$ and ${\cal{E}}$ as formally having the same order than $e$, both could be somewhat larger than $e$, and
in some cases they can reach almost the order of $e^{1/2}$. Consequently, the superposition of the opposite sign $\cos3{\cal{G}}$ 
curve onto the $\cos{\cal{G}}$ may cause further flattening almost in the order of $e$.
These areas might cover even the 40--50\% of the total curve. It is trivial, that if this system were observed within
this time interval, the apsidal motion period would be found significantly longer than the theoretically expected value.
On the other side, there is no region, where the slope of the difference curve is significantly larger than the ``unperturbed''
reference curve. The consequences of the fact above will be discussed in Sect.~\ref{sec:discussion}.
 
We now study numerically the eclipsing system \object{AS~Camelopardalis}, 
where only about a 10\% or less of the total period is covered by observations.

\section{Numerical studies \label{sec:numstudies}}

We carried out several sets of integrations with the numeric integrator described in \citet{borkoetal04}.
These runs partly serve as numerical support for the analytical calculations described in the previous
section, and partly serve as study on the dynamical evolution of eccentric hierarchical triple systems
with and without dissipation. 
The initial parameters of the binary were taken from \citet{khodykinvedeneyev97}, with the exception of the
inner structure constants which were set according to the tables of \citet{claretgimenez91}.
The physical parameters of the three stars as well as those initial orbital elements which 
are the same in the different runs are listed in Table~\ref{tab:fixparams}, while those orbital elements
which differ in the individual runs are listed in Table~\ref{tab:inputparams}. In this latter
table we give the angular orbital elements both in the observational ($\omega$, $\omega'$, $\Omega'$, $i'$),
and the dynamical ($g$, $h$, $i_1$, $h'$, $i_2$) coordinate systems. Note, our input parameters for the
integrator are only the observational angular elements, and the others are only derived quantities. 
All columns of this latter table refer to four individual runs. One of them is the dissipationless case, starting
the binary from its periastron, and three of them with dissipation 
($\lambda_1=\lambda_2=10^{-4}$), with initial position of the binary in
its periastron, apastron, and at true anomaly $v\approx104\degr$, where
the instantaneous orbital angular velocity in the present configurations is nearly equal to the averaged one
(as we used the option of our code which makes the angular velocity vectors of the stellar rotation for both stars 
equally long and parallel to the initial instantaneous orbital angular velocity vector in an iterative way, 
it means that the initial rotation of the stars were synchronized in these three different ways.)
We emphasize again, that as we were not interested in it exclusively dynamical evolution, but for
observational consequences too, we carried out some runs where the dynamics of the triple were
the same, but its orientation with respect to the Earth were different (e.g. AS3, AS4 and AS3b, AS4b runs).

\subsection{Non-dissipative runs}

\begin{figure}
\centering
\includegraphics[width=3.8cm]{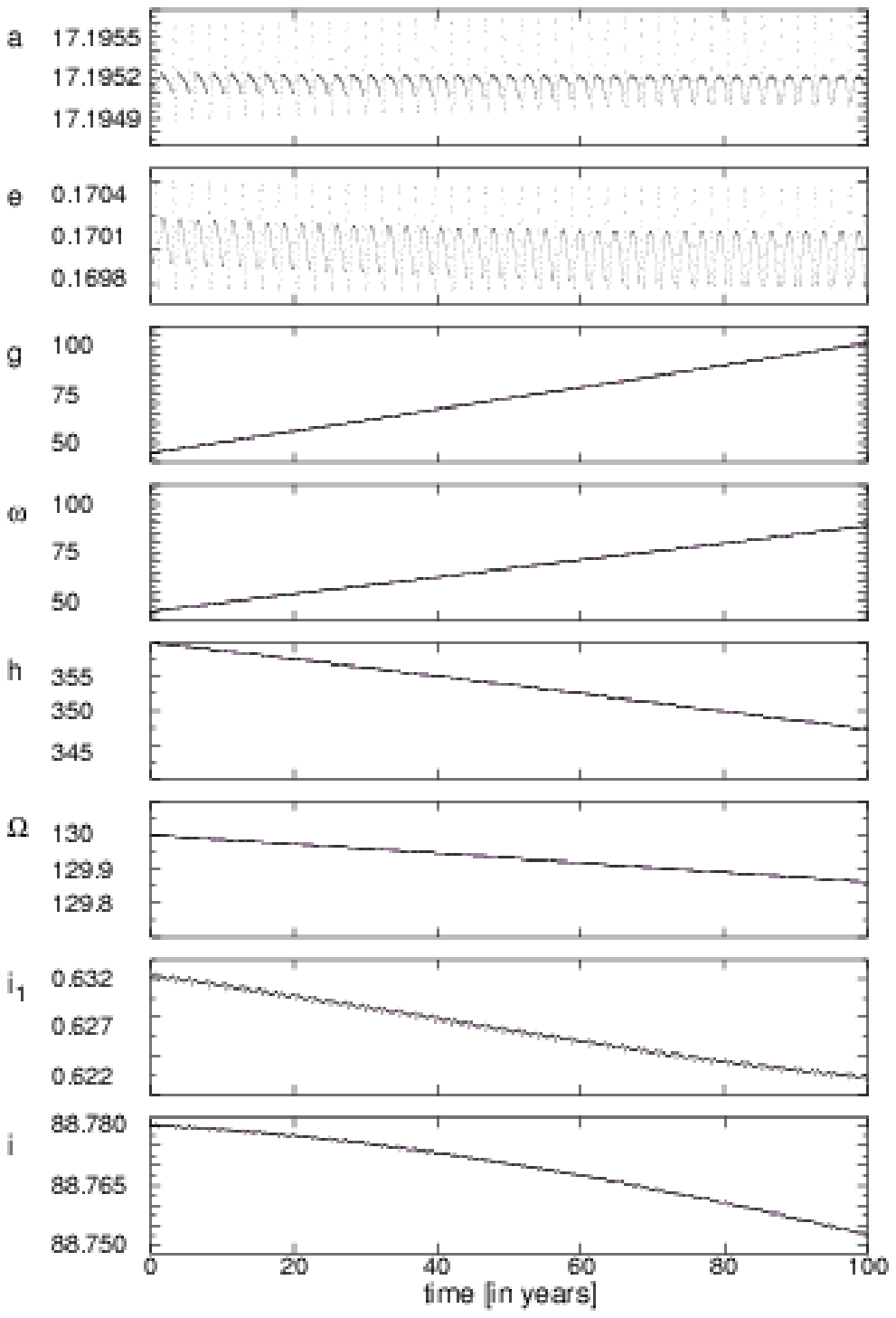}~~~~\includegraphics[width=3.8cm]{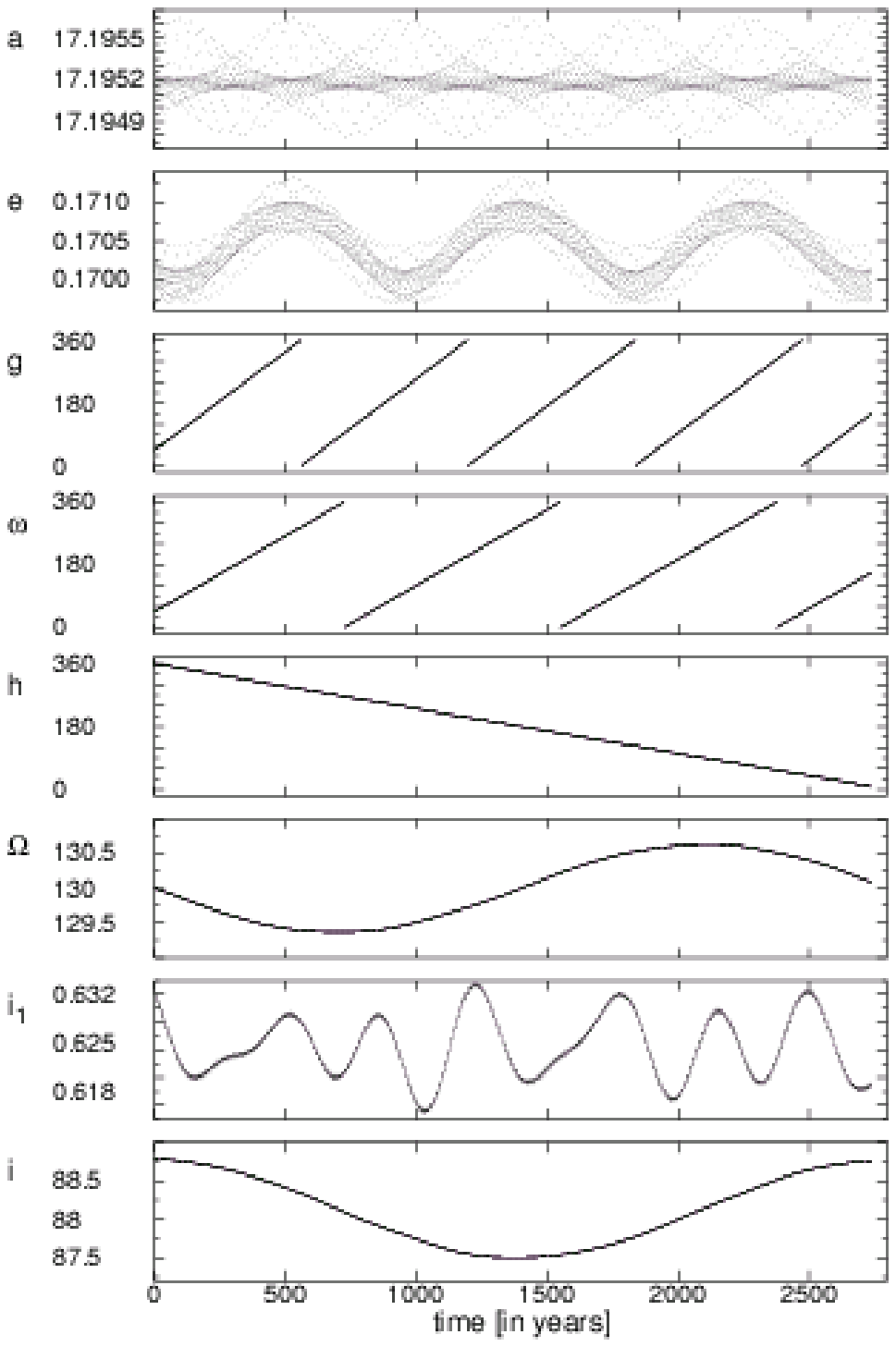}
\caption{The variation of both the dynamical and the observational orbital elements of the binary 
during 100 and approx. 2800 years (1 million days). Semi-major axis is given in $R_\odot$,
while the angular elements in degrees. The mutual inclination is $i_\mathrm{m}\approx0\fdg8$.
See Col.~AS1 of Table~\ref{tab:inputparams} for the initial values.}
\label{fig:AS1}
\end{figure}

\begin{figure}
\centering
\includegraphics[width=3.8cm]{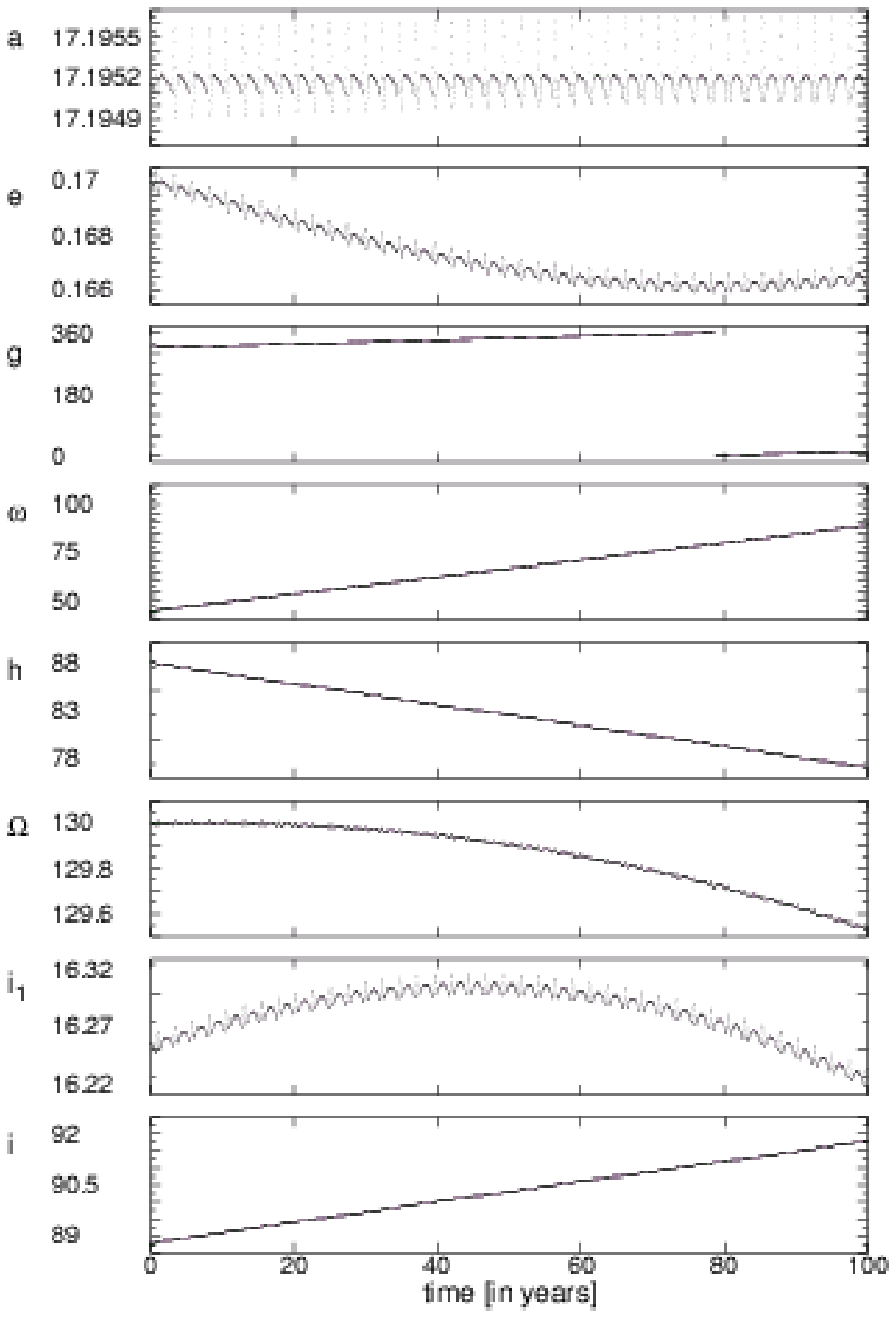}~~~~\includegraphics[width=3.8cm]{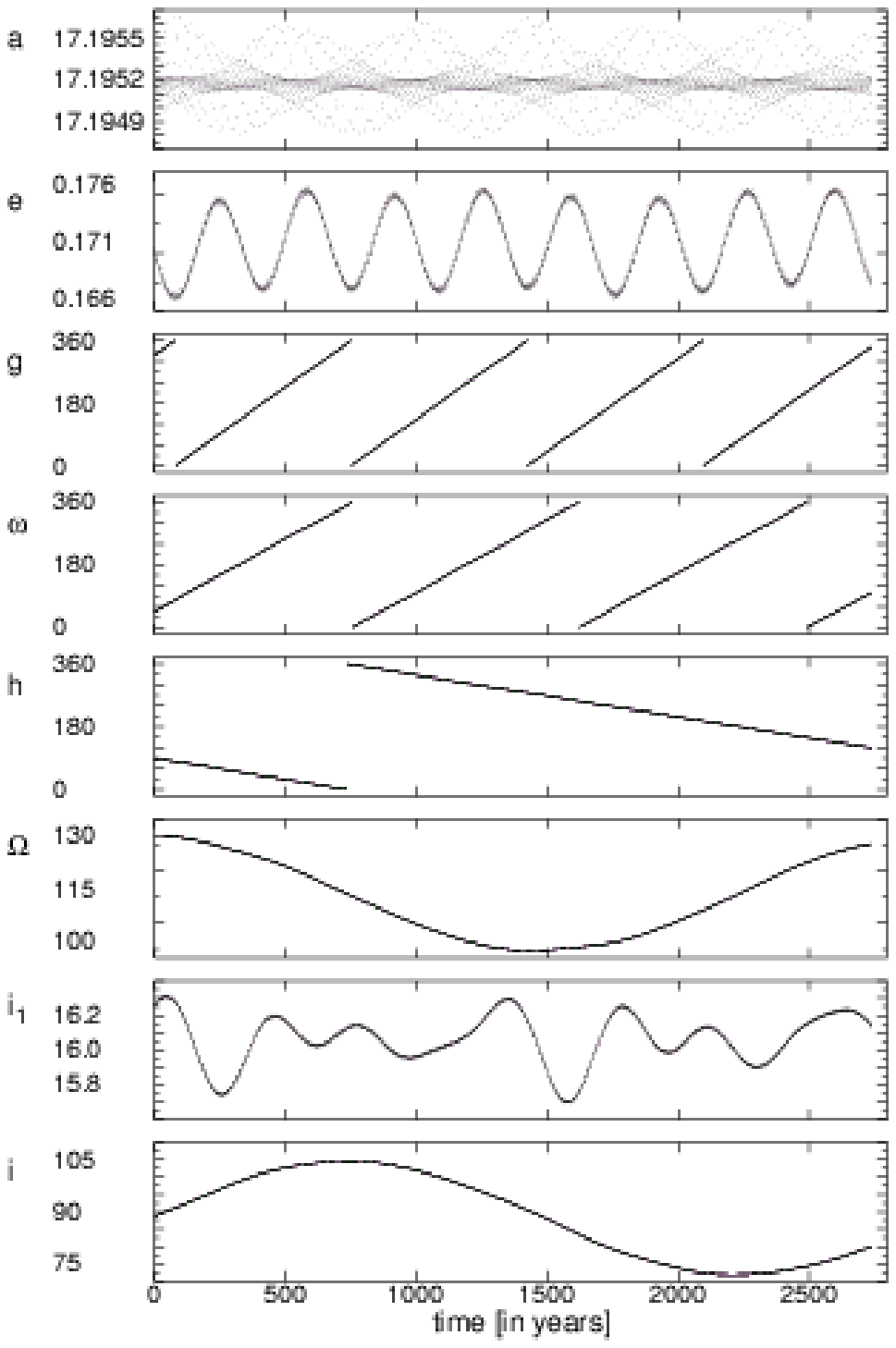}
\caption{The variation of both the dynamical and the observational orbital elements of the binary 
during 100 and approx. 2800 years (1 million days). Semi-major axis is given in $R_\odot$,
while the angular elements in degrees. The mutual inclination is $i_\mathrm{m}\approx20\degr$.
See Col.~AS2 of Table~\ref{tab:inputparams} for the initial values.}
\label{fig:AS2}
\end{figure}

\begin{figure}
\centering
\includegraphics[width=3.8cm]{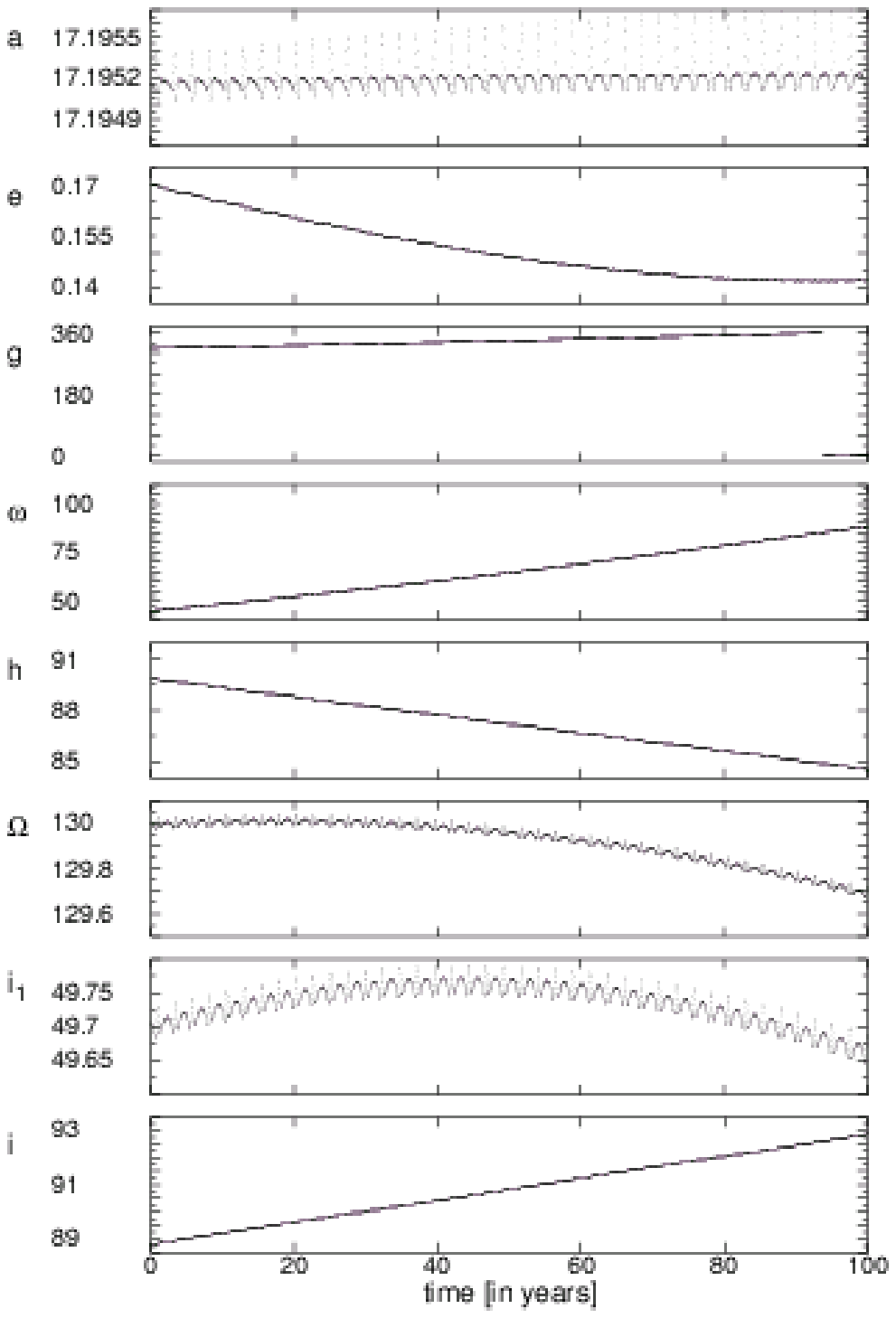}~~~~\includegraphics[width=3.8cm]{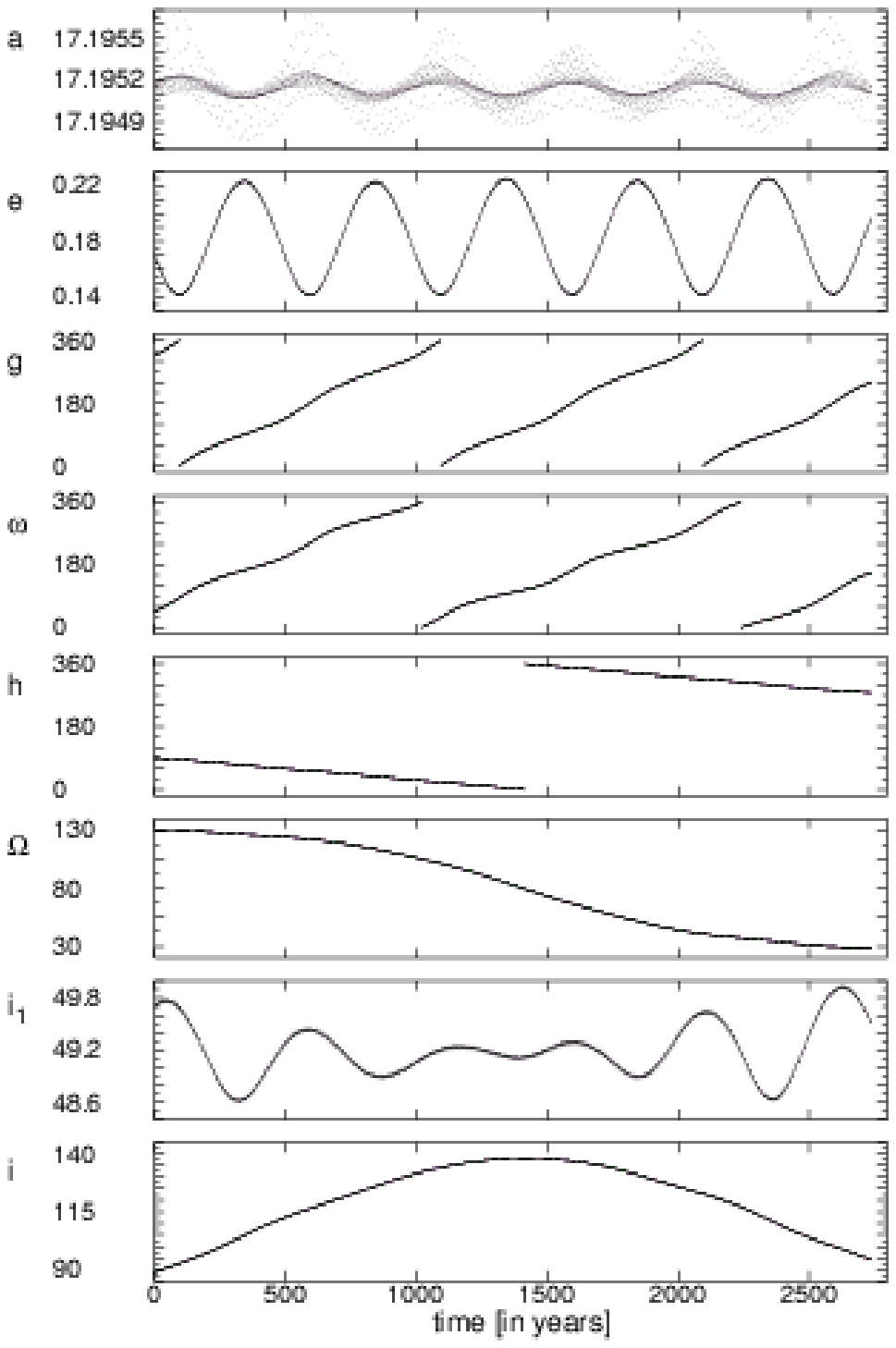}
\caption{The variation of both the dynamical and the observational orbital elements of the binary 
during 100 and approx. 2800 years (1 million days). Semi-major axis is given in $R_\odot$,
while the angular elements in degrees. The mutual inclination is $i_\mathrm{m}\approx60\degr$.
See Col.~AS3 of Table~\ref{tab:inputparams} for the initial values.}
\label{fig:AS3}
\end{figure}

\begin{figure}
\centering
\includegraphics[width=3.8cm]{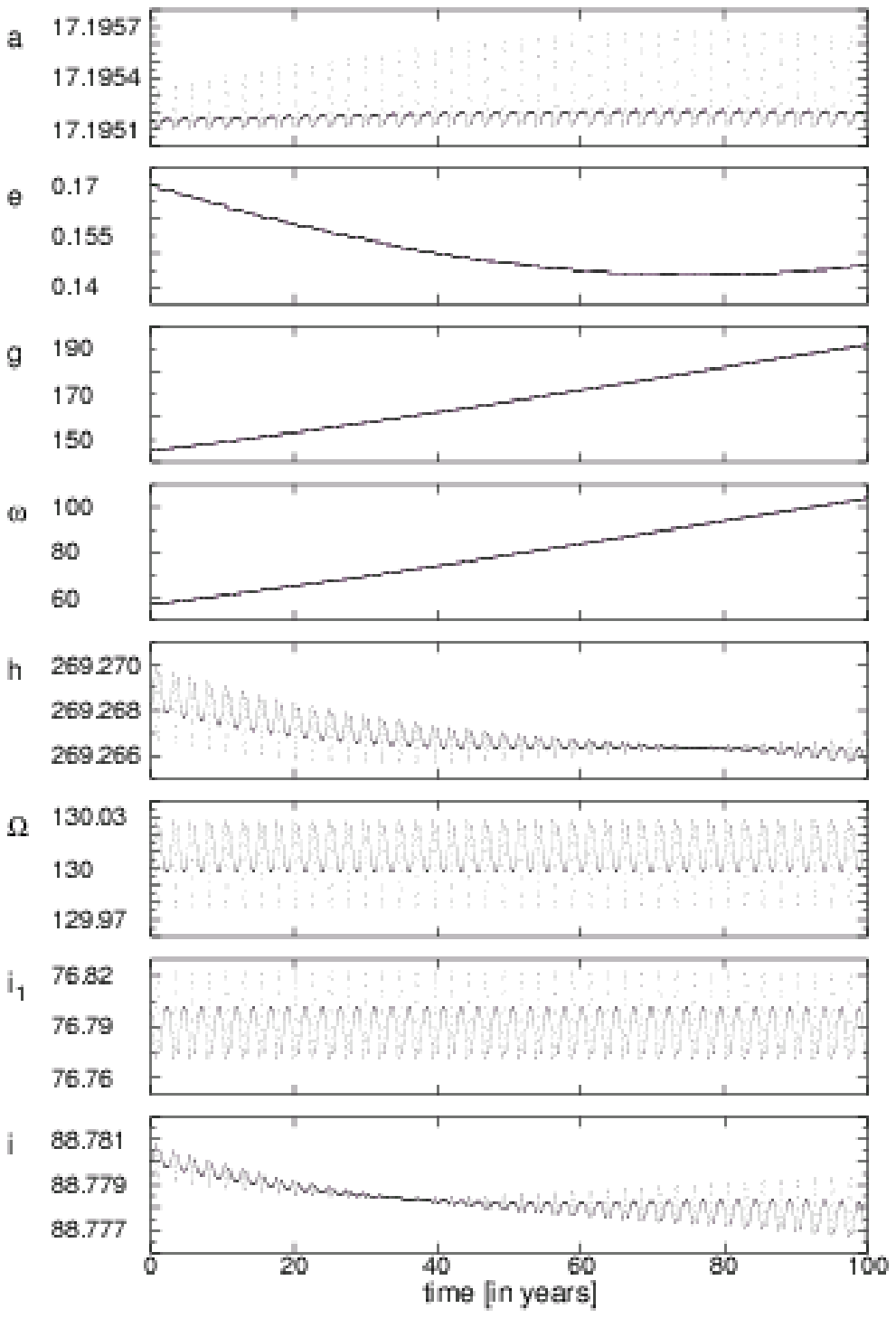}~~~~\includegraphics[width=3.8cm]{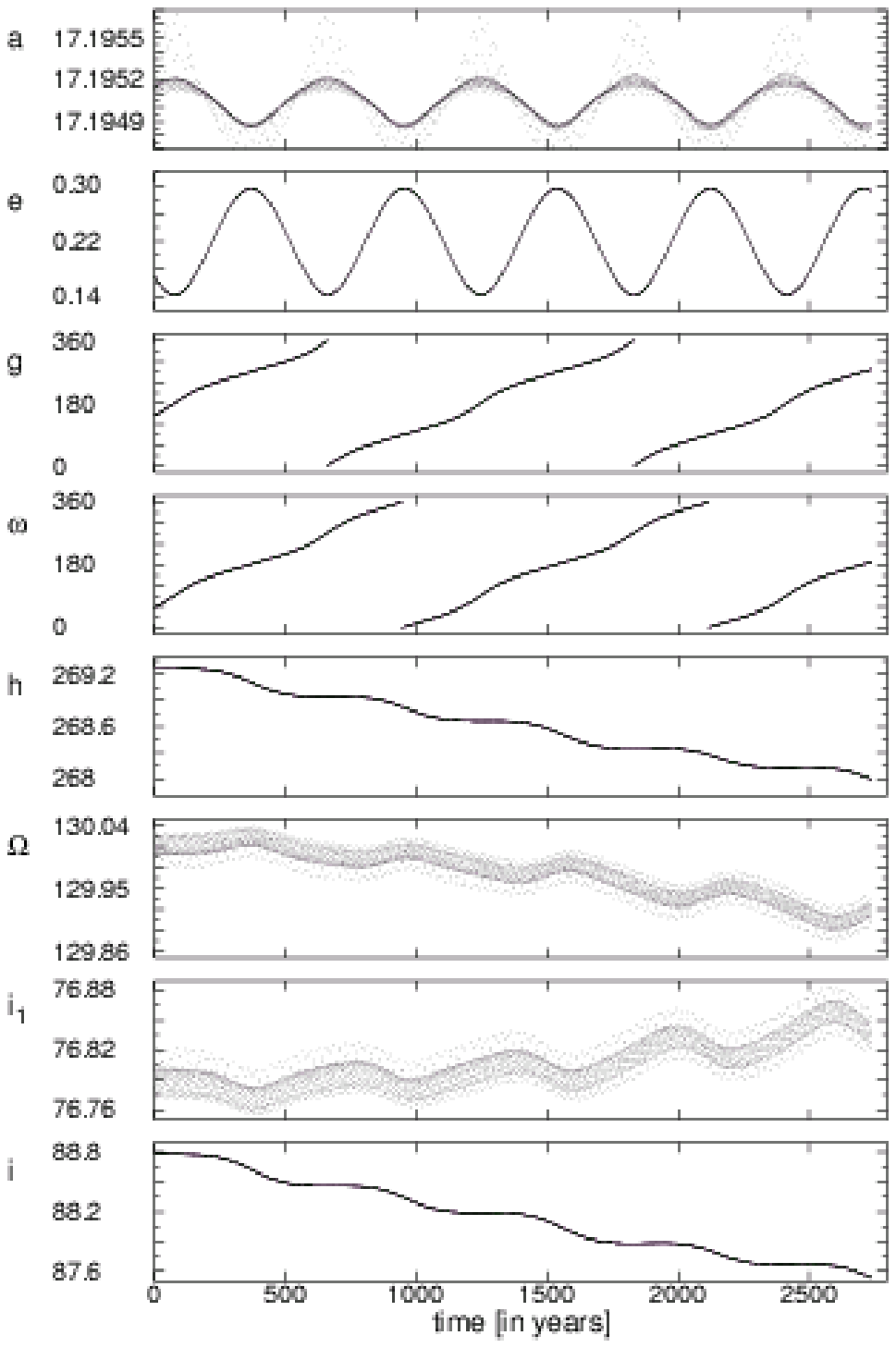}
\caption{The variation of both the dynamical and the observational orbital elements of the binary 
during 100 and approx. 2800 years (1 million days). Semi-major axis is given in $R_\odot$,
while the angular elements in degrees. The mutual inclination is $i_\mathrm{m}\approx90\degr$.
See Col.~AS4 of Table~\ref{tab:inputparams} for the initial values.}
\label{fig:AS4}
\end{figure}

The variation of the orbital elements during the integrations can be seen in Figs.~\ref{fig:AS1}-\ref{fig:AS4}.
These figures represent the dissipationless, departed from periastron runs. The left panels cover a time interval
of 100 years, which is naturally briefer than a moment from dynamical point of
view, but this is the time interval (in the best scenarios) which can usually be reached 
for the present studies. The numerically generated O--C curve, as well as the variations of the orbital elements $e$
and $\omega$ of the AS2--AS4b runs were also used in the previous section as an illustration for theoretical
thinkings (see Figs.~\ref{fig:numanal2060}--\ref{fig:ketO-C}).
Additional integrations were performed for the close binary without the tertiary component both in the quasi-synchronized rotator, and the precessing 
secondary component case (this latter is not presented here).

\begin{table}
\caption[]{Fixed initial parameters of the binary as well as of the tertiary for every run.
The $k_j^{(i)}$ constants are taken from the tables of \citet{claretgimenez91}.
The other parameters for the binary (with the exception of $\Omega$ which has an arbitrary value) 
are from \citet{khodykinvedeneyev97}. The parameters of the third companion (primed quantities) are partly
from our earlier light-time solution \citep[see][]{borko03}.}
\label{tab:fixparams}
$$
\begin{array}{ll|ll}
\hline
\hline
\noalign{\smallskip}
a&17.195\mathrm{R}_\odot &a'&736.98\mathrm{R}_\odot \\
e&0.17 &e'&0.41\\
\tau^a&50\,000 \mathrm{HJD} &\tau'&52\,085\mathrm{HJD} \\
&50\,000.82&&\\
&50\,001.7&&\\
i&88\fdg78&&\\
\Omega&130\degr&& \\
\noalign{\smallskip}
\hline
\noalign{\smallskip}
m_1&3.3 \mathrm{M}_\odot & m_2&2.5 \mathrm{M}_\odot\\
m_3&1.1\mathrm{M}_\odot&&\\
R_1&2.60 \mathrm{R}_\odot & R_2&1.96 \mathrm{R}_\odot\\
k_2^{(1)}&0.0049 & k_2^{(2)}&0.0038\\
k_3^{(1)}&0.0011 & k_3^{(2)}&0.0008\\
\lambda_1&0.0 &\lambda_2&0.0 \\
&0.0001&&0.0001\\
\noalign{\smallskip}
\hline
\noalign{\smallskip}
\end{array}
$$
{\small $a$: The three values refer to departure from periastron, 
instantaneous-angular-velocity-equal-to-average, and apastron positions, respectively.}
\end{table}
\begin{table}
\caption[]{Initial orbital elements, and some derived quantities, which were different at individual integration runs.}
\label{tab:inputparams}
$$
\begin{array}{lllllll}
\hline
\hline
\noalign{\smallskip}
&\mathrm{AS}1&\mathrm{AS}2&\mathrm{AS}3&\mathrm{AS3b}&\mathrm{AS}4&\mathrm{AS4b}\\
\hline
\noalign{\smallskip}
g&45\fdg0&317\fdg5&316\fdg6&57\fdg1&145\fdg1&327\fdg1\\
\omega&45\degr&45\degr&45\degr&57\fdg1&57\fdg1&147\fdg1 \\
h&0\fdg00&87\fdg94&89\fdg89&0\fdg00&269\fdg27&180\fdg00\\
i_1&0\fdg63&16\fdg25&49\fdg71&49\fdg74&76\fdg80&76\fdg76\\
\hline
\noalign{\smallskip}
g'&81\fdg00&352\fdg93&350\fdg75&351\fdg00&172\fdg22&351\fdg00\\
\omega'&261\degr&261\degr&261\degr&171\degr&261\degr&351\degr\\
h'&179\fdg87&267\fdg95&269\fdg89&180\fdg00&89\fdg27&0\fdg00\\
\Omega'&130\degr&110\degr&70\degr&130\degr&220\degr&130\degr\\
i_2&0\fdg15&3\fdg75&10\fdg27&10\fdg26&13\fdg16&13\fdg16\\
i'&88\fdg0&88\fdg0&88\fdg0&28\fdg78&88\fdg0&178\fdg7\\
\hline
\noalign{\smallskip}
i_\mathrm{m}&0\fdg78&20\fdg01&59\fdg98&60\fdg00&89\fdg96&89\fdg92\\
\hline
\noalign{\smallskip}
\end{array}
$$
\end{table}

In the quasi-synchronized binary case $U=381\,800\,\mathrm{d}\approx1\,046\,\mathrm{y}$ was found for the
period of the apsidal motion without the relativistic contribution. 
This yields $\dot\omega_\mathrm{cl}=\dot g_\mathrm{cl}=34\fdg4/100\,\mathrm{y}$ for the classical part
of the apsidal advance rate, which is in good agreement with the theoretical value of
\citet{maloneyetal89} $\dot{g}_\mathrm{cl}=35\fdg8\pm5\fdg8/100\,\mathrm{y}$.

\begin{figure*}
\centering
\includegraphics[width=8cm]{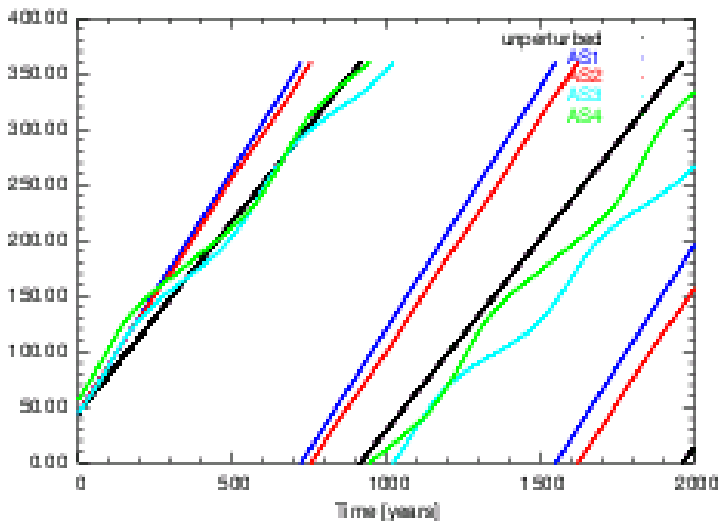}~~~\includegraphics[width=8cm]{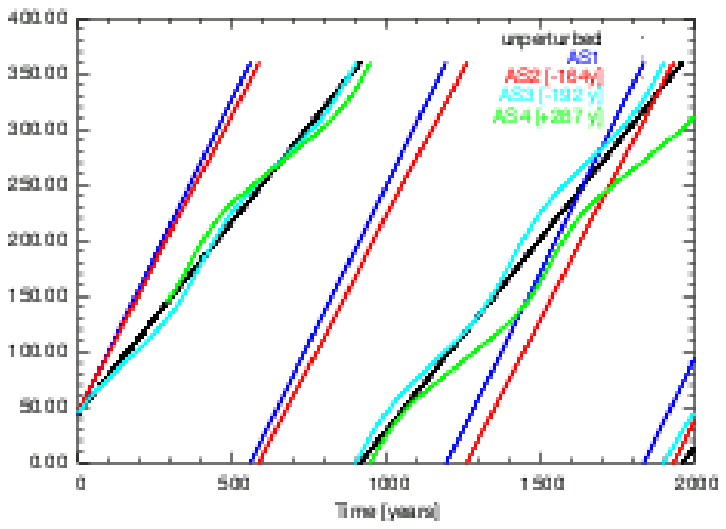}
\caption{The variation of the argument of periastron in the observational (left panel) and the dynamical (right panel) 
frame of reference. In the right panel we shifted the individual curves in time to start the $g$-curves
from the same initial value.}
\label{fig:omega0g0-k}
\end{figure*}

In Fig.~\ref{fig:omega0g0-k} we plotted the variation of the apsidal line both in the observational ($\omega$), 
and in the dynamical ($g$) frame of references for five different configurations of the system. 
Comparing the four third-body perturbed runs with the quasi-synchronized
binary configuration, one can see that there are no cases where the apsidal motion period would be significantly longer
than in the reference case. However, on the contrary, for the low mutual inclination cases (AS1, AS2), the apsidal motion period is
found to be remarkably smaller. This is in good correspondance with our discussion on $(\Pi^*)^{-1}$ in Sect.~\ref{subsubsect:egh}.
Consequently, we can conclude again, that the observational effects due to the specific spatial configuration of the orbital plane of
the perturbing third star should play a more important role in the explanation of the anomalously slow apsidal motion
than the dynamical perturbations themselves.

\begin{figure*}
\centering
\includegraphics[width=5cm]{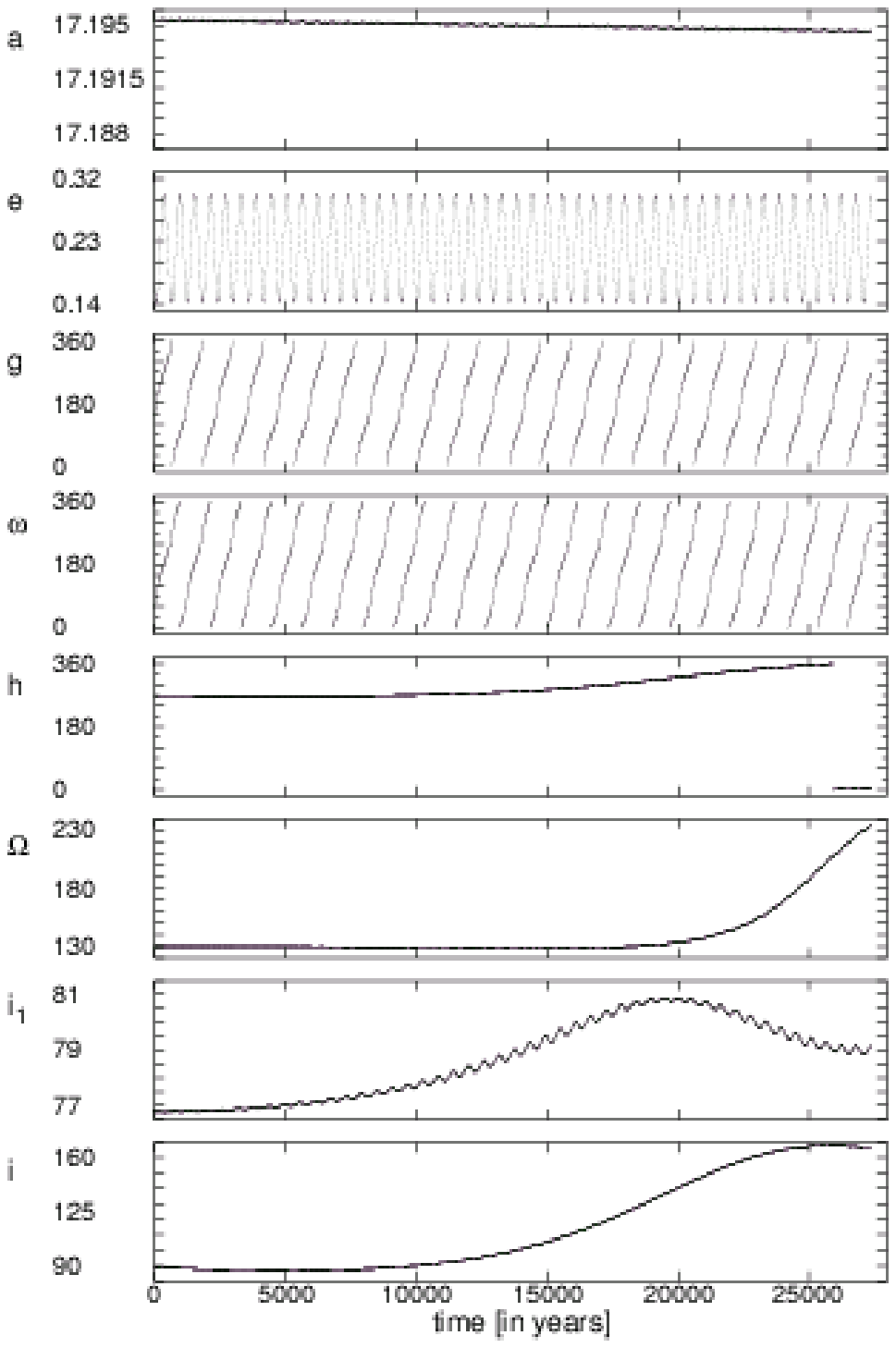}~~\includegraphics[width=5cm]{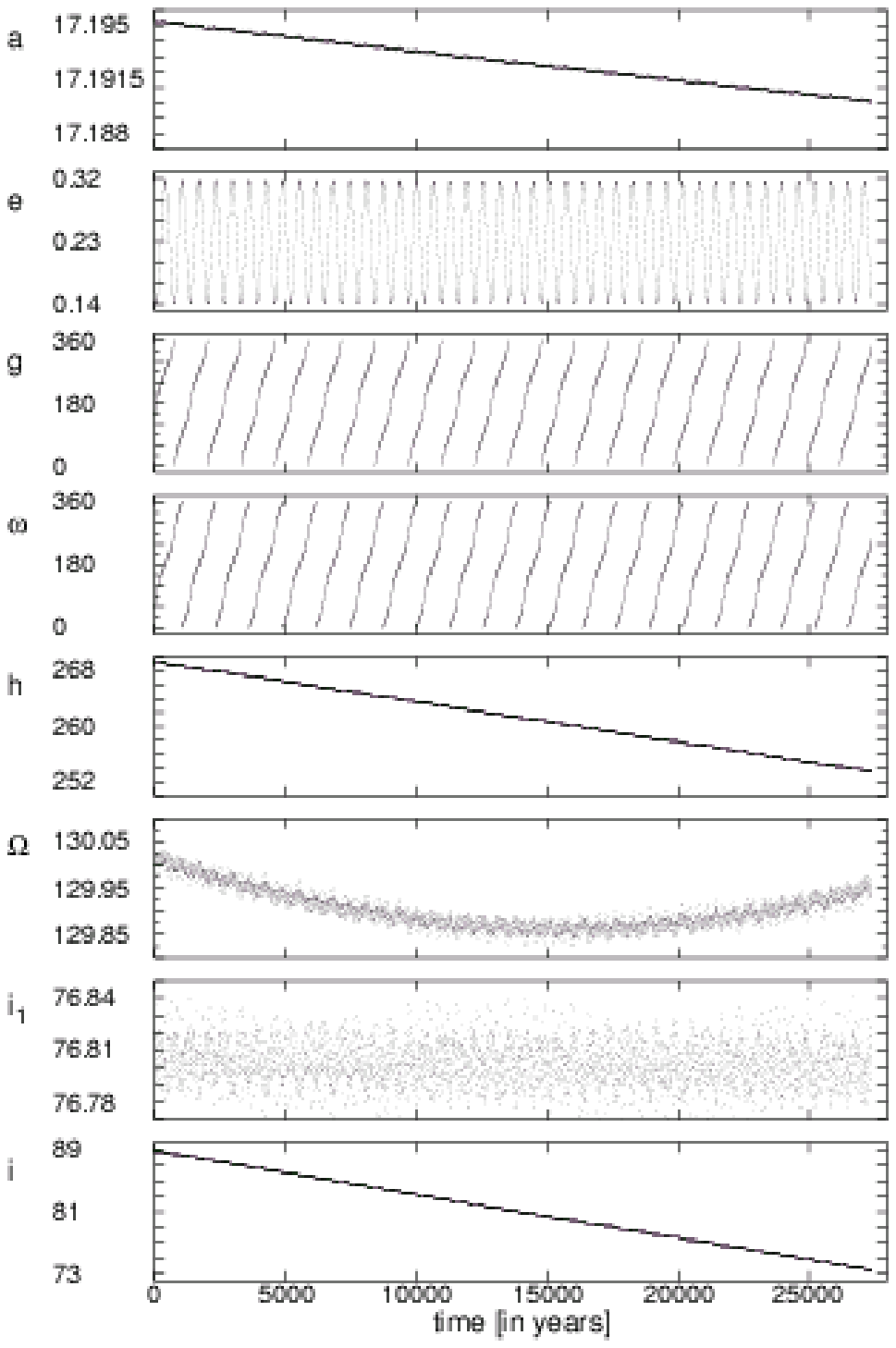}~~\includegraphics[width=5cm]{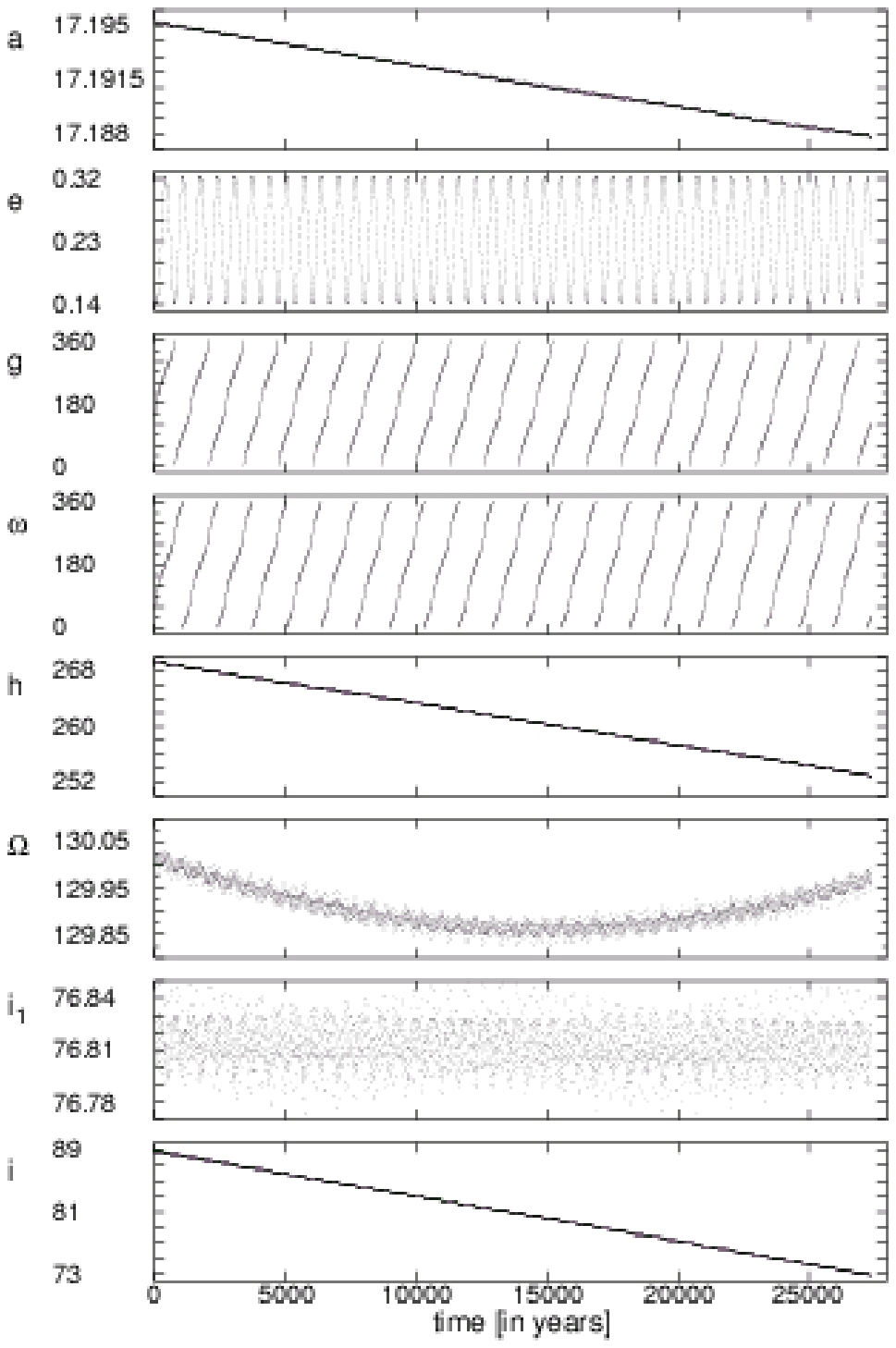}
\caption{The variation of the orbital elements of the binary in the dissipative case.
The initial parameters are listed in Table~\ref{tab:inputparams}, Col.~AS4. Left: The stellar rotation was
initially synchronized in the periastron. Medium: The stellar rotation
was synchronized to the average value of the Keplerian angular velocity. Right: The stellar
rotation was synchronized in the apastron.}
\label{fig:AS4dis3X}
\end{figure*}

\subsection{Dissipative runs}

Considering the dissipative runs, in Fig.~\ref{fig:AS4dis3X}
we give the variation of the orbital elements in the case of the AS4 configuration for all the three
initial synchronizations, i.e. when the synchronization was done to the periastron (left panel), the average value 
of the orbital angular velocity (medium panel), and the apastron value (right panel). The following 
differences can be observed. 

The shrinkage of the binary orbit is more than three times faster
during the total integration time in the apastron-synchronized case. For a qualitative explanation we
recall here the simplified dissipative part of the equation of motion of the binary from Eq.~(23) of 
\citet{borkoetal04}.
According to this
\begin{equation}
\vec{f}_\mathrm{dis}\sim-\lambda_i\left(3\frac{\vec{\rho}_1\cdot\dot{\vec{\rho}}_1}{\rho_1^2}\vec{\rho}_1-\vec{p}_i\times\vec{\rho}_1\right),
\end{equation}
where $\vec{\rho_1}$ is the Jacobian vector directed from the primary to the secondary, and $\vec{p}_i$ is
the difference vector of the rotational angular velocity vector of the $i$-th star ($i=1,2$) and the orbital angular
velocity vector.
As one can see easily, in the apastron-synchronized case the $constant\cdot\vec{p}_i\times\vec{\rho}_1$
terms give a continuously negative transversal force component (which has a maximum amplitude at periastron, and almost
zero only around the apastron). 

The amplitude of the $e$-cycles is also somewhat larger
in these latter cases, and the period of the apsidal advance is longer. Nevertheless,
the difference in the period between the periastron-synchronized and apastron-synchronized
case is about 15\%, so it remains clearly under the amount of the observed discrepancy.
Furthermore, we note that, although the orbital shrinking as well as the synchronization of
the angular rotation are evident on the time-scale of the present integration, 
it can be clearly seen that there is no evidence for the circularization of the inner orbit. This
suggests that the presence of a not so distant third body may basically modify the dissipative
circularization process. Naturally, further investigations are needed in this question.

In contradiction to these smaller differences in the behaviour of the above-mentioned orbital
elements, the angular elements (both in the observational and the dynamical frames) show completely
different variations in these latter two cases than in the periastron-synchronized one.
The reason can be found in the special initial configuration, i.e. in the (almost) perpendicular
position of the inner and outer orbital plane. In this position the difference in
the angular momentum stored initially in the stellar rotation was able to change the
direction of the orbital precession. 

\begin{figure*}
\centering
\includegraphics[width=8cm]{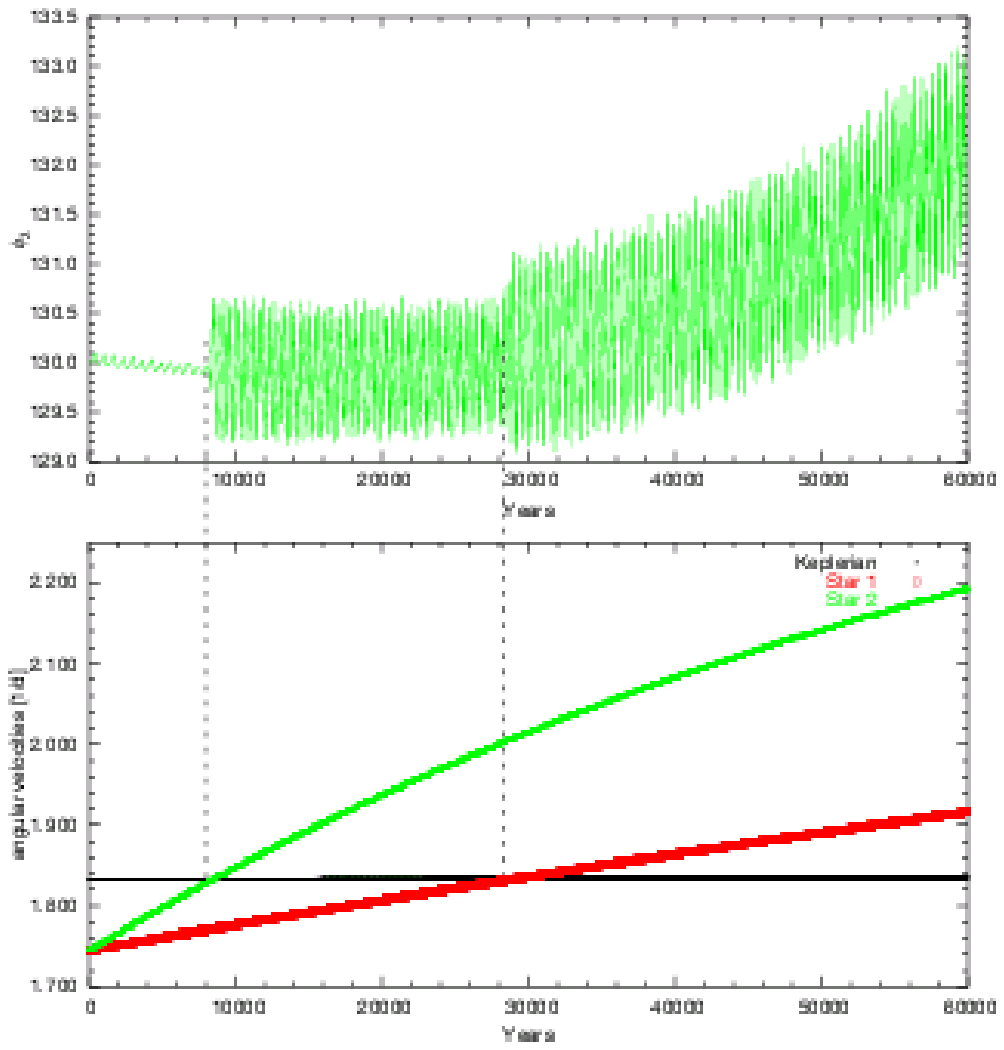}~~~\includegraphics[width=8cm]{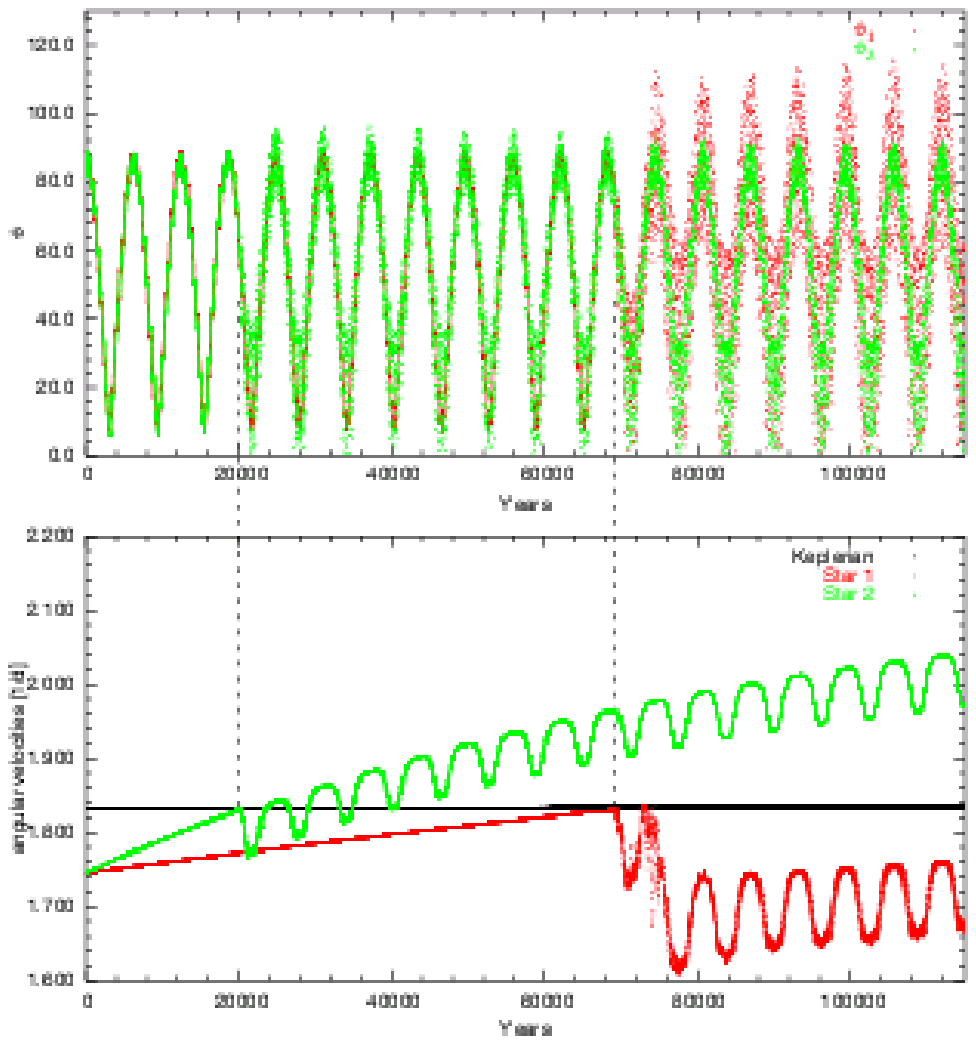}
\caption{Resonance phenomena between the orbital revolution and the stellar rotation. When the stellar rotation rate reaches the
average orbital revolution rate, the amplitude of the precession of the stellar rotational axes suddenly grows. 
This figure shows the dissipative AS4 (left) and AS3b (right) runs where the rotational angular velocities initially were set to close
to the averaged orbital one.}
\label{fig:rotresonances}
\end{figure*}

Finally, we consider the stellar rotation affected by the dissipation. As is well known, the phase space
of the rotation of oblate bodies usually contains large chaotic regions \citep[in the sense of our Solar system see e.g.][]{laskarrobutel93}.
In Fig.~\ref{fig:rotresonances} we show some interesting resonances. We found in several different runs, that when
the angular velocity of the rotation of one of the stars becomes temporarily equal to the average orbital
angular velocity of the binary, typical resonance phenomena occur, i.e. the amplitude of the stellar precession
suddenly increases, or even some fluctuations arise in the stellar rotational rate, and, consequently, this can
manifest even in some similar fluctuations in the binary's semi-major axis, as happens in our AS3b run.
Detailed investigations of such resonance phenomena may be the subject of future studies.
   
\section{Discussion and conclusion \label{sec:discussion}}
In this paper we gave analytically the net effect of the perturbations of a third companion and the
mutual tidal and rotational distortion of the binary members for the orbital elements of an
eccentric binary star. We investigated primarily how the presence of the third body can affect the O--C
curve, which is the most important source of the apsidal motion information in relativistic eccentric 
eclipsing binaries. For this purpose we have chosen one particular eclipsing system, namely \object{AS Camelopardalis}.
Nevertheless, naturally, most of our results are valid in general, independently of this specific
system, although we stress, that in the case of other systems, or even in the same system with different
third body parameters, the magnitude of some parameters can differ from our assumptions.

Our work on this topic is not the first, but one of its main novelties is that we calculated simultaneously
the effect of the tidal distortion and the tertiary perturbations. There is a strong interaction 
between the two physical processes. Here we refer to the strong dependence of the tidal forces on the eccentricity.
Consequently, the third body forced eccentricity variation influences notably the tidal phenomena, and besides
the pure, third star forced apsidal motion rate variation, this effect produces further remarkable change 
in the apsidal motion speed. 
At this point we emphasize again the necessity of the common treatment of the two different physical processes,
as their simultaneous presence can basically change the evolution of a close binary system. We mainly refer to
the so-called Kozai resonance, but further consideration of this topic can be found in \citet{eggletonetal98}. Note that from our 
Eq.~(\ref{eq:dedg}) naturally almost the same condition for the tidal prevention of the 
Kozai resonance arises\footnote{We said ``almost'', because it is not clear for us
why the authors use in their analytical stability investigation such orbital elements which refer to the observable
system. It should be clear that there is no connection between the physical instability in the system and its
orientation with respect to the Earth. So in their Eq.~(37) which defines the stability criterium they should 
use (according to their notation) $\phi$ (this is $-g$ in the present paper), instead of $\omega$.}, which was given in \citet{khodykinetal04} by Eq.~(38). 

The other novelty, as we already emphasized, is the calculation the mathematical shape of the O--C curve,
and we thoroughly investigated the reliability of the data which can be obtained from such a distorted O--C when
only a small portion of the total apsidal motion period is covered.

We found that there is a significant domain of the possible hierarchical triple body system configurations where
both the dynamical and the observational effects appear corresponding the tendency to measure a longer, and even
remarkably longer apsidal advance rate, than is theoretically expected according to the known physical parameters,
and the measured eccentricity and visible argument of periastron. This happens when the mutual inclination
of the close and the wide orbits is large, i.e. the orbits are nearly perpendicular to each other, and, furthermore,
the orbital plane of the tertiary almost coincides with the plane of the sky. As the observable inclination of the anomalously slow apsidal motion eclipsing binaries is 
necessarily close to $i=90\degr$, this means that almost every second near a perpendicular triple system might belong
to the ideal case. Note, that the first condition which mainly
(but not exclusively) comes from the dynamical calculations is in accordance with the results of the previously
mentioned studies. However, we have shown that it has a notably lower significance alone as stated previously.
So, we concluded, that the observational effects can play a remarkably more important role in the detection of anomalously
slow apsidal motion rate than the dynamical ones.

This result is all the more impressive, because it offers a natural answer for a further serious question:
why have we not observed these third companions yet? 
We consider this question in detail. Several direct and indirect methods can be used now to
detect a not so distant tertiary in a binary system. A relatively detailed, but not complete listing of these
methods is found in \citet{pribullarucinski06}. However, despite the diversity of the methods, most
were applied only for a few binary systems, and we do not know that any of the direct methods would have been
applied for any anomalously slow apsidal motion system. So we mainly concentrate here on the most usual, 
perhaps we can say: ``classical'', although in several cases very uncertain way of the detection of a third 
component in an eclipsing binary system, the method of which is based on the light-time effect (LITE) manifesting on the O--C diagram. 
However, the light-time effect arises from the varying distance of the eclipsing system from the observer, due to
the revolution of the binary around the centre of mass of the triple system. Consequently, when this motion takes place
nearly in the plane of the sky, the distance of the eclipsing binary remains almost constant, so the amplitude of
the LITE (related to $a'\sin{i'}$) might remain under the detection limit. (The same can be said about the detection via
the systematic radial velocity variation.) Unfortunately, this is exactly the situation in our ideal case for miscalculating the real
apsidal motion period from a small fraction of a complete revolution. 

At this point we have to stop for a short interplay, and note, that 
in the case of \object{AS~Camelopardalis}, \citet{ibvs4690} really reported that the O--C curve
shows small amplitude periodic variations after the removal of the apsidal motion terms. They fitted light-time solution,
and found a third body of mass ($m_3=1.1-1.7\,\mathrm{M}_\odot$) orbiting in an 
eccentric ($e'=0.5$) orbit with period $P'=805\,\mathrm{d}$. However, we also carried out the analysis of 
the O--C diagram, but only a very poor, and consequently, very questionable fit was found  \citep[see][]{borko03}. 
Nevertheless, we used our calculated orbital elements as input parameters in our numerical studies.

The question naturally arises of what could be an efficient method to detect perturbing bodies (if they
exist at all) in such configurations. There are several methods listed in the above mentioned paper of \citet{pribullarucinski06}
which could be satisfactory for this purpose. Instead of repeating them, we would like to suggest a further dynamical
method, which is based on the direct detection of the perturbations of the tertiary. As one can see from our results,
the variation of the eccentricity $e$ of the binary may reach even some $10^{-2}$ during some decades in our sample system.
Such a variation could have been detected by the present accuracy of radial velocity measurements. 
This is why we suppose that the different values of the eccentricity obtained for \object~{AS Cam} from the
1969--1971 and 1981 observations reflect real variation instead of simple observational errors.
Note when \citealt{maloneyetal91} analyzed the 1969--1971 observations of \citet{hilditch72a,hilditch72b}
and \citet{padaliasrivastava75}, they found that the orbital eccentricity of the binary was $e\approx0.14$,
while \citet{khaliullinkozyreva83} deduced from the December 1981 photometry already $e\approx0.16$.
Furthermore, according to recent O--C studies (which were cited earlier) $\omega\approx220\degr-240\degr$
were found. If we assume that the third body revolves close to the plane of the sky, the dynamical argument of
periastron ($g$) should have a similar value. Then, according to our Eq.~(\ref{eq:eeconst}), $\Delta{e}$ is positive,
which is also in good correspondence with the measured tendency. In conclusion, we stress, that it is very important 
to obtain new precise light curves, as well as radial velocity measurements from this, and other anomalously slow
apsidal motion systems. Moreover, as we predict the motion of the tertiary in the plane of
the sky, we can expect positive results mainly from the newest interferometric equipments in the near future.

Finally, we also obtained new numerical results on the interaction of the orbital evolution and stellar rotation in
close hierarchical triple stellar systems containing an eccentric binary. The most important fact is that 
resonances might occur due to the variation of the stellar rotational rate during the dissipation-driven synchronization process.
Such resonances were found for example in the case when the rotational rate of one of the stars reached the average Keplerian angular
velocity of the orbital revolution. It is necessary to investigate these phenomena further in detail.

\begin{acknowledgements} 
This work was supported in part by the French-Hungarian bilateral project Balaton, Grant No.~F36/04.
The authors would like to thank Sz. Csizmadia and A. P\'al for their valuable comments and
suggestions, and Rita Borkovits-J\'ozsef and L. Szabados for the linguistic corrections.
This research has made use of NASA's Astrophysics Data System Bibliographic Services.
\end{acknowledgements}

\begin{appendix}
\onecolumn
\section{Calculation of higher order perturbative terms \label{appendix}}
As mentioned in the text, in the most interesting cases the amplitude of the eccentricity variation
on the apse-node time scale reaches the unity. Consequently, the eccentricity no longer can be considered as
constant on the rhs of the perturbation equations. Consequently, first the eccentricity $e$ should be calculated as a
function of $g$ in an iterative manner.

The corresponding perturbation equation can be written in the following Taylor-expansion:
\begin{eqnarray}
\frac{\mathrm{d}e}{\mathrm{d}g}&=&\left(\frac{\mathrm{d}e}{\mathrm{d}g}\right)_{e_0}+\left(\frac{\partial}{\partial{e}}\frac{\mathrm{d}e}{\mathrm{d}g}\right)_{e_0}\Delta{e}+\frac{1}{2}\left(\frac{\partial^2}{\partial{e^2}}\frac{\mathrm{d}e}{\mathrm{d}g}\right)_{e_0}\left(\Delta{e}\right)^2+... \label{dudeTaylor} \\
&=&\left\{\left[{\cal{E}}\right]_0\Sigma_0+\left\{\left[{\cal{E}}+B_1+N_1-{\cal{E}}A_1\right]\Sigma_0+\left[{\cal{E}}(B_1-EA_1)\right]\Sigma_1\right\}_0\frac{1}{e_0}\Delta{e}\right.\nonumber \\
&&+\frac{1}{2}\left\{\left[(2-2A_1)(B_1+N_1-{\cal{E}}A_1)+B_2+N_2-{\cal{E}}A_2\right]\Sigma_0\right. \nonumber \\
&&+\left[2(B_1+N_1)(B_1-EA_1)+{\cal{E}}(B2-EA_2+2B_1-2EA_1-4A_1B_1+4EA_1^2)\right]\Sigma_1\nonumber \\
&&\left.\left.+\frac{1}{2}\left[{\cal{E}}(B_1^2+E^2A_1^2-2EA_1B_1)\right]\Sigma_2\right\}_0\frac{1}{e_0^2}(\Delta{e})^2\right\}, \label{eq:dedgtaylor}
\end{eqnarray}
where
\begin{eqnarray}
E&=&\frac{B}{A}, \\
{\cal{E}}&=&\frac{A_\mathrm{t}}{A}, \\
A_1&=&e\frac{1}{A}\frac{\mathrm{d}{A}}{\mathrm{d}{e}}, \\
A_2&=&e^2\frac{1}{A}\frac{\mathrm{d}^2A}{\mathrm{d}{e^2}}, \\
B_1&=&e\frac{1}{A}\frac{\mathrm{d}{B}}{\mathrm{d}{e}}, \\
B_2&=&e^2\frac{1}{A}\frac{\mathrm{d}^2B}{\mathrm{d}{e^2}}, \\
N_1&=&e\frac{1}{A}\frac{\mathrm{d}{A_\mathrm{n2}}}{\mathrm{d}{e}}, \\
N_2&=&e^2\frac{1}{A}\frac{\mathrm{d}^2{A_\mathrm{n2}}}{\mathrm{d}{e^2}}, \\
\Sigma_0&=&\frac{\sin2g}{1+E\cos2g}, \\
\Sigma_1&=&\frac{\mathrm{d}\Sigma_0}{\mathrm{d}E}, \\
\Sigma_2&=&\frac{\mathrm{d}^2{\Sigma_{0}}}{\mathrm{d}E^2}.
\end{eqnarray}
(In the calculation of the derivatives above, it should also be considered, that, as it can be seen easily
\begin{eqnarray}
\frac{\mathrm{d}I}{\mathrm{d}e}&=&\frac{e}{1-e^2}\left(I+\frac{C_1}{C_2}\right), \\
\frac{\mathrm{d}C_1}{\mathrm{d}e}&=&-\frac{e}{1-e^2}C_1, \\
\frac{\mathrm{d}C_2}{\mathrm{d}e}&=&0,
\end{eqnarray}
as far as the present approximation is used.)
As the tidal terms depend on the fifth power of the separation, and, consequently, this produces a very strong eccentricity dependence,
the $A_1$, $A_2$ derivatives may produce some numerical problems even in the case of medium eccentricities. Nevertheless, in the present
situation they have the same order of magnitude as $E$ or ${\cal{E}}$, at least at the near-perpendicular configurations.
Carrying out the iterative integrations, and writing $e$ into the following Fourier-series
\begin{equation}
e=e_0(1+\epsilon_0+\epsilon_2\cos2g+\epsilon_4\cos4g+\epsilon_6\cos6g+\epsilon_8\cos8g)+....
\end{equation}
the coefficients up to fifth order in $E,{\cal{E}},A_1,A_2$ are as follows:
\begin{eqnarray}
\epsilon_0&=&\left(\frac{1}{8}-\frac{1}{8}A_1\right){\cal{E}}^2+\left(\frac{3}{512}-\frac{33}{512}A_1-\frac{3}{128}A_2\right){\cal{E}}^4+\left(-\frac{9}{256}A_1-\frac{3}{256}A_2\right)E{\cal{E}}^3+\left(\frac{9}{128}-\frac{9}{64}A_1\right)E^2{\cal{E}}^2 \nonumber \\
&&+\left(\frac{1}{8}B_1+\frac{1}{8}N_1\right){\cal{E}} \nonumber \\
&&+\left[\frac{1}{2}+\left(\frac{3}{64}-\frac{3}{16}A_1-\frac{3}{64}A_2+\frac{9}{64}A_1^2\right){\cal{E}}^2-\frac{1}{32}E{\cal{E}}+\frac{1}{8}E^2+\frac{5}{3072}{\cal{E}}^4-\frac{1}{768}E{\cal{E}}^3+\frac{31}{768}E^2{\cal{E}}^2-\frac{5}{192}E^3{\cal{E}}+\frac{1}{16}E^4\right. \nonumber \\
&&\left.+\left(\frac{7}{32}B_1+\frac{3}{16}N_1+\frac{3}{64}B_2+\frac{3}{64}N_2\right){\cal{E}}-\frac{1}{32}(B_1+N_1)E\right]{\cal{E}}\cos2g_0 \nonumber \\
&&+\left[\left(\frac{1}{16}-\frac{1}{16}A_1\right){\cal{E}}-\frac{1}{8}E+\left(\frac{1}{192}-\frac{11}{192}A_1-\frac{1}{48}A_2\right){\cal{E}}^3+\left(-\frac{1}{64}+\frac{5}{96}A_1+\frac{1}{96}A_2\right)E{\cal{E}}^2+\left(\frac{1}{24}-\frac{1}{16}A_1\right)E^2{\cal{E}}-\frac{1}{16}E^3\right. \nonumber \\
&&\left.+\frac{1}{16}(B_1+N_1)\right]{\cal{E}}\cos4g_0\nonumber \\
&&+\left[\left(\frac{1}{192}-\frac{1}{48}A_1-\frac{1}{192}A_2+\frac{1}{64}A_1^2\right){\cal{E}}^2+\left(-\frac{1}{32}+\frac{1}{24}A_1\right)E{\cal{E}}+\frac{1}{24}E^2+\frac{5}{12288}{\cal{E}}^4-\frac{3}{1024}E{\cal{E}}^3+\frac{31}{3072}E^2{\cal{E}}^2\right.\nonumber \\
&&\left.-\frac{7}{256}E^3{\cal{E}}+\frac{1}{32}E^4+\left(\frac{1}{96}B_1+\frac{1}{48}N_1+\frac{1}{192}B_2+\frac{1}{192}N_2\right){\cal{E}}-\frac{1}{32}(B_1+N_1)E\right]{\cal{E}}\cos6g_0 \nonumber \\
&&+\left[\left(\frac{1}{3072}-\frac{11}{3072}A_1-\frac{1}{768}A_2\right){\cal{E}}^3+\left(-\frac{1}{256}+\frac{29}{1536}A_1+\frac{7}{1536}A_2\right)E{\cal{E}}^2+\left(\frac{11}{768}-\frac{3}{128}A_1\right)E^2{\cal{E}}-\frac{1}{64}E^3\right]{\cal{E}}\cos8g_0 \nonumber \\
&&+\left[\frac{1}{61440}{\cal{E}}^4-\frac{1}{3072}E{\cal{E}}^3+\frac{7}{3072}E^2{\cal{E}}^2-\frac{5}{768}E^3{\cal{E}}+\frac{1}{160}E^4\right]{\cal{E}}\cos10g_0,
\end{eqnarray}
\begin{eqnarray}
\epsilon_2&=&-\frac{1}{2}{\cal{E}}+\left(\frac{1}{64}+\frac{1}{64}A_2-\frac{1}{64}A_1^2\right){\cal{E}}^3+\left(-\frac{1}{32}+\frac{3}{32}A_1\right)E{\cal{E}}^2-\frac{1}{8}E^2{\cal{E}}-\frac{1}{1536}{\cal{E}}^5+\frac{1}{384}E{\cal{E}}^4+\frac{1}{96}E^2{\cal{E}}^3-\frac{5}{192}E^3{\cal{E}}^2-\frac{1}{16}E^4{\cal{E}}\nonumber \\
&&-\frac{1}{64}(4B_1+B_2+N_2){\cal{E}}^2-\frac{1}{32}(B_1+N_1)E{\cal{E}} \nonumber \\
&&+\left[\left(-\frac{1}{2}+\frac{1}{2}A_1\right){\cal{E}}+\left(\frac{1}{64}-\frac{3}{64}A_1\right){\cal{E}}^3+\left(-\frac{1}{32}+\frac{9}{32}A_1+\frac{3}{32}A_2\right)E{\cal{E}}^2+\left(-\frac{1}{8}+\frac{3}{8}A_1\right)E^2{\cal{E}}-\frac{1}{2}(B_1+N_1)\right]\epsilon_0 \nonumber \\
&&+\left[\left(\frac{1}{2}A_1+\frac{1}{4}A_2-\frac{1}{2}A_1^2\right){\cal{E}}-\frac{1}{2}(B_1+N_1)-\frac{1}{4}(B_2+N_2)\right]\epsilon_0^2,
\end{eqnarray}
\begin{eqnarray}
\epsilon_4&=&\left(\frac{1}{16}-\frac{1}{16}A_1\right){\cal{E}}^2+\frac{1}{8}E{\cal{E}}\nonumber \\
&&+\left(-\frac{1}{384}+\frac{1}{192}A_1-\frac{1}{768}A_2\right){\cal{E}}^4+\left(-\frac{7}{192}A_1-\frac{1}{64}A_2\right)E{\cal{E}}^3+\left(\frac{1}{24}-\frac{5}{48}A_1\right)E^2{\cal{E}}^2+\frac{1}{16}E^3{\cal{E}}+\frac{1}{16}(B_1+N_1){\cal{E}} \nonumber \\
&&+\left[\left(\frac{1}{16}-\frac{1}{4}A_1-\frac{1}{16}A_2+\frac{3}{16}A_1^2\right){\cal{E}}^2+\left(\frac{1}{8}-\frac{1}{4}A_1\right)E{\cal{E}}-\frac{1}{384}{\cal{E}}^4+\frac{1}{24}E^2{\cal{E}}^2+\frac{1}{16}E^3{\cal{E}}\right. \nonumber \\
&&\left.+\frac{1}{16}(6B_1+4N_1+B_2+N_2){\cal{E}}+\frac{1}{8}(B_1+N_1)E\right]\epsilon_0+\left[\left(-\frac{3}{16}A_1-\frac{3}{32}A_2\right){\cal{E}}^2+\left(-\frac{1}{4}A_1-\frac{1}{8}A_2\right)E{\cal{E}}\right]\epsilon_0^2,
\end{eqnarray}
\begin{eqnarray}
\epsilon_6&=&\left(-\frac{1}{192}+\frac{1}{48}A_1+\frac{1}{192}A_2-\frac{1}{64}A_1^2\right){\cal{E}}^3+\left(-\frac{1}{32}+\frac{5}{96}A_1\right)E{\cal{E}}^2-\frac{1}{24}E^2{\cal{E}}\nonumber \\
&&+\frac{1}{4096}{\cal{E}}^5+\frac{1}{1024}E^4{\cal{E}}-\frac{5}{1024}E^2{\cal{E}}^3-\frac{7}{256}E^3{\cal{E}}^2-\frac{1}{32}E^4{\cal{E}}-\frac{1}{192}(8B_1+4N_1+B_2+N_2){\cal{E}}^2-\frac{1}{32}(B_1+N_1)E{\cal{E}} \nonumber \\
&&+\left[\left(-\frac{1}{192}+\frac{11}{192}A_1+\frac{1}{48}A_2\right){\cal{E}}^3+\left(-\frac{1}{32}+\frac{19}{96}A_1+\frac{5}{96}A_2\right)E{\cal{E}}^2+\left(-\frac{1}{24}+\frac{1}{8}A_1\right)E^2{\cal{E}}\right]\epsilon_0,
\end{eqnarray}
\begin{eqnarray}
\epsilon_8&=&\left(\frac{1}{3072}-\frac{11}{3072}A_1-\frac{1}{768}A_2\right){\cal{E}}^4+\left(\frac{1}{256}-\frac{35}{1536}A_1-\frac{3}{512}A_2\right){\cal{E}}^3E+\left(\frac{11}{768}-\frac{13}{384}A_1\right)E^2{\cal{E}}^2+\frac{1}{64}E^3{\cal{E}} \nonumber \\
&&+\left(\frac{1}{3072}{\cal{E}}^4+\frac{1}{256}E{\cal{E}}^3+\frac{11}{768}E^2{\cal{E}}^2+\frac{1}{64}E^3{\cal{E}}\right)\epsilon_0,
\end{eqnarray}
\begin{eqnarray}
\epsilon_{10}&=&-\frac{1}{61440}{\cal{E}}^5-\frac{1}{3072}E{\cal{E}}^4-\frac{7}{3072}E^2{\cal{E}}^3-\frac{5}{768}E^3{\cal{E}}^2-\frac{1}{160}E^4{\cal{E}}.
\end{eqnarray}

For $\frac{\mathrm{d}u}{\mathrm{d}g}$ a similar equation can be written as Eq.~(\ref{eq:dedgtaylor}) and this gives the $u(g)$ relation as follows:
\begin{equation}
\Pi^*(u-u_0^*)=g+\gamma_2\sin2g+\gamma_4\sin4g+\gamma_6\sin6g+\gamma_8\sin8g+\gamma_{10}\sin10g+...,
\label{eq:Picsillagudef}
\end{equation}
where $(\Pi^*)^{-1}$ gives the apsidal motion period in the invariable plane in the unit of the inner orbital period, i.e.
\begin{eqnarray}
(\Pi^*)^{-1}&=&\Pi^{-1}\left\{1+\left(-\frac{1}{16}A_2+\frac{1}{8}A_1^2\right){\cal{E}}^2-\frac{1}{2}A_1E{\cal{E}}+\left(\frac{3}{1024}A_2-\frac{3}{512}A_1^2+\frac{1}{512}A_1A_2+\frac{1}{256}A_2^2\right){\cal{E}}^4\right. \nonumber \\
&&+\left(\frac{1}{64}A_1-\frac{7}{256}A_2+\frac{9}{128}A_1^2+\frac{15}{256}A_1A_2\right)E{\cal{E}}^3+\left(-\frac{5}{64}A_1-\frac{45}{256}A_2+\frac{103}{128}A_1^2\right)E^2{\cal{E}}^2-\frac{23}{32}A_1E^3{\cal{E}} \nonumber \\
&&+\frac{1}{4}B_1{\cal{E}}+\frac{1}{128}(-B_1+B_2){\cal{E}}^3+\left(\frac{3}{64}B_1+\frac{7}{64}B_2\right)E{\cal{E}}^2+\frac{9}{16}B_1E^2{\cal{E}} \nonumber \\
&&+\left[-A_1+\left(-\frac{1}{8}A_2+\frac{1}{4}A_1^2+\frac{1}{8}A_1A_2-\frac{1}{4}A_1^3\right){\cal{E}}^2+\left(-\frac{1}{2}A_1-\frac{1}{2}A_2+2A_1^2\right)E{\cal{E}}-A_1E^2+\frac{3}{512}A_2{\cal{E}}^4\right. \nonumber \\
&&+\left(\frac{1}{64}A_1-\frac{5}{128}A_2\right)E{\cal{E}}^3+\left(-\frac{5}{64}A_1-\frac{55}{128}A_2\right)E^2{\cal{E}}^2-\frac{23}{32}(A_1+A_2)E^3{\cal{E}}-A_1E^4\nonumber \\
&&\left.+\left(\frac{1}{4}(B_1+B_2)-\frac{5}{4}A_1B_1-\frac{1}{8}A_2(B_1+N_1)\right){\cal{E}}+\left(B_1-\frac{1}{2}A_1(B_1+N_1)\right)E\right]\epsilon_0 \nonumber \\
&&+\left[-\frac{1}{2}A_2+A_1^2+\left(-\frac{1}{16}A_2+\frac{1}{8}A_1^2+\frac{1}{4}A_1A_2+\frac{1}{16}A_2^2\right){\cal{E}}^2+\left(-\frac{1}{2}A_2+2A_1^2+\frac{3}{4}A_1A_2\right)E{\cal{E}}+\left(-\frac{1}{2}A_2+\frac{5}{2}A_1^2\right)E^2\right. \nonumber \\
&&\left.\left.+\frac{1}{4}B_2{\cal{E}}+\frac{1}{2}B_2E\right]\epsilon_0^2\right\},
\label{eq:Picsillag-1}
\end{eqnarray}
where as before
\begin{equation}
\Pi=A\sqrt{1-E^2}.
\end{equation}
Furthermore,
\begin{eqnarray}
\gamma_2&=&-\frac{1}{2}E-\frac{1}{8}E^3-\frac{1}{16}E^5\nonumber \\
&&+\frac{1}{4}A_1{\cal{E}}-\frac{1}{128}\left(A_1-A_2+2A_1^2\right){\cal{E}}^3+\frac{1}{64}\left(3A_1+5A_2-13A_1^2\right)E{\cal{E}}^2+\frac{5}{16}A_1E^2{\cal{E}}-\frac{1}{64}\left(B_1+3B_2\right){\cal{E}}^2-\frac{9}{32}B_1E{\cal{E}}\nonumber \\
&&+\left[\frac{1}{4}\left(A_1+A_2-2A_1^2\right){\cal{E}}-\frac{1}{128}\left(A_1-A_2\right){\cal{E}}^3+\frac{1}{64}\left(3A_1+13A_2\right)E{\cal{E}}^2+\frac{5}{16}\left(A_1+A_2\right)E^2{\cal{E}}+\frac{3}{8}A_1E^3\right. \nonumber \\
&&\left.-\frac{1}{2}B_1+\frac{1}{4}A_1(B_1+N_1)\right]\epsilon_0+\left[\frac{1}{4}\left(A_2-2A_1^2\right){\cal{E}}+\frac{1}{4}\left(A_2-2A_1^2\right)E-\frac{1}{4}B_2\right]\epsilon_0^2,
\end{eqnarray}
\begin{eqnarray}
\gamma_4&=&\frac{1}{8}E^2+\frac{1}{16}E^4\nonumber \\
&&-\frac{1}{64}(A_1+A_2+3A_1^2){\cal{E}}^2-\frac{5}{32}A_1E{\cal{E}}+\frac{1}{1536}(A_1+A_2){\cal{E}}^4+\frac{1}{768}(2A_1-9A_2)E{\cal{E}}^3-\frac{1}{192}(8A_1+11A_2)E^2{\cal{E}}^2-\frac{41}{192}A_1E^3{\cal{E}}\nonumber \\
&&+\left(\frac{1}{16}B_1-\frac{1}{64}A_1(B_1+N_1)\right){\cal{E}}\nonumber \\
&&+\left[-\frac{1}{64}(A_1+3A_2+9A_1^2+3A_1A_2){\cal{E}}^2-\frac{5}{32}(A_1+A_2-3A_1^2)E{\cal{E}}-\frac{1}{4}A_1E^2+\frac{1}{16}(B_1+B_2){\cal{E}}+\frac{1}{4}B_1E\right]\epsilon_0 \nonumber \\
&&+\left[-\frac{1}{32}A_2{\cal{E}}^2-\frac{5}{32}A_2E{\cal{E}}-\frac{1}{8}A_2E^2\right]\epsilon_0^2,
\end{eqnarray}
\begin{eqnarray}
\gamma_6&=&-\frac{1}{24}E^3-\frac{1}{32}E^5\nonumber \\
&&+\frac{1}{1152}(A_1+3A_2-10A_1^2-4A_1A_2){\cal{E}}^3+\frac{1}{576}(9A_1+9A_2-35A_1^2)E{\cal{E}}^2+\frac{13}{144}A_1E^2{\cal{E}}-\frac{1}{192}(B_1+B_2){\cal{E}}^2-\frac{5}{96}B_1E{\cal{E}}\nonumber \\
&&+\left[\frac{1}{1152}(A_1+7A_2){\cal{E}}^3+\frac{1}{64}(A_1+3A_2)E{\cal{E}}^2+\frac{13}{144}(A_1+A_2)E^2{\cal{E}}+\frac{1}{8}A_1E^3\right]\epsilon_0,
\end{eqnarray}
\begin{eqnarray}
\gamma_8&=&\frac{1}{64}E^4-\frac{1}{24576}(A_1+7A_2){\cal{E}}^4-\frac{7}{6144}(A_1+3A_2)E{\cal{E}}^3-\frac{71}{6144}(A_1+A_2)E^2{\cal{E}}^2-\frac{77}{1536}A_1E^3{\cal{E}}, \\
\gamma_{10}&=&-\frac{1}{160}E^5,
\end{eqnarray}
and, finally,
\begin{equation}
\Pi^*u_0^*=\Pi^*u_0-(g_0+\gamma_2\sin2g_0+\gamma_4\sin4g_0+\gamma_6\sin6g_0+\gamma_8\sin8g_0+\gamma_{10}\sin10g_0)+{\cal{O}}(e^6,E^6).
\end{equation}
As the next step we carry out the inverse transformation. Introducing the variable
\begin{equation}
{\cal{G}}=\Pi^*(u-u_0^*),
\end{equation}
the Fourier coefficients of the following equation
\begin{equation}
g={\cal{G}}+G_2\sin2{\cal{G}}+G_4\sin4{\cal{G}}+G_6\sin6{\cal{G}}+G_8\sin8{\cal{G}}+G_{10}\sin10{\cal{G}}+...
\label{eq:gGvel}
\end{equation}
can be calculated as e.g.
\begin{equation}
G_n=\frac{1}{\pi}\int_0^{2\pi}(g-{\cal{G}})\sin n{\cal{G}}(g)\frac{\mathrm{d}{\cal{G}}}{\mathrm{d}g}\mathrm{d}g,
\label{eq:GnFourier}
\end{equation}
where both $\sin n{\cal{G}}(g)$ and $\frac{\mathrm{d}{\cal{G}}}{\mathrm{d}g}$ can easily be calculated from Eq.~(\ref{eq:Picsillagudef}).
The individual coefficients are as follows:
\begin{eqnarray}
G_2&=&\frac{1}{2}E+\frac{1}{8}E^3+\frac{1}{16}E^5\nonumber \\
&&-\frac{1}{4}A_1{\cal{E}}+\frac{1}{256}(2A_1-2A_2+5A_1^2+A_1A_2){\cal{E}}^3-\frac{1}{128}(7A_1+11A_2-28A_1^2)E{\cal{E}}^2-\frac{21}{64}A_1E^2{\cal{E}}+\frac{1}{64}(B_1+3B_2){\cal{E}}^2+\frac{5}{16}B_1E{\cal{E}}\nonumber \\
&&+\left[-\frac{1}{4}\left(A_1+A_2-2A_1^2\right){\cal{E}}-\frac{1}{2}A_1E+\frac{1}{128}(A_1-A_2){\cal{E}}^3-\frac{1}{128}(7A_1+29A_2)E{\cal{E}}^2-\frac{21}{64}(A_1+A_2)E^2{\cal{E}}-\frac{3}{8}A_1E^3\right. \nonumber \\
&&\left.+\frac{1}{2}B_1-\frac{1}{4}A_1(B_1+N_1)\right]\epsilon_0+\left[-\frac{1}{4}(A_2-2A_1^2){\cal{E}}-\frac{1}{4}(A_2-2A_1^2)E+\frac{1}{4}B_2\right]\epsilon_0^2,
\end{eqnarray}
\begin{eqnarray}
G_4&=&\frac{1}{8}E^2+\frac{1}{16}E^4\nonumber \\
&&+\frac{1}{64}(A_1+A_2+A_1^2){\cal{E}}^2-\frac{3}{32}A_1E{\cal{E}}-\frac{1}{1536}(A_1+A_2){\cal{E}}^4+\frac{1}{2304}(14A_1+15A_2)E{\cal{E}}^3-\frac{1}{192}(A_1+4A_2)E^2{\cal{E}}^2-\frac{95}{576}A_1E^3{\cal{E}}\nonumber \\
&&-\frac{1}{16}B_1{\cal{E}}+\frac{1}{64}A_1(B_1+N_1){\cal{E}} \nonumber \\
&&+\left[\frac{1}{64}(A_1+3A_2-A_1^2+5A_1A_2){\cal{E}}^2-\frac{3}{32}(A_1+A_2+3A_1^2)E{\cal{E}}-\frac{1}{4}A_1E^2-\frac{1}{16}(B_1+B_2){\cal{E}}+\frac{1}{4}B_1E\right]\epsilon_0 \nonumber \\
&&+\left[\frac{1}{32}A_2{\cal{E}}^2-\frac{3}{32}A_2E{\cal{E}}-\frac{1}{8}A_2E^2\right]\epsilon_0^2,
\end{eqnarray}
\begin{eqnarray}
G_6&=&\frac{1}{24}E^3+\frac{1}{32}E^5\nonumber \\
&&-\frac{1}{2304}(2A_1+6A_2+7A_1^2+19A_1A_2){\cal{E}}^3+\frac{1}{1152}(9A_1+9A_2+16A_1^2)E{\cal{E}}^2-\frac{25}{576}A_1E^2{\cal{E}}+\frac{1}{192}(B_1+B_2){\cal{E}}^2-\frac{1}{24}B_1E{\cal{E}} \nonumber \\
&&+\left[-\frac{1}{1152}(A_1+7A_2){\cal{E}}^3+\frac{1}{128}(A_1+3A_2)E{\cal{E}}^2-\frac{25}{576}(A_1+A_2)E^2{\cal{E}}-\frac{1}{8}A_1E^3\right]\epsilon_0,
\end{eqnarray}
\begin{eqnarray}
G_8&=&\frac{1}{64}E^4\nonumber \\
&&+\frac{1}{24576}(A_1+7A_2){\cal{E}}^4-\frac{11}{18432}(A_1+3A_2)E{\cal{E}}^3+\frac{23}{6144}(A_1+A_2)E^2{\cal{E}}^2-\frac{97}{4608}A_1E^3{\cal{E}}, \\
\end{eqnarray}
\begin{eqnarray}
G_{10}&=&\frac{1}{160}E^5.
\end{eqnarray}
By the use of Eq.~(\ref{eq:gGvel}) the other orbital elements can also be easily expressed as a function of $u$, namely
\begin{eqnarray}
e&=&e_0(1+E_{00}+E_2\cos2{\cal{G}}+E_4\cos4{\cal{G}}+E_6\cos6{\cal{G}}+E_8\cos8{\cal{G}}+E_{10}\cos10{\cal{G}})+...,
\end{eqnarray}
where
\begin{eqnarray}
E_{00}&=&-\frac{1}{8}A_1{\cal{E}}^2+\frac{1}{4}E{\cal{E}}+\frac{3}{512}(A_1-A_2){\cal{E}}^4-\frac{1}{256}(2+14A_1+13A_2)E{\cal{E}}^3+\frac{1}{128}(4-37A_1)E^2{\cal{E}}^2+\frac{5}{32}E^3{\cal{E}} \nonumber \\
&&+\left[1-\frac{1}{8}(2A_1+A_2-3A_1^2){\cal{E}}^2+\frac{1}{4}(1-2A_1)E{\cal{E}}-\frac{1}{128}E{\cal{E}}^3+\frac{1}{32}E^2{\cal{E}}^2+\frac{5}{32}E^3{\cal{E}}+\frac{1}{4}B_1{\cal{E}}+\frac{1}{4}(B_1+N_1)E\right]\epsilon_0\nonumber \\
&&+\left[-\frac{1}{8}(A_1+2A_2){\cal{E}}^2-\frac{1}{4}(2A_1+A_2)E{\cal{E}}\right]\epsilon_0^2,
\end{eqnarray}
\begin{eqnarray}
E_2&=&-\frac{1}{2}{\cal{E}}+\frac{1}{128}(2+5A_1+3A_2-3A_1^2){\cal{E}}^3-\frac{1}{64}(6-7A_1)E{\cal{E}}^2-\frac{1}{8}E^2{\cal{E}}-\frac{1}{1536}{\cal{E}}^5+\frac{1}{192}E{\cal{E}}^4+\frac{1}{384}E^2{\cal{E}}^3-\frac{1}{12}E^3{\cal{E}}^2\nonumber \\
&&-\frac{1}{16}E^4{\cal{E}}-\frac{1}{64}(6B_1+B_2+N_2){\cal{E}}^2-\frac{3}{32}(B_1+N_1)E{\cal{E}}\nonumber \\
&&+\left[-\frac{1}{2}(1-A_1){\cal{E}}+\frac{1}{64}(1+2A_1+4A_2){\cal{E}}^3-\frac{1}{64}(6-32A_1-7A_2)E{\cal{E}}^2-\frac{1}{8}(1-3A_1)E^2{\cal{E}}-\frac{1}{2}(B_1+N_1)\right]\epsilon_0\nonumber \\
&&+\left[\frac{1}{4}(2A_1+A_2-2A_1^2){\cal{E}}-\frac{1}{2}(B_1+N_1)-\frac{1}{4}(B_2+N_2)\right]\epsilon_0^2,
\end{eqnarray}
\begin{eqnarray}
E_4&=&\frac{1}{16}(1+A_1){\cal{E}}^2-\frac{1}{8}E{\cal{E}}-\frac{1}{2304}(6+16A_1-3A_2){\cal{E}}^4+\frac{1}{192}(3+2A_1+6A_2)E{\cal{E}}^3+\frac{1}{288}(3+38A_1)E^2{\cal{E}}^2-\frac{1}{16}E^3{\cal{E}}\nonumber \\
&&+\frac{1}{16}(B_1+N_1){\cal{E}}+\left[\frac{1}{16}(1+A_2-3A_1^2){\cal{E}}^2-\frac{1}{8}(1-2A_1)E{\cal{E}}-\frac{1}{384}{\cal{E}}^4+\frac{1}{64}E{\cal{E}}^3+\frac{1}{96}E^2{\cal{E}}^2-\frac{1}{16}E^3{\cal{E}}\right. \nonumber \\
&&\left.+\frac{1}{16}(2B_1+4N_1+B_2+N_2)-\frac{1}{8}(B_1+N_1)E\right]\epsilon_0+\left[\frac{1}{32}(-2A_1+5A_2){\cal{E}}^2+\frac{1}{8}(2A_1+A_2)E{\cal{E}}\right]\epsilon_0^2,
\end{eqnarray}
\begin{eqnarray}
E_6&=&-\frac{1}{384}(2+7A_1+A_2+3A_1^2){\cal{E}}^3+\frac{1}{192}(6+7A_1)E{\cal{E}}^2-\frac{1}{24}E^2{\cal{E}}+\frac{1}{4096}{\cal{E}}^5-\frac{7}{3072}E{\cal{E}}^4+\frac{3}{1024}E^2{\cal{E}}^3+\frac{13}{768}E^3{\cal{E}}^2\nonumber \\
&&-\frac{1}{32}E^4{\cal{E}}-\frac{1}{192}(2B_1+4N_1+B_2+N_2){\cal{E}}^2+\frac{1}{32}(B_1+N_1)E{\cal{E}}\nonumber \\
&&+\left[-\frac{1}{192}(1+4A_1+8A_2){\cal{E}}^3+\frac{1}{192}(6-4A_1+7A_2)E{\cal{E}}^2-\frac{1}{24}(1-3A_1)E^2{\cal{E}}\right]\epsilon_0,
\end{eqnarray}
\begin{eqnarray}
E_8&=&\frac{1}{9216}(3+25A_1+18A_2){\cal{E}}^4-\frac{1}{1536}(6+23A_1+3A_2)E{\cal{E}}^3+\frac{11}{768}E^2{\cal{E}}^2-\frac{1}{64}E^3{\cal{E}}\nonumber \\
&&+\left[\frac{1}{3072}{\cal{E}}^4-\frac{1}{256}E{\cal{E}}^3+\frac{11}{2304}(3+4A_1)E^2{\cal{E}}^2-\frac{1}{64}E^3{\cal{E}}\right]\epsilon_0,
\end{eqnarray}
\begin{eqnarray}
E_{10}&=&-\frac{1}{61440}{\cal{E}}^5+\frac{1}{3072}E{\cal{E}}^4-\frac{7}{3072}E^2{\cal{E}}^3+\frac{5}{768}E^3{\cal{E}}^2-\frac{1}{160}E^4{\cal{E}}.
\end{eqnarray}

The perturbations of the other orbital elements can be calculated in a similar way, but as our final purpose is to obtain the
analytical form of the O--C in the function of the eclipsing cycle number (which is highly related to $u$, as well as ${\cal{G}}$),
now we use the following direct relation,
\begin{equation}
\frac{\mathrm{d}X}{\mathrm{d}{\cal{G}}}=\left(\frac{\mathrm{d}X}{\mathrm{d}{\cal{G}}}\right)_{e_0}+\left(\frac{\partial}{\partial{e}}\frac{\mathrm{d}X}{\mathrm{d}{\cal{G}}}\right)_{e_0}\Delta{e}+\frac{1}{2}\left(\frac{\partial^2}{\partial{e^2}}\frac{\mathrm{d}X}{\mathrm{d}{\cal{G}}}\right)_{e_0}\left(\Delta{e}\right)^2+..., \label{eq:dXdGTaylor} \\
\end{equation}
where $X$ means any of the remaining orbital elements or related quantities.
So, for the node ($h$) in the dynamical system, as well as for $u_\mathrm{m1}$ (which the latter denotes
that part of $u_\mathrm{m}$ which can be derived from $\mathrm{d}h\cos{i_1}$):
\begin{eqnarray}
h&=&h_0^*+H_0{\cal{G}}+H_2\sin2{\cal{G}}+H_4\sin4{\cal{G}}+H_6\sin6{\cal{G}}+H_8\sin8{\cal{G}}+..., \\
u_\mathrm{m1}&=&(u_\mathrm{m1})_0^*+U_0{\cal{G}}+U_2\sin2{\cal{G}}+U_4\sin4{\cal{G}}+U_6\sin6{\cal{G}}+U_8\sin8{\cal{G}}+...,
\end{eqnarray}
where
\begin{eqnarray}
H_0&=&J+K+\left[\frac{1}{16}(K_2+J_2)-\frac{1}{8}A_1(J_1+K_1)\right]{\cal{E}}^2+\frac{1}{4}(J_1+K_1)E{\cal{E}} \nonumber \\
&&+\frac{1}{4}(-A_1K+K_1){\cal{E}}+\frac{1}{2}KE-\frac{1}{128}(K_1-K_2){\cal{E}}^3+\frac{1}{32}(K_1+2K_2)E{\cal{E}}^2+\frac{5}{32}K_1E^2{\cal{E}}+\frac{1}{8}KE^3 \nonumber \\
&&+\left[J_1+K_1+\frac{1}{4}(K_1+K_2-2A_1K_1-A_1K-A_2K){\cal{E}}+\frac{1}{2}(K_1-A_1K)E+\frac{1}{8}(J_2+K_2){\cal{E}}^2+\frac{1}{4}(J_1+K_1+J_2+K_2)E{\cal{E}}\right]\epsilon_0\nonumber \\
&&+\left[\frac{1}{2}(J_2+K_2)+\frac{1}{4}K_2{\cal{E}}+\frac{1}{4}K_2E\right]\epsilon_0^2,
\end{eqnarray}
\begin{eqnarray}
H_2&=&-\frac{1}{2}K+\left[\frac{1}{128}(A_1+A_2)K-\frac{1}{64}(1-5A_1)K_1-\frac{3}{64}K_2\right]{\cal{E}}^2-\left[\frac{7}{64}A_1K+\frac{5}{32}K_1\right]E{\cal{E}}+\frac{1}{8}KE^2 \nonumber \\
&&-\frac{1}{4}(J_1+K_1){\cal{E}}+\frac{1}{128}(J_1+K_1-J_2-K_2){\cal{E}}^3-\frac{3}{64}(J_1+K_1+J_2+K_2)E{\cal{E}}^2-\frac{1}{16}(J_1+K_1)E^2{\cal{E}}\nonumber \\
&&+\left[-\frac{1}{2}K_1-\frac{1}{4}(J_1+K_1+J_2+K_2-A_1J_1-A_1K_1){\cal{E}}-\frac{1}{64}(K_1+7K_2){\cal{E}}^2-\frac{5}{32}(K_1+K_2)E{\cal{E}}+\frac{1}{8}K_1E^2\right]\epsilon_0\nonumber \\
&&+\left[-\frac{1}{4}K_2-\frac{1}{4}(J_2+K_2){\cal{E}}\right]\epsilon_0^2,
\end{eqnarray}
\begin{eqnarray}
H_4&=&\frac{1}{16}(K_1+A_1K){\cal{E}}-\frac{1}{8}KE-\frac{1}{768}(K_1-3K_2){\cal{E}}^3+\frac{1}{128}(2K_1+K_2)E{\cal{E}}^2-\frac{5}{192}K_1E^2{\cal{E}}\nonumber \\
&&+\frac{1}{64}(J_1+K_1+J_2+K_2+A_1J_1+A_1K_1){\cal{E}}^2-\frac{1}{32}(J_1+K_1)E{\cal{E}}\nonumber \\
&&+\left[\frac{1}{16}(K_1+K_2+A_1K+A_2K){\cal{E}}-\frac{1}{8}(K_1-A_1K)E+\frac{1}{64}(J_1+K_1+3J_2+3K_2){\cal{E}}^2-\frac{1}{32}(J_1+K_1+J_2+K_2)E{\cal{E}}\right]\epsilon_0\nonumber \\
&&+\left[\frac{1}{16}K_2{\cal{E}}-\frac{1}{16}K_2E\right]\epsilon_0^2,
\end{eqnarray}
\begin{eqnarray}
H_6&=&-\left[\frac{1}{384}(A_1+A_2)K+\frac{1}{192}(1+3A_1)K_1+\frac{1}{192}K_2\right]{\cal{E}}^2+\frac{1}{192}(7A_1K+6K_1)E{\cal{E}}-\frac{1}{24}KE^2 \nonumber \\
&&-\frac{1}{1152}(J_1+K_1+3J_2+3K_2){\cal{E}}^3+\frac{1}{192}(J_1+K_1+J_2+K_2)E{\cal{E}}^2-\frac{1}{144}(J_1+K_1)E^2{\cal{E}} \nonumber \\
&&+\left[-\frac{1}{192}(K_1+3K_2){\cal{E}}^2+\frac{1}{32}(K_1+K_2)E{\cal{E}}-\frac{1}{24}K_1E^2\right]\epsilon_0,
\end{eqnarray}
\begin{eqnarray}
H_8&=&\frac{1}{3072}(K_1+3K_2){\cal{E}}^3-\frac{1}{256}(K_1+K_2)E{\cal{E}}^2+\frac{11}{768}K_1{\cal{E}}E^2-\frac{1}{64}KE^3,
\end{eqnarray}
where
\begin{eqnarray}
J&=&\frac{1}{\Pi^*}A_\mathrm{h1}, \\
J_1&=&e\frac{1}{\Pi^*}\frac{\mathrm{d}{A_\mathrm{h1}}}{\mathrm{d}{e}}, \\
J_2&=&e^2\frac{1}{\Pi^*}\frac{\mathrm{d}^2{A_\mathrm{h1}}}{\mathrm{d}^2{e}}, \\
K&=&\frac{1}{\Pi^*}A_\mathrm{h2}, \\
K_1&=&e\frac{1}{\Pi^*}\frac{\mathrm{d}{A_\mathrm{h2}}}{\mathrm{d}{e}}, \\
K_2&=&e^2\frac{1}{\Pi^*}\frac{\mathrm{d}^2{A_\mathrm{h2}}}{\mathrm{d}^2{e}},
\end{eqnarray}
and
\begin{eqnarray}
A_\mathrm{h1}&=&-\frac{1}{\cos{i_1}}A_\mathrm{n1} \nonumber \\
&=&-\frac{2}{5}A_\mathrm{G}(1-e^2)^{1/2}I\frac{C}{C_2},
\end{eqnarray}
while $A_\mathrm{h2}$ could be derived from $A_\mathrm{n2}$ in a similar manner.
Here $J$ was treated as second order in $e$, while $J_{1,2}$, $K$, and $K_{1,2}$-s
were considered as third order.
To obtain the corresponding expressions for $u_\mathrm{m1}$, $A_\mathrm{h1,2}$ and its derivatives should be simply replaced by $-A_\mathrm{n1,2}$ 
and derivatives. As these quantities will be used later, for the sake of the clarity we define them here:
\begin{eqnarray}
L&=&-\frac{1}{\Pi^*}A_\mathrm{n1}, \\
L_1&=&-e\frac{1}{\Pi^*}\frac{\mathrm{d}{A_\mathrm{n1}}}{\mathrm{d}{e}}, \\
L_2&=&-e^2\frac{1}{\Pi^*}\frac{\mathrm{d}^2{A_\mathrm{n1}}}{\mathrm{d}^2{e}}, \\
M&=&-\frac{1}{\Pi^*}A_\mathrm{n2}, \\
M_1&=&-e\frac{1}{\Pi^*}\frac{\mathrm{d}{A_\mathrm{n2}}}{\mathrm{d}{e}}, \\
M_2&=&-e^2\frac{1}{\Pi^*}\frac{\mathrm{d}^2{A_\mathrm{n2}}}{\mathrm{d}^2{e}},
\end{eqnarray}

Formally, similar expression can be written for the direct perturbative terms in the orbital motion ($\delta$).
Nevertheless, in this case the magnitude of the derivatives differ from those above,
so we rewrite the results according to the orders of the direct terms as follows:
\begin{equation}
\delta=\delta_0^*+D_0{\cal{G}}+D_2\sin2{\cal{G}}+D_{4}\sin4{\cal{G}}+D_6\sin6{\cal{G}}+D_8\sin8{\cal{G}}+..., 
\label{eq:vkozvapp}
\end{equation}
where
\begin{eqnarray}
D_0&=&V+\left(\frac{1}{16}V_2-\frac{1}{8}A_1V_1\right){\cal{E}}^2+\frac{1}{4}V_1E{\cal{E}}-\frac{3}{1024}V_2{\cal{E}}^4-\frac{1}{256}(2V_1-5V_2)E{\cal{E}}^3+\frac{1}{256}(8V_1+17V_2)E^2{\cal{E}}^2+\frac{5}{32}V_1E^3{\cal{E}}\nonumber \\
&&+\frac{1}{4}\left(-W_1+A_1W_0\right){\cal{E}}-\frac{1}{2}WE+\frac{1}{128}(W_1-W_2){\cal{E}}^3-\frac{1}{32}(W_1+2W_2)E{\cal{E}}^2-\frac{5}{32}W_1E^2{\cal{E}}-\frac{1}{8}WE^3\nonumber \\
&&+\left[V_1+\frac{1}{8}(V_2-2A_1V_1-A_2V_1-2A_1V_2){\cal{E}}^2+\frac{1}{4}\left(V+1+V_2-2A_1V_1\right)E{\cal{E}}\right. \nonumber \\
&&\left.-\frac{1}{4}(W_1+W_2-A_1W-A_2W-2A_1W_1){\cal{E}}-\frac{1}{2}(W_1-A_1W)E\right]\epsilon_0\nonumber \\
&&+\left[\frac{1}{2}V_2+\frac{1}{16}V_2{\cal{E}}^2+\frac{1}{4}V_2E{\cal{E}}-\frac{1}{4}W_2{\cal{E}}-\frac{1}{4}W_2E\right]\epsilon_0^2,
\end{eqnarray}
\begin{eqnarray}
D_2&=&-\frac{1}{4}V_1{\cal{E}}+\frac{1}{256}(2V_1-2V_2+5A_1V_1+3A_2V_1+6A_1V_2){\cal{E}}^3-\frac{1}{128}(6V_1+6V_2-7A_1V_1)E{\cal{E}}^2-\frac{1}{16}V_1E^2{\cal{E}}\nonumber \\
&&+\frac{1}{2}W+\frac{1}{128}(2W_1+6W_2-A_1W-A_2W-10A_1W_1){\cal{E}}^2+\frac{1}{64}(10W_1+7A_1W)E{\cal{E}}-\frac{1}{8}WE^2\nonumber \\
&&+\left[-\frac{1}{4}(V_1+V_2+A_1V_1){\cal{E}}+\frac{1}{128}(V_1-V_2){\cal{E}}^3-\frac{3}{64}(V_1+3V_2)E{\cal{E}}^2-\frac{1}{16}(V_1+V_2)E^2{\cal{E}}-\frac{1}{4}V_1(B_1+N_1)\right. \nonumber \\
&&\left.+\frac{1}{2}W_1+\frac{1}{64}(W_1+7W_2){\cal{E}}^2+\frac{5}{32}(W_1+W_2)E{\cal{E}}-\frac{1}{8}W_1E^2\right]\epsilon_0\nonumber \\
&&+\left[-\frac{1}{8}(2V_2-2A_1V_1-A_2V_1-2A_1V_2){\cal{E}}+\frac{1}{4}W_2\right]\epsilon_0^2,
\end{eqnarray}
\begin{eqnarray}
D_4&=&\frac{1}{64}(V_1+V_2+A_1V_1){\cal{E}}^2-\frac{1}{32}V_1E{\cal{E}}-\frac{1}{1536}(V_1+V_2){\cal{E}}^4+\frac{1}{256}(V_1+2V_2)E{\cal{E}}^3+\frac{1}{384}(V_1+V_2)E^2{\cal{E}}^2-\frac{1}{64}V_1E^3{\cal{E}}\nonumber \\
&&+\frac{1}{64}V_1(B_1+N_1){\cal{E}}-\frac{1}{16}(W_1+A_1W){\cal{E}}+\frac{1}{8}WE+\frac{1}{768}(W_1-3W_2){\cal{E}}^3-\frac{1}{128}(2W_1+W_2)E{\cal{E}}^2+\frac{5}{192}W_1E^2{\cal{E}} \nonumber \\
&&+\left[\frac{1}{64}(V_1+3V_2+A_2V_1-A_1V_2){\cal{E}}^2-\frac{1}{32}(V_1+V_2-2A_1V_1)E{\cal{E}}-\frac{1}{16}(W_1+W_2+A_1W+A_2W){\cal{E}}+\frac{1}{8}(W_1-A_1W)E\right]\epsilon_0 \nonumber \\
&&+\left[\frac{1}{32}V_2{\cal{E}}^2-\frac{1}{32}V_2E{\cal{E}}-\frac{1}{16}W_2{\cal{E}}+\frac{1}{16}W_2E\right]\epsilon_0^2,
\end{eqnarray}
\begin{eqnarray}
D_6&=&-\frac{1}{2304}(2V_1+6V_2+7A_1V_1+1A_2V_1+6A_1V_2){\cal{E}}^3+\frac{1}{1152}(6V_1+6V_2+7A_1V_1)E{\cal{E}}^2-\frac{1}{144}V_1E^2{\cal{E}} \nonumber \\
&&+\frac{1}{384}(2W_1+2W_2+A_1W+A_2W+6A_1W_1){\cal{E}}^2-\frac{1}{192}(6W_1+7A_1W)E{\cal{E}}+\frac{1}{24}E^2W\nonumber \\
&&+\left[-\frac{1}{1152}(V_1+7V_2){\cal{E}}^3+\frac{1}{192}(V_1+3V_2)E{\cal{E}}^2-\frac{1}{144}(V_1+V_2)E^2{\cal{E}}\right.\nonumber \\
&&\left.+\frac{1}{192}(W_1+3W_2){\cal{E}}^2-\frac{1}{32}(W_1+W_2)E{\cal{E}}+\frac{1}{24}W_1E^2\right]\epsilon_0,
\end{eqnarray}
\begin{eqnarray}
D_8&=&\frac{1}{24576}(V_1+7V_2){\cal{E}}^4-\frac{1}{2048}(V_1+3V_2)E{\cal{E}}^3+\frac{11}{6144}(V_1+V_2)E^2{\cal{E}}^2-\frac{1}{512}V_1E^3{\cal{E}}\nonumber \\
&&-\frac{1}{3072}(W_1+3W_2){\cal{E}}^3+\frac{1}{256}(W_1+W_2)E{\cal{E}}^2-\frac{11}{768}W_1E^2{\cal{E}}+\frac{1}{64}WE^3.
\label{eq:D8}
\end{eqnarray}
In the above equations
\begin{eqnarray}
V&=&\frac{1}{\Pi^*}A_\mathrm{d}, \\
V_1&=&e\frac{1}{\Pi^*}\frac{\mathrm{d}A_\mathrm{d}}{\mathrm{d}e}, \\
V_2&=&e^2\frac{1}{\Pi^*}\frac{\mathrm{d}^2A_\mathrm{d}}{\mathrm{d}e^2}, \\
W&=&\frac{1}{\Pi^*}B_\mathrm{d},\\
W_1&=&e\frac{1}{\Pi^*}\frac{\mathrm{d}B_\mathrm{d}}{\mathrm{d}e}, \\
W_2&=&e^2\frac{1}{\Pi^*}\frac{\mathrm{d}^2B_\mathrm{d}}{\mathrm{d}e^2},
\end{eqnarray}
where in our computations $V$s were considered as first, while $W$s as second order quantities in $E$ (or ${\cal{E}}$, $e$).

As a next step we calculate the angular elements in the observational system of references. 
We apply the following relations from the theory of spherical geometry:
\begin{eqnarray}
\cos{i}&=&\cos{I_0}\cos{i_1}-\sin{I_0}\sin{i_1}\cos{h}, \label{eq:cosi}\\
\mathrm{d}u_\mathrm{m}&=&\mathrm{d}{h}\frac{\cos i_1-\cos{I_0}\cos{i}}{1-\cos^2i}-\mathrm{d} i_1\frac{\sin I_0\sin h\cos i}{1-\cos^2i} , \label{eq:dumapp}\\
\mathrm{d}\Omega&=&\mathrm{d}{h}\frac{\cos I_0-\cos i_1\cos i}{1-\cos^2 i}+\mathrm{d} i_1\frac{\sin I_0\sin h}{1-\cos^2 i}, \label{eq:dOmegaapp}\\
\mathrm{d} i&=&\mathrm{d}{i_1}\frac{\cos{I_0}\sin{i_1}+\cos{i_1}\sin{I_0}\cos{h}}{\sin{i}}-\mathrm{d}{h}\frac{\sin{h}\sin{I_0}\sin{i_1}}{\sin{i}}. \label{eq:diapp}
\end{eqnarray}
By the use of these relations, after some algebra and the Taylorian expansion of the $1-\cos^2i$ denominator we obtain
\begin{eqnarray}
-\mathrm{d}\Omega\cos{i}&=&\mathrm{d}h\cos{i_1}(C_{00}+C_{02}\cos2i_1+C_{04}\cos4i_1+C_{06}\cos6i_1) \nonumber \\
&&+\mathrm{d}h\cos{h}\sin{i_1}(C_{10}+C_{12}\cos2i_1+C_{14}\cos4i_1+C_{16}\cos6i_1) \nonumber \\
&&+\mathrm{d}h\cos2h\cos{i_1}(C_{20}+C_{22}\cos2i_1+C_{24}\cos4i_1+C_{26}\cos6i_1) \nonumber \\
&&+\mathrm{d}h\cos3h\sin{i_1}(C_{30}+C_{32}\cos2i_1+C_{34}\cos4i_1+C_{36}\cos6i_1) \nonumber \\
&&+\mathrm{d}h\cos4h\cos{i_1}(C_{40}+C_{42}\cos2i_1+C_{44}\cos4i_1+C_{46}\cos6i_1) \nonumber \\
&&+\mathrm{d}h\cos5h\sin{i_1}(C_{50}+C_{52}\cos2i_1+C_{54}\cos4i_1+C_{56}\cos6i_1) \nonumber \\
&&+\mathrm{d}h\cos6h\cos{i_1}(C_{60}+C_{62}\cos2i_1+C_{64}\cos4i_1+C_{66}\cos6i_1) \nonumber \\
&&-\mathrm{d{i_1}}\sin{h}\cos{i_1}(S_{10}+S_{12}\cos2i_1+S_{14}\cos4i_1)+ ...,
\label{eq:dOmegacosi}
\end{eqnarray}
where
\begin{eqnarray}
C_{00}&=&\frac{765}{4096}-\frac{4725}{8192}\cos2I_0+\frac{357}{4096}\cos4I_0-\frac{91}{8192}\cos6I_0,  \\
C_{02}&=&\frac{1271}{8192}+\frac{8295}{16384}\cos2I_0-\frac{1743}{8192}\cos4I_0+\frac{329}{16384}\cos6I_0, \\
C_{04}&=&\frac{117}{4096}+\frac{525}{8192}\cos2I_0+\frac{483}{4096}\cos4I_0-\frac{189}{8192}\cos6I_0,\\
C_{06}&=&\frac{25}{8192}+\frac{105}{16384}\cos2I_0+\frac{63}{8192}\cos4I_0+\frac{231}{16384}\cos6I_0, \\
C_{12}&=&-\frac{3449}{4096}\sin2I_0-\frac{31}{1024}\sin4I_0-\frac{5}{4096}\sin6I_0, \\
C_{14}&=&-\frac{31}{512}\sin2I_0-\frac{27}{128}\sin4I_0-\frac{3}{512}\sin6I_0, \\
C_{16}&=&-\frac{15}{4096}\sin2I_0-\frac{9}{1024}\sin4I_0-\frac{99}{4096}\sin6I_0,\\
C_{20}&=&\frac{1271}{8192}-\frac{4549}{16384}\cos2I_0+\frac{1121}{8192}\cos4I_0-\frac{235}{16384}\cos6I_0, \\
C_{22}&=&-\frac{1893}{16384}+\frac{10767}{32768}\cos2I_0-\frac{4003}{16384}\cos4I_0+\frac{1025}{32768}\cos6I_0, \\
C_{24}&=&-\frac{287}{8192}-\frac{707}{16384}\cos2I_0+\frac{903}{8192}\cos4I_0-\frac{525}{16384}\cos6I_0, \\
C_{26}&=&-\frac{75}{16384}-\frac{255}{32768}\cos2I_0-\frac{45}{16384}\cos4I_0+\frac{495}{32768}\cos6I_0, \\
C_{32}&=&-\frac{501}{8192}\sin2I_0+\frac{57}{2048}\sin4I_0+\frac{15}{8192}\sin6I_0, \\
C_{34}&=&\frac{57}{1024}\sin2I_0-\frac{9}{256}\sin4I_0+\frac{5}{1024}\sin6I_0, \\
C_{36}&=&\frac{45}{8192}\sin2I_0+\frac{15}{2048}\sin4I_0-\frac{55}{8192}\sin6I_0, \\
C_{40}&=&\frac{117}{4096}-\frac{387}{8192}\cos2I_0+\frac{99}{4096}\cos4I_0-\frac{45}{8192}\cos6I_0, \\
C_{42}&=&-\frac{287}{8192}+\frac{1057}{16384}\cos2I_0-\frac{329}{8192}\cos4I_0+\frac{175}{16384}\cos6I_0, \\
C_{44}&=&\frac{19}{4096}-\frac{149}{8192}\cos2I_0+\frac{85}{4096}\cos4I_0-\frac{59}{8192}\cos6I_0, \\
C_{46}&=&\frac{15}{8192}+\frac{15}{16384}\cos2I_0-\frac{39}{8192}\cos4I_0+\frac{33}{16384}\cos6I_0, \\
C_{52}&=&-\frac{25}{8192}\sin2I_0+\frac{5}{2048}\sin4I_0-\frac{5}{8192}\sin6I_0, \nonumber \\
C_{54}&=&\frac{5}{1024}\sin2I_0-\frac{1}{256}\sin4I_0+\frac{1}{1024}\sin6I_0, \\
C_{56}&=&-\frac{15}{8192}\sin2I_0+\frac{3}{2048}\sin4I_0-\frac{3}{8192}\sin6I_0, \\
C_{60}&=&\frac{25}{8192}-\frac{75}{16384}\cos2I_0+\frac{15}{8192}\cos4I_0-\frac{5}{16384}\cos6I_0, \\
C_{62}&=&-\frac{75}{16382}+\frac{225}{32768}\cos2I_0-\frac{45}{16384}\cos4I_0+\frac{15}{32768}\cos6I_0, \\
C_{64}&=&\frac{15}{8192}-\frac{45}{16384}\cos2I_0+\frac{9}{8192}\cos4I_0-\frac{3}{16384}\cos6I_0, \\
C_{66}&=&-\frac{5}{16384}+\frac{15}{32768}\cos2I_0-\frac{3}{16384}\cos4I_0+\frac{1}{32768}\cos6I_0, \\
S_{10}&=&-\frac{1727}{2048}\sin2I_0-\frac{17}{512}\sin4I_0-\frac{19}{2048}\sin6I_0, \\
S_{12}&=&-\frac{31}{512}\sin2I_0-\frac{27}{128}\sin4I_0-\frac{3}{512}\sin6I_0, \\
S_{14}&=&-\frac{5}{2048}\sin2I_0-\frac{3}{512}\sin4I_0-\frac{33}{2048}\sin6I_0
\label{eq:CSnndef}
\end{eqnarray}
As Eq.~(\ref{eq:di1dgdedgvel}) reveals there is only a very small cyclic variation in $i_1$. Consequently,
in Eq.~(\ref{eq:dOmegacosi}) $i_1$ can be considered as constant. With this approximation the integration
is a very easy task, and we do not feel the necessity to report its result here. (Nevertheless, it can be read
directly -- with the exclusion of the secular term -- from the $V_{0nm}$ coefficients [Eqs.~(\ref{eq:V_020})--(\ref{eq:V_005})] 
of the final form of the O--C given below [Eq.~(\ref{eq:O-Cfinalaotig})].)
Instead we list the analytical form of the argument of periastron in the observational frame of reference ($\omega$):
\begin{eqnarray}
\omega&=&g+u_\mathrm{m}, \nonumber \\
&=&\omega_0^*+O_0{\cal{G}}+O_2\sin2{\cal{G}}+O_4\sin4{\cal{G}}+O_6\sin6{\cal{G}}+O_8\sin8{\cal{G}}\nonumber \\
&&+O_{n0}\sin(nh_0^*+nH_0{\cal{G}})+O_{n\mp m}\sin[nh_0^*+(nH_0\mp m){\cal{G}}],
\label{eq:omegaapp}
\end{eqnarray}
where
\begin{eqnarray}
O_0&=&1+U_0(1+{\cal{C}}_0), \\
O_n&=&G_n+U_n(1+{\cal{C}}_0), \\
O_{n0}&=&\frac{1}{n}{\cal{C}}_n, \\
O_{n\mp2}&=&\pm{\cal{C}}_{n}\left\{\frac{1}{4}K+\frac{1}{8}(J_1+K_1){\cal{E}}+\frac{1}{128}(K_1+3K_2){\cal{E}}^2+\frac{5}{64}K_1E{\cal{E}}-\frac{1}{16}KE^2\right.\nonumber \\
&&\left.+\left[\frac{1}{4}K_1+\frac{1}{8}(J_1+J_2+K_1+K_2){\cal{E}}\right]\epsilon_0+\frac{1}{8}K_2\epsilon_0^2\right\}, \\
O_{n\mp4}&=&\pm{\cal{C}}_n\left\{-\frac{1}{32}(K_1+A_1K){\cal{E}}+\frac{1}{16}KE-\frac{1}{128}(J_1+J_2+K_1+K_2){\cal{E}}^2+\frac{1}{64}(J_1+K_1)E{\cal{E}}\right. \nonumber \\
&&\left.+\left[-\frac{1}{32}(K_1+K_2){\cal{E}}+\frac{1}{16}K_1E\right]\epsilon_0\right\}, \\
O_{n\mp6}&=&\pm{\cal{C}}_n\left[\frac{1}{384}(K_1+K_2){\cal{E}}^2-\frac{1}{64}K_1E{\cal{E}}+\frac{1}{48}KE^2\right],
\label{eq:Onmp6}
\end{eqnarray}
and, furthermore, we applied the following abbreviations:
\begin{equation}
{\cal{C}}_0=C_{00}+C_{02}\cos2i_1+C_{04}\cos4i_1+C_{06}\cos6i_1,
\label{eq:C0def}
\end{equation}
but e.g.
\begin{equation}
{\cal{C}}_1=\sin{i_1}(C_{10}+C_{12}\cos2i_1+C_{14}\cos4i_1+C_{16}\cos6i_1),
\label{eq:C1def}
\end{equation}
i.e. at the trigonometric terms the $\sin{i_1}$ or $\cos{i_1}$ multiplicators outside
the parenthesis are also included.
(Note, in the following for the sake of the simplicity we consider ${\cal{C}}_n$s
as being in the order of $e^n$. This is not necessarily exactly correct, but as ${\cal{C}}_n$s come from the Taylorian
expansion of $1-\cos^2{i}$ it may be partly verified.)

We are now in the position to give the final analytic form of the O--C curve up to the fifth order.
This is given as follows:
\begin{eqnarray}
\frac{2\pi}{P}O-C&=&V_{100}\cos(\omega_0^*+O_0{\cal{G}})+V_{200}\sin(2\omega_0^*+2O_0{\cal{G}})+V_{300}\cos(3\omega^*+3O_0{\cal{G}}) \nonumber \\
&&+V_{400}\sin(4\omega_0^*+4O_0{\cal{G}})+V_{500}\cos(5\omega^*+5O_0{\cal{G}}) \nonumber \\
&&+V_{120}\cos[\omega_0^*+(O_0+2){\cal{G}}]+V_{1-20}\cos[\omega_0^*+(O_0-2){\cal{G}}]\nonumber \\
&&+V_{140}\cos[\omega_0^*+(O_0+4){\cal{G}}]+V_{1-40}\cos[\omega_0^*+(O_0-4){\cal{G}}] \nonumber \\
&&+V_{160}\cos[\omega_0^*+(O_0+6){\cal{G}}]+V_{1-60}\cos[\omega_0^*+(O_0-6){\cal{G}}]\nonumber \\
&&+V_{180}\cos[\omega_0^*+(O_0+8){\cal{G}}]+V_{1-80}\cos[\omega_0^*+(O_0-8){\cal{G}}] \nonumber \\
&&+V_{220}\sin[2\omega_0^*+(2O_0+2){\cal{G}}]+V_{2-20}\sin[2\omega_0^*+(2O_0-2){\cal{G}}]\nonumber \\
&&+V_{240}\sin[2\omega_0^*+(2O_0+4){\cal{G}}]+V_{2-40}\sin[2\omega_0^*+(2O_0-4){\cal{G}}]\nonumber \\
&&+V_{260}\sin[2\omega_0^*+(2O_0+6){\cal{G}}]+V_{2-60}\sin[2\omega_0^*+(2O_0-6){\cal{G}}]\nonumber \\
&&+V_{320}\sin[3\omega_0^*+(3O_0+2){\cal{G}}]+V_{2-20}\sin[3\omega_0^*+(3O_0-2){\cal{G}}]\nonumber \\
&&+V_{340}\sin[3\omega_0^*+(3O_0+4){\cal{G}}]+V_{2-40}\sin[3\omega_0^*+(3O_0-4){\cal{G}}]\nonumber \\
&&+V_{420}\sin[4\omega_0^*+(4O_0+2){\cal{G}}]+V_{4-20}\sin[4\omega_0^*+(4O_0-2){\cal{G}}]\nonumber \\
&&+V_{101}\cos[\omega_0^*+h_0^*+(O_0+H_0){\cal{G}}]+V_{10-1}\cos[\omega_0^*-h_0^*+(O_0-H_0){\cal{G}}] \nonumber \\
&&+V_{102}\cos[\omega_0^*+2h_0^*+(O_0+2H_0){\cal{G}}]+V_{10-2}\cos[\omega_0^*-2h_0^*+(O_0-2H_0){\cal{G}}] \nonumber \\
&&+V_{103}\cos[\omega_0^*+3h_0^*+(O_0+3H_0){\cal{G}}]+V_{10-3}\cos[\omega_0^*-3h_0^*+(O_0-3H_0){\cal{G}}] \nonumber \\
&&+V_{104}\cos[\omega_0^*+4h_0^*+(O_0+4H_0){\cal{G}}]+V_{10-4}\cos[\omega_0^*-4h_0^*+(O_0-4H_0){\cal{G}}] \nonumber \\
&&+V_{121}\cos[\omega_0^*+h_0^*+(O_0+2+H_0){\cal{G}}]+V_{12-1}\cos[\omega_0^*-h_0^*+(O_0+2-H_0){\cal{G}}] \nonumber \\
&&+V_{1-21}\cos[\omega_0^*+h_0^*+(O_0-2+H_0){\cal{G}}]+V_{1-2-1}\cos[\omega_0^*-h_0^*+(O_0-2-H_0){\cal{G}}] \nonumber \\
&&+V_{122}\cos[\omega_0^*+2h_0^*+(O_0+2+2H_0){\cal{G}}]+V_{12-2}\cos[\omega_0^*-2h_0^*+(O_0+2-2H_0){\cal{G}}] \nonumber \\
&&+V_{1-22}\cos[\omega_0^*+2h_0^*+(O_0-2+2H_0){\cal{G}}]+V_{1-2-2}\cos[\omega_0^*-2h_0^*+(O_0-2-2H_0){\cal{G}}] \nonumber \\
&&+V_{123}\cos[\omega_0^*+3h_0^*+(O_0+2+3H_0){\cal{G}}]+V_{12-3}\cos[\omega_0^*-3h_0^*+(O_0+2-3H_0){\cal{G}}] \nonumber \\
&&+V_{1-23}\cos[\omega_0^*+3h_0^*+(O_0-2+3H_0){\cal{G}}]+V_{1-2-3}\cos[\omega_0^*-3h_0^*+(O_0-2-3H_0){\cal{G}}] \nonumber \\
&&+V_{141}\cos[\omega_0^*+h_0^*+(O_0+4+H_0){\cal{G}}]+V_{14-1}\cos[\omega_0^*-h_0^*+(O_0+4-H_0){\cal{G}}] \nonumber \\
&&+V_{1-41}\cos[\omega_0^*+h_0^*+(O_0-4+H_0){\cal{G}}]+V_{1-2-1}\cos[\omega_0^*-h_0^*+(O_0-4-H_0){\cal{G}}] \nonumber \\
&&+V_{142}\cos[\omega_0^*+2h_0^*+(O_0+4+2H_0){\cal{G}}]+V_{12-2}\cos[\omega_0^*-2h_0^*+(O_0+4-2H_0){\cal{G}}] \nonumber \\
&&+V_{1-42}\cos[\omega_0^*+2h_0^*+(O_0-4+2H_0){\cal{G}}]+V_{1-2-2}\cos[\omega_0^*-2h_0^*+(O_0-4-2H_0){\cal{G}}] \nonumber \\
&&+V_{161}\cos[\omega_0^*+h_0^*+(O_0+6+H_0){\cal{G}}]+V_{16-1}\cos[\omega_0^*-h_0^*+(O_0+6-H_0){\cal{G}}] \nonumber \\
&&+V_{1-61}\cos[\omega_0^*+h_0^*+(O_0-6+H_0){\cal{G}}]+V_{1-6-1}\cos[\omega_0^*-h_0^*+(O_0-6-H_0){\cal{G}}] \nonumber \\
&&+V_{201}\sin[2\omega_0^*+h_0^*+(2O_0+H_0){\cal{G}}]+V_{20-1}\sin[2\omega_0^*-h_0^*+(2O_0-H_0){\cal{G}}] \nonumber \\
&&+V_{202}\sin[2\omega_0^*+2h_0^*+(2O_0+2H_0){\cal{G}}]+V_{20-1}\sin[2\omega_0^*-2h_0^*+(2O_0-2H_0){\cal{G}}] \nonumber \\
&&+V_{203}\sin[2\omega_0^*+3h_0^*+(2O_0+3H_0){\cal{G}}]+V_{20-1}\sin[2\omega_0^*-3h_0^*+(2O_0-3H_0){\cal{G}}] \nonumber \\
&&+V_{221}\sin[2\omega_0^*+h_0^*+(2O_0+2+H_0){\cal{G}}]+V_{22-1}\sin[2\omega_0^*-h_0^*+(2O_0+2-H_0){\cal{G}}] \nonumber \\
&&+V_{2-21}\sin[2\omega_0^*+h_0^*+(2O_0-2+H_0){\cal{G}}]+V_{2-2-1}\sin[2\omega_0^*-h_0^*+(2O_0-2-H_0){\cal{G}}] \nonumber \\
&&+V_{222}\sin[2\omega_0^*+2h_0^*+(2O_0+2+2H_0){\cal{G}}]+V_{22-2}\sin[2\omega_0^*-2h_0^*+(2O_0+2-2H_0){\cal{G}}] \nonumber \\
&&+V_{2-22}\sin[2\omega_0^*+2h_0^*+(2O_0-2+2H_0){\cal{G}}]+V_{2-2-2}\sin[2\omega_0^*-2h_0^*+(2O_0-2-2H_0){\cal{G}}] \nonumber \\
&&+V_{241}\sin[2\omega_0^*+h_0^*+(2O_0+4+H_0){\cal{G}}]+V_{24-1}\sin[2\omega_0^*-h_0^*+(2O_0+4-H_0){\cal{G}}] \nonumber \\
&&+V_{2-41}\sin[2\omega_0^*+h_0^*+(2O_0-4+H_0){\cal{G}}]+V_{2-4-1}\sin[2\omega_0^*-h_0^*+(2O_0-4-H_0){\cal{G}}] \nonumber \\
&&+V_{301}\cos[3\omega_0^*+h_0^*+(3O_0+H_0){\cal{G}}]+V_{30-1}\cos[3\omega_0^*-h_0^*+(3O_0-H_0){\cal{G}}] \nonumber \\
&&+V_{302}\cos[3\omega_0^*+2h_0^*+(3O_0+2H_0){\cal{G}}]+V_{30-2}\cos[3\omega_0^*-2h_0^*+(3O_0-2H_0){\cal{G}}] \nonumber \\
&&+V_{321}\cos[3\omega_0^*+h_0^*+(3O_0+2+H_0){\cal{G}}]+V_{32-1}\cos[3\omega_0^*-h_0^*+(3O_0+2-H_0){\cal{G}}] \nonumber \\
&&+V_{3-21}\cos[3\omega_0^*+h_0^*+(3O_0-2+H_0){\cal{G}}]+V_{3-2-1}\cos[3\omega_0^*-h_0^*+(3O_0-2-H_0){\cal{G}}] \nonumber \\
&&+V_{401}\sin[4\omega_0^*+h_0^*+(4O_0+H_0){\cal{G}}]+V_{40-1}\sin[4\omega_0^*-h_0^*+(4O_0-H_0){\cal{G}}] \nonumber \\
&&+V_{020}\sin2{\cal{G}}+V_{040}\sin4{\cal{G}}+V_{060}\sin6{\cal{G}}+V_{080}\sin8{\cal{G}}\nonumber \\
&&+V_{001}\sin(h_0^*+H_0{\cal{G}})+V_{002}\sin(2h_0^*+2H_0{\cal{G}})+V_{003}\sin(3h_0^*+3H_0{\cal{G}}) \nonumber \\
&&+V_{004}\sin(4h_0^*+4H_0{\cal{G}})+V_{005}\sin(5h_0^*+5H_0{\cal{G}}) \nonumber \\
&&+V_{021}\sin[h_0^*+(H_0+2){\cal{G}}]+V_{0-21}\sin[h_0^*+(H_0-2){\cal{G}}] \nonumber \\
&&+V_{022}\sin[2h_0^*+(2H_0+2){\cal{G}}]+V_{0-22}\sin[2h_0^*+(2H_0-2){\cal{G}}]+\ \nonumber \\
&&+V_{041}\sin[h_0^*+(H_0+4){\cal{G}}]+V_{0-41}\sin[h_0^*+(H_0-4){\cal{G}}]+{\cal{O}}(e^6,E^6,{\cal{E}}^6),
\label{eq:O-Cfinalaotig}
\end{eqnarray}
where
\begin{eqnarray}
V_{100}&=&je_0\left\{-2+\frac{1}{32}(8A_1+A_1^2){\cal{E}}^2-\frac{1}{8}(4+A_1)E{\cal{E}}+\frac{1}{8}E^2+\frac{1}{64}E{\cal{E}}^3-\frac{17}{256}E^2{\cal{E}}^2-\frac{37}{128}E^3{\cal{E}}+\frac{35}{512}E^4\right. \nonumber \\
&&+\left(\frac{1}{2}+\frac{1}{8}E{\cal{E}}-\frac{1}{32}E^2\right){\cal{C}}_1^2-\frac{1}{32}{\cal{C}}_1^4+\frac{1}{8}{\cal{C}}_2^2-\frac{1}{4}ME(1+{\cal{C}}_0) \nonumber \\
&&\left.+\left[-2+\frac{1}{4}(2A_1+A_2)E^2-\frac{1}{8}(4-6A_1+A_2)E{\cal{E}}+\frac{1}{8}(1-2A_1)E^2+\frac{1}{2}{\cal{C}}_1^2\right]\epsilon_0\right\},
\label{eq:V_100}
\end{eqnarray}
\begin{eqnarray}
V_{1-20}&=&je_0\left\{\frac{1}{4}(2-A_1){\cal{E}}+\frac{1}{2}E-\frac{1}{256}(4+7A_1+9A_2){\cal{E}}^3+\frac{1}{256}(20-71A_1-21A_2)E{\cal{E}}^2+\frac{1}{64}(15-22A_1)E^2{\cal{E}}+\frac{9}{64}E^3\right. \nonumber \\
&&-\left[\frac{1}{16}(2-A_1){\cal{E}}+\frac{1}{8}E\right]{\cal{C}}_1^2-\left[\frac{1}{2}M+\frac{1}{4}(L_1+M_1){\cal{E}}\right](1+{\cal{C}}_0) \nonumber \\
&&+\left[\frac{1}{4}(2-4A_1-A_2+2A_1^2){\cal{E}}+\frac{1}{2}(1-A_1)E-\frac{1}{64}{\cal{E}}^3+\frac{5}{64}E{\cal{E}}^2+\frac{15}{64}E^2{\cal{E}}+\frac{9}{64}E^3+B_1+\frac{1}{2}N_1\right. \nonumber \\
&&\left.\left.-\left(\frac{1}{8}{\cal{E}}+\frac{1}{8}E\right){\cal{C}}_1^2-\frac{1}{2}(M+M_1)(1+{\cal{C}}_0)\right]\epsilon_0+\left[-\frac{3}{4}(A_1+A_2){\cal{E}}-\frac{1}{4}(2A_1+A_2)E\right]\epsilon_0^2\right\}, \\
V_{120}&=&je_0\left\{\frac{1}{4}(2+A_1){\cal{E}}-\frac{1}{2}E-\frac{1}{256}(4+13A_1+3A_2){\cal{E}}^3+\frac{1}{256}(28+25A_1+23A_2)E{\cal{E}}^2-\frac{1}{64}(1-17A_1)E^2{\cal{E}}-\frac{5}{64}E^3\right. \nonumber \\
&&+\left[-\frac{1}{16}(2+A_1){\cal{E}}+\frac{1}{8}E\right]{\cal{C}}_1^2+\left[\frac{1}{2}M+\frac{1}{4}(L_1+M_1){\cal{E}}\right](1+{\cal{C}}_0) \nonumber \\
&&+\left[\frac{1}{4}(2+A_2-2A_1^2){\cal{E}}-\frac{1}{2}(1-A_1)E-\frac{1}{64}{\cal{E}}^3+\frac{7}{64}E{\cal{E}}^2-\frac{1}{64}E^2{\cal{E}}-\frac{5}{64}E^3+\frac{1}{2}N_1\right. \nonumber \\
&&\left.\left.+\left(-\frac{1}{8}{\cal{E}}+\frac{1}{8}E\right){\cal{C}}_1^2+\frac{1}{2}(M+M_1)(1+{\cal{C}}_0)\right]\epsilon_0+\left[\frac{1}{4}(-A_1+A_2){\cal{E}}+\frac{1}{4}(2A_1+A_2)E\right]\epsilon_0^2\right\},
\end{eqnarray}
\begin{eqnarray}
V_{1-40}&=&je_0\left\{\frac{1}{64}(-4+A_1+A_2){\cal{E}}^2-\frac{1}{32}A_1{\cal{E}}E+\frac{1}{16}E^2+\frac{1}{384}{\cal{E}}^4-\frac{1}{96}E{\cal{E}}^3-\frac{29}{768}E^2{\cal{E}}^2+\frac{1}{96}E^3{\cal{E}}+\frac{9}{256}E^4\right.\nonumber \\
&&-\left(\frac{1}{8}B_1+\frac{1}{16}N_1\right){\cal{E}}+\left(\frac{1}{64}{\cal{E}}^2-\frac{1}{64}E^2\right){\cal{C}}_1^2+\left(\frac{1}{8}M+\frac{1}{16}M_1\right){\cal{E}}(1+{\cal{C}}_0) \nonumber \\
&&\left.+\left[\frac{1}{32}(-2+5A_1+2A_2){\cal{E}}^2-\frac{1}{32}(2A_1+A_2)E{\cal{E}}+\frac{1}{16}(1-2A_1)E^2\right]\epsilon_0\right\}, \\
V_{140}&=&je_0\left\{-\frac{1}{64}(4+9A_1+A_2+2A_1^2){\cal{E}}^2+\frac{1}{32}(8+5A_1)E{\cal{E}}-\frac{3}{16}E^2+\frac{1}{384}{\cal{E}}^4-\frac{1}{48}E{\cal{E}}^3+\frac{19}{768}E^2{\cal{E}}^2+\frac{13}{192}E^3{\cal{E}}-\frac{19}{256}E^4\right. \nonumber \\
&&-\frac{1}{16}{\cal{E}}N_1+\left(\frac{1}{64}{\cal{E}}^2-\frac{1}{16}E{\cal{E}}+\frac{3}{64}E^2\right){\cal{C}}_1^2+\left[-\left(\frac{1}{8}M+\frac{1}{16}M_1\right){\cal{E}}+\frac{1}{4}ME\right](1+{\cal{C}}_0)\nonumber \\
&&\left.+\left[-\frac{1}{32}(2+5A_1+6A_2){\cal{E}}^2+\frac{1}{32}(8-6A_1+5A_2)E{\cal{E}}+\frac{3}{16}(-1+2A_1)E^2\right]\epsilon_0\right\},
\end{eqnarray}
\begin{eqnarray}
V_{1-60}&=&je_0\left\{\frac{1}{2304}\left(12+13A_1-9A_2\right){\cal{E}}^3+\frac{1}{768}(-12+5A_1+3A_2){\cal{E}}^2E-\frac{1}{576}(3+7A_1){\cal{E}}E^2+\frac{1}{64}E^3\right.\nonumber \\
&&\left.+\left[\frac{1}{192}{\cal{E}}^3-\frac{1}{64}{\cal{E}}^2E-\frac{1}{192}{\cal{E}}E^2+\frac{1}{64}E^3\right]\epsilon_0\right\}, \\
V_{160}&=&je_0\left\{\frac{1}{2304}(12+71A_1+21A_2){\cal{E}}^3-\frac{1}{768}(36+91A_1+9A_2){\cal{E}}^2E+\frac{1}{576}(69+52A_1){\cal{E}}E^2-\frac{5}{64}E^3\right. \nonumber \\
&&\left.+\left[\frac{1}{192}{\cal{E}}^3-\frac{3}{64}{\cal{E}}^2E+\frac{23}{192}{\cal{E}}E^2-\frac{5}{64}E^3\right]\epsilon_0\right\},
\end{eqnarray}
\begin{eqnarray}
V_{1-80}&=&je_0\left\{-\frac{1}{3072}{\cal{E}}^4+\frac{1}{384}{\cal{E}}^3E-\frac{7}{1536}E^2{\cal{E}}^2-\frac{1}{384}{\cal{E}}E^3+\frac{5}{1024}E^4\right\}, \\
V_{180}&=&je_0\left\{-\frac{1}{3072}{\cal{E}}^4+\frac{1}{192}{\cal{E}}^3E-\frac{43}{1536}E^2{\cal{E}}^2+\frac{11}{192}{\cal{E}}E^3-\frac{35}{1024}E^4\right\}, 
\end{eqnarray}
\begin{eqnarray}
V_{10-1}&=&je_0{\cal{C}}_1\left\{1-\frac{1}{8}A_1{\cal{E}}^2+\frac{1}{16}(4+A_1){\cal{E}}E-\frac{1}{16}E^2-\frac{1}{8}{\cal{C}}_1^2+\frac{1}{4}{\cal{C}}_2\right. \nonumber \\
&&\left.+\left[1+\frac{1}{4}{\cal{E}}E-\frac{1}{16}E^2-\frac{1}{8}{\cal{C}}_1^2+\frac{1}{4}{\cal{C}}_2\right]\epsilon_0\right\},  \\
V_{101}&=&je_0{\cal{C}}_1\left\{-1+\frac{1}{8}A_1{\cal{E}}^2-\frac{1}{16}(4+A_1){\cal{E}}E+\frac{1}{16}E^2+\frac{1}{8}{\cal{C}}_1^2+\frac{1}{4}{\cal{C}}_2\right. \nonumber \\
&&\left.+\left[-1-\frac{1}{4}{\cal{E}}E+\frac{1}{16}E^2+\frac{1}{8}{\cal{C}}_1^2+\frac{1}{4}{\cal{C}}_2\right]\epsilon_0\right\},
\end{eqnarray}
\begin{eqnarray}
V_{1-2-1}&=&je_0{\cal{C}}_1\left\{\frac{1}{8}(-2+A_1){\cal{E}}-\frac{1}{4}E+\frac{1}{128}{\cal{E}}^3-\frac{5}{128}{\cal{E}}^2E-\frac{15}{128}{\cal{E}}E^2-\frac{9}{128}E^3\right. \nonumber \\
&&+\left(\frac{1}{32}{\cal{E}}+\frac{1}{32}E\right){\cal{C}}_1^2-\left(\frac{1}{16}{\cal{E}}+\frac{1}{16}E\right){\cal{C}}_2-\frac{1}{4}K+\frac{1}{4}M(1+{\cal{C}}_0)\nonumber \\
&&\left.+\left[\frac{1}{8}(-2+4A_1+A_2){\cal{E}}+\frac{1}{4}(-1+A_1)E\right]\epsilon_0\right\}, \\
V_{1-21}&=&je_0{\cal{C}}_1\left\{\frac{1}{8}(2-A_1){\cal{E}}+\frac{1}{4}E-\frac{1}{128}{\cal{E}}^3+\frac{5}{128}{\cal{E}}^2E+\frac{15}{128}{\cal{E}}E^2+\frac{9}{128}E^3\right. \nonumber \\
&&-\left(\frac{1}{32}{\cal{E}}+\frac{1}{32}E\right){\cal{C}}_1^2-\left(\frac{1}{16}{\cal{E}}+\frac{1}{16}E\right){\cal{C}}_2-\frac{1}{4}K-\frac{1}{4}M(1+{\cal{C}}_0)\nonumber \\
&&\left.+\left[\frac{1}{8}(2-4A_1-A_2){\cal{E}}+\frac{1}{4}(1-A_1)E\right]\epsilon_0\right\}, \\
V_{12-1}&=&je_0{\cal{C}}_1\left\{-\frac{1}{8}(2+A_1){\cal{E}}+\frac{1}{4}E+\frac{1}{128}{\cal{E}}^3-\frac{7}{128}{\cal{E}}^2E+\frac{1}{128}{\cal{E}}E^2+\frac{5}{128}E^3\right. \nonumber \\
&&+\left(\frac{1}{32}{\cal{E}}-\frac{1}{32}E\right){\cal{C}}_1^2+\left(-\frac{1}{16}{\cal{E}}+\frac{1}{16}E\right){\cal{C}}_2+\frac{1}{4}K-\frac{1}{4}M(1+{\cal{C}}_0)\nonumber \\
&&\left.+\left[-\frac{1}{8}(2+A_2){\cal{E}}+\frac{1}{4}(1-A_1)E\right]\epsilon_0\right\}, \\
V_{121}&=&je_0{\cal{C}}_1\left\{\frac{1}{8}(2+A_1){\cal{E}}-\frac{1}{4}E-\frac{1}{128}{\cal{E}}^3+\frac{7}{128}{\cal{E}}^2E-\frac{1}{128}{\cal{E}}E^2-\frac{5}{128}E^3\right. \nonumber \\
&&+\left(-\frac{1}{32}{\cal{E}}+\frac{1}{32}E\right){\cal{C}}_1^2+\left(-\frac{1}{16}{\cal{E}}+\frac{1}{16}E\right){\cal{C}}_2+\frac{1}{4}K+\frac{1}{4}M(1+{\cal{C}}_0)\nonumber \\
&&\left.+\left[\frac{1}{8}(2+A_2){\cal{E}}+\frac{1}{4}(-1+A_1)E\right]\epsilon_0\right\},
\end{eqnarray}
\begin{eqnarray}
V_{1-4-1}&=&je_0{\cal{C}}_1\left\{\frac{1}{128}(4-A_1-A_2){\cal{E}}^2+\frac{1}{64}A_1{\cal{E}}E-\frac{1}{32}E^2+\left(\frac{1}{32}{\cal{E}}^2-\frac{1}{32}E^2\right)\epsilon_0\right\}, \\
V_{1-41}&=&je_0{\cal{C}}_1\left\{\frac{1}{128}(-4+A_1+A_2){\cal{E}}^2-\frac{1}{64}A_1{\cal{E}}E+\frac{1}{32}E^2+\left(-\frac{1}{32}{\cal{E}}^2+\frac{1}{32}E^2\right)\epsilon_0\right\}, \\
V_{14-1}&=&je_0{\cal{C}}_1\left\{\frac{1}{128}(4+9A_1+A_2){\cal{E}}^2-\frac{1}{64}(8+5A_1){\cal{E}}E+\frac{3}{32}E^2+\left(\frac{1}{32}{\cal{E}}^2-\frac{1}{8}{\cal{E}}E+\frac{3}{32}E^2\right)\epsilon_0\right\}, \\
V_{141}&=&je_0{\cal{C}}_1\left\{-\frac{1}{128}(4+9A_1+A_2){\cal{E}}^2+\frac{1}{64}(8+5A_1){\cal{E}}E-\frac{3}{32}E^2+\left(-\frac{1}{32}{\cal{E}}^2+\frac{1}{8}{\cal{E}}E+\frac{3}{32}E^2\right)\epsilon_0\right\},
\end{eqnarray}
\begin{eqnarray}
V_{1-6-1}&=&je_0{\cal{C}}_1\left\{-\frac{1}{384}{\cal{E}}^3+\frac{1}{128}{\cal{E}}^2E+\frac{1}{384}{\cal{E}}E-\frac{1}{128}E^3\right\}, \\
V_{1-61}&=&je_0{\cal{C}}_1\left\{\frac{1}{384}{\cal{E}}^3-\frac{1}{128}{\cal{E}}^2E-\frac{1}{384}{\cal{E}}E+\frac{1}{128}E^3\right\}, \\
V_{16-1}&=&je_0{\cal{C}}_1\left\{-\frac{1}{384}{\cal{E}}^3+\frac{3}{128}{\cal{E}}^2E-\frac{23}{384}{\cal{E}}E^2+\frac{5}{128}E^3\right\}, \\
V_{161}&=&je_0{\cal{C}}_1\left\{\frac{1}{384}{\cal{E}}^3-\frac{3}{128}{\cal{E}}^2E+\frac{23}{384}{\cal{E}}E^2-\frac{5}{128}E^3\right\},
\end{eqnarray}
\begin{eqnarray}
V_{10-2}&=&je_0\left\{\left(\frac{1}{2}{\cal{C}}_2-\frac{1}{4}{\cal{C}}_1^2\right)\left(1+\epsilon_0+\frac{1}{4}{\cal{E}}E-\frac{1}{16}E^2\right)+\frac{1}{6}{\cal{C}}_1{\cal{C}}_3-\frac{1}{8}{\cal{C}}_1^2{\cal{C}}_2+\frac{1}{48}{\cal{C}}_1^4\right\}, \\
V_{102}&=&je_0\left\{\left(\frac{1}{2}{\cal{C}}_2+\frac{1}{4}{\cal{C}}_1^2\right)\left(-1-\epsilon_0-\frac{1}{4}{\cal{E}}E+\frac{1}{16}E^2\right)+\frac{1}{6}{\cal{C}}_1{\cal{C}}_3+\frac{1}{8}{\cal{C}}_1^2{\cal{C}}_2+\frac{1}{48}{\cal{C}}_1^4\right\},
\end{eqnarray}
\begin{eqnarray}
V_{1-2-2}&=&je_0({\cal{C}}_1^2-2{\cal{C}}_2)\left\{\frac{1}{32}(2-A_1){\cal{E}}+\frac{1}{16}E+\left(\frac{1}{16}{\cal{E}}+\frac{1}{16}E\right)\epsilon_0\right\}, \\
V_{1-22}&=&je_0({\cal{C}}_1^2+2{\cal{C}}_2)\left\{\frac{1}{32}(2-A_1){\cal{E}}+\frac{1}{16}E+\left(\frac{1}{16}{\cal{E}}+\frac{1}{16}E\right)\epsilon_0\right\}, \\
V_{12-2}&=&je_0({\cal{C}}_1^2-2{\cal{C}}_2)\left\{\frac{1}{32}(2+A_1){\cal{E}}-\frac{1}{16}E+\left(\frac{1}{16}{\cal{E}}-\frac{1}{16}E\right)\epsilon_0\right\}, \\
V_{122}&=&je_0({\cal{C}}_1^2+2{\cal{C}}_2)\left\{\frac{1}{32}(2+A_1){\cal{E}}-\frac{1}{16}E+\left(\frac{1}{16}{\cal{E}}-\frac{1}{16}E\right)\epsilon_0\right\},
\end{eqnarray}
\begin{eqnarray}
V_{1-4-2}&=&je_0({\cal{C}}_1^2-2{\cal{C}}_2)\left\{-\frac{1}{128}{\cal{E}}^2+\frac{1}{128}E^2\right\}, \\
V_{1-42}&=&je_0({\cal{C}}_1^2+2{\cal{C}}_2)\left\{-\frac{1}{128}{\cal{E}}^2+\frac{1}{128}E^2\right\}, \\
V_{14-2}&=&je_0({\cal{C}}_1^2-2{\cal{C}}_2)\left\{-\frac{1}{128}{\cal{E}}^2+\frac{1}{32}{\cal{E}}E-\frac{3}{128}E^2\right\}, \\
V_{142}&=&je_0({\cal{C}}_1^2+2{\cal{C}}_2)\left\{-\frac{1}{128}{\cal{E}}^2+\frac{1}{32}{\cal{E}}E-\frac{3}{128}E^2\right\},
\end{eqnarray}
\begin{eqnarray}
V_{10-3}&=&\frac{1}{24}je_0({\cal{C}}_1^3-6{\cal{C}}_1{\cal{C}}_2+8{\cal{C}}_3)(1+\epsilon_0), \\
V_{103}&=&-\frac{1}{24}je_0({\cal{C}}_1^3+6{\cal{C}}_1{\cal{C}}_2+8{\cal{C}}_3)(1+\epsilon_0), 
\end{eqnarray}
\begin{eqnarray}
V_{1-2-3}&=&-\frac{1}{96}je_0({\cal{C}}_1^3-6{\cal{C}}_1{\cal{C}}_2+8{\cal{C}}_3)({\cal{E}}+E), \\
V_{1-23}&=&\frac{1}{96}je_0({\cal{C}}_1^3+6{\cal{C}}_1{\cal{C}}_2+8{\cal{C}}_3)({\cal{E}}+E), \\
V_{12-3}&=&-\frac{1}{96}je_0({\cal{C}}_1^3-6{\cal{C}}_1{\cal{C}}_2+8{\cal{C}}_3)({\cal{E}}-E), \\
V_{123}&=&\frac{1}{96}je_0({\cal{C}}_1^3+6{\cal{C}}_1{\cal{C}}_2+8{\cal{C}}_3)({\cal{E}}-E),
\end{eqnarray}
\begin{eqnarray}
V_{10-4}&=&\frac{1}{192}je_0(-{\cal{C}}_1^4+12{\cal{C}}_1^2{\cal{C}}_2-32{\cal{C}}_1{\cal{C}}_3-12{\cal{C}}_2^2+48{\cal{C}}_4), \\
V_{104}&=&-\frac{1}{192}je_0({\cal{C}}_1^4+12{\cal{C}}_1^2{\cal{C}}_2+32{\cal{C}}_1{\cal{C}}_3+12{\cal{C}}_2^2+48{\cal{C}}_4),
\end{eqnarray}
\begin{eqnarray}
V_{200}&=&\frac{3}{4}e_0^2\left\{1+\frac{1}{6}e_0^2+\frac{1}{8}(1-2A_1){\cal{E}}^2+\frac{1}{4}(2+A_1)E{\cal{E}}-\frac{1}{4}E^2-{\cal{C}}_1^2\right.\nonumber \\
&&\left.+\left[2+\frac{2}{3}e_0^2+\frac{1}{4}{\cal{E}}^2+E{\cal{E}}-\frac{1}{2}E^2-2{\cal{C}}_1^2\right]\epsilon_0+\epsilon_0^2\right\}
\end{eqnarray}
\begin{eqnarray}
V_{2-20}&=&e_0^2\left\{\frac{1}{16}(-6+3A_1-2e_0^2){\cal{E}}-\frac{1}{16}(6+e_0^2)E-\frac{9}{64}{\cal{E}}^2E-\frac{15}{64}{\cal{E}}E^2-\frac{3}{32}E^3\right.\nonumber \\
&&+\left(\frac{3}{8}{\cal{E}}+\frac{3}{8}E\right){\cal{C}}_1^2+\frac{3}{8}M(1+{\cal{C}}_0) \nonumber \\
&&\left.+\left[\frac{3}{16}(-4+5A_1+3A_2){\cal{E}}+\frac{3}{8}(-2+A_1)E\right]\epsilon_0+\left(-\frac{3}{8}{\cal{E}}-\frac{3}{8}E\right)\epsilon_0^2\right\}, \\
V_{220}&=&e_0^2\left\{-\frac{1}{16}(6+3A_1+2e_0^2){\cal{E}}+\frac{1}{16}(6+e_0^2)E-\frac{9}{64}{\cal{E}}^2E+\frac{9}{64}{\cal{E}}E^2\right. \nonumber \\
&&+\left(\frac{3}{8}{\cal{E}}-\frac{3}{8}E\right)C_1^2-\frac{3}{8}M(1+{\cal{C}}_0) \nonumber \\
&&\left.+\left[-\frac{3}{16}(4+A_1+A_2){\cal{E}}+\frac{3}{8}(2-A_1)E\right]\epsilon_0+\left(-\frac{3}{8}{\cal{E}}+\frac{3}{8}E\right)\epsilon_0^2\right\},
\end{eqnarray}
\begin{eqnarray}
V_{2-40}&=&e_0^2\left\{\frac{3}{256}(8-5A_1-A_2){\cal{E}}^2+\frac{3}{128}(4-A_1){\cal{E}}E+\left(\frac{3}{16}{\cal{E}}^2+\frac{3}{16}{\cal{E}}E\right)\epsilon_0\right\},\\
V_{240}&=&e_0^2\left\{\frac{3}{256}(8+13A_1+A_2){\cal{E}}^2-\frac{3}{128}(12+7A_1){\cal{E}}E+\frac{3}{16}E^2+\left(\frac{3}{16}{\cal{E}}^2-\frac{9}{16}{\cal{E}}E+\frac{3}{8}E^2\right)\epsilon_0\right\},
\end{eqnarray}
\begin{eqnarray}
V_{2-60}&=&e_0^2\left\{-\frac{1}{64}{\cal{E}}^3+\frac{1}{64}{\cal{E}}E^2\right\}, \\
V_{260}&=&e_0^2\left\{-\frac{1}{64}{\cal{E}}^3+\frac{3}{32}{\cal{E}}^2E-\frac{11}{64}{\cal{E}}E^2+\frac{3}{32}E^3\right\},
\end{eqnarray}
\begin{eqnarray}
V_{20-1}&=&-e_0^2{\cal{C}}_1\left\{\frac{3}{4}(1+\epsilon_0)^2+\frac{1}{8}e_0^2+\frac{3}{32}{\cal{E}}^2+\frac{3}{8}{\cal{E}}E-\frac{3}{16}E^2-\frac{3}{8}{\cal{C}}_1^2+\frac{3}{8}{\cal{C}}_2\right\},\\
V_{201}&=&e_0^2{\cal{C}}_1\left\{\frac{3}{4}(1+\epsilon_0)^2+\frac{1}{8}e_0^2+\frac{3}{32}{\cal{E}}^2+\frac{3}{8}{\cal{E}}E-\frac{3}{16}E^2-\frac{3}{8}{\cal{C}}_1^2-\frac{3}{8}{\cal{C}}_2\right\},
\end{eqnarray}
\begin{eqnarray}
V_{2-2-1}&=&\frac{3}{8}e_0^2{\cal{C}}_1\left[\left(1-\frac{1}{2}A_1\right){\cal{E}}+E+2({\cal{E}}+E)\epsilon_0\right], \\
V_{2-21}&=&-\frac{3}{8}e_0^2{\cal{C}}_1\left[\left(1-\frac{1}{2}A_1\right){\cal{E}}+E+2({\cal{E}}+E)\epsilon_0\right], \\
V_{22-1}&=&\frac{3}{8}e_0^2{\cal{C}}_1\left[\left(1+\frac{1}{2}A_1\right){\cal{E}}-E+2({\cal{E}}-E)\epsilon_0\right], \\
V_{2-21}&=&-\frac{3}{8}e_0^2{\cal{C}}_1\left[\left(1+\frac{1}{2}A_1\right){\cal{E}}-E+2({\cal{E}}-E)\epsilon_0\right], 
\end{eqnarray}
\begin{eqnarray}
V_{2-4-1}&=&-\frac{3}{32}e_0^2{\cal{C}}_1({\cal{E}}^2+{\cal{E}}E), \\
V_{2-41}&=&\frac{3}{32}e_0^2{\cal{C}}_1({\cal{E}}^2+{\cal{E}}E), \\
V_{24-1}&=&-\frac{3}{32}e_0^2{\cal{C}}_1({\cal{E}}^2-3{\cal{E}}E+2E^2), \\
V_{241}&=&\frac{3}{32}e_0^2{\cal{C}}_1({\cal{E}}^2-3{\cal{E}}E+2E^2), 
\end{eqnarray}
\begin{eqnarray}
V_{20-2}&=&\frac{3}{8}e_0^2({\cal{C}}_1^2-{\cal{C}}_1)(1+2\epsilon_0), \\
V_{202}&=&\frac{3}{8}e_0^2({\cal{C}}_1^2+{\cal{C}}_1)(1+2\epsilon_0),
\end{eqnarray}
\begin{eqnarray}
V_{2-2-2}&=&-\frac{3}{16}e_0^2({\cal{C}}_1^2-{\cal{C}}_2)({\cal{E}}+E), \\
V_{2-22}&=&-\frac{3}{16}e_0^2({\cal{C}}_1^2+{\cal{C}}_2)({\cal{E}}+E), \\
V_{22-2}&=&-\frac{3}{16}e_0^2({\cal{C}}_1^2-{\cal{C}}_2)({\cal{E}}-E), \\
V_{2-22}&=&-\frac{3}{16}e_0^2({\cal{C}}_1^2+{\cal{C}}_2)({\cal{E}}-E),
\end{eqnarray}
\begin{eqnarray}
V_{20-3}&=&-\frac{1}{8}e_0^2({\cal{C}}_1^3-3{\cal{C}}_1{\cal{C}}_2+2{\cal{C}}_3), \\
V_{203}&=&\frac{1}{8}e_0^2({\cal{C}}_1^3+3{\cal{C}}_1{\cal{C}}_2+2{\cal{C}}_3),
\end{eqnarray}
\begin{eqnarray}
V_{300}&=&\frac{1}{3}je_0^3\left\{1+\frac{3}{8}e_0^2+\frac{3}{8}{\cal{E}}^2+\frac{3}{4}E{\cal{E}}-\frac{9}{16}E^2-\frac{9}{4}{\cal{C}}_1^2+3\epsilon_0+3\epsilon_0^2\right\},
\end{eqnarray}
\begin{eqnarray}
V_{3-20}&=&je_0^3\left\{\frac{1}{8}(-2+A_1){\cal{E}}-\frac{1}{4}E+\left(-\frac{3}{4}{\cal{E}}-\frac{3}{4}E\right)\epsilon_0\right\}, \\
V_{320}&=&je_0^3\left\{-\frac{1}{8}(2+A_1){\cal{E}}+\frac{1}{4}E+\left(-\frac{3}{4}{\cal{E}}+\frac{3}{4}E\right)\epsilon_0\right\},
\end{eqnarray}
\begin{eqnarray}
V_{3-40}&=&e_0^3\left\{\frac{3}{32}{\cal{E}}^2+\frac{1}{8}E{\cal{E}}+\frac{1}{32}E^2\right\}, \\
V_{340}&=&e_0^3\left\{\frac{3}{32}{\cal{E}}^2-\frac{1}{4}E{\cal{E}}+\frac{5}{32}E^2\right\},
\end{eqnarray}
\begin{eqnarray}
V_{30-1}&=&-\frac{1}{2}je_0^3{\cal{C}}_1(1+3\epsilon_0), \\
V_{301}&=&\frac{1}{2}je_0^3{\cal{C}}_1(1+3\epsilon_0), \\
\end{eqnarray}
\begin{eqnarray}
V_{3-2-1}&=&\frac{3}{8}jE_0^3{\cal{C}}_1({\cal{E}}+E), \\
V_{3-21}&=&-\frac{3}{8}jE_0^3{\cal{C}}_1({\cal{E}}+E), \\
V_{32-1}&=&\frac{3}{8}jE_0^3{\cal{C}}_1({\cal{E}}-E), \\
V_{321}&=&-\frac{3}{8}jE_0^3{\cal{C}}_1({\cal{E}}-E),
\end{eqnarray}
\begin{eqnarray}
V_{30-2}&=&\frac{1}{8}je_0^3(3{\cal{C}}_1^2-2{\cal{C}}_2), \\
V_{302}&=&\frac{1}{8}je_0^3(3{\cal{C}}_1^2+2{\cal{C}}_2),
\end{eqnarray}
\begin{eqnarray}
V_{400}&=&-\frac{5}{32}e_0^4(1+4\epsilon_0), \\
\end{eqnarray}
\begin{eqnarray}
V_{4-20}&=&e_0^4\left\{\frac{5}{32}{\cal{E}}+\frac{5}{32}E\right\}, \\
V_{420}&=&e_0^4\left\{\frac{5}{32}{\cal{E}}-\frac{5}{32}E\right\},
\end{eqnarray}
\begin{eqnarray}
V_{40-1}&=&\frac{5}{16}e_0^4{\cal{C}}_1, \\
V_{401}&=&-\frac{5}{16}e_0^4{\cal{C}}_1,
\end{eqnarray}
\begin{eqnarray}
V_{500}&=&-\frac{3}{40}je_0^5.
\end{eqnarray}
\begin{eqnarray}
V_{020}&=&-\frac{1}{4}V_1{\cal{E}}+\frac{1}{256}(2V_1-2V_2+5A_1V_1+3A_2V_1+6A_1V_2){\cal{E}}^3-\frac{1}{128}(6V_1+6V_2-7A_1V_1)E{\cal{E}}^2-\frac{1}{16}V_1E^2{\cal{E}}^2\nonumber \\
&&+\frac{1}{2}W+\frac{1}{128}(2W_1+6W_2-A_1W-A_2W-10A_1W_1){\cal{E}}^2+\frac{1}{64}(10W_1+7A_1W)E{\cal{E}}-\frac{1}{8}WE^2\nonumber \\
&&+\left[-\frac{1}{4}(V_1+V_2+A_1V_1){\cal{E}}+\frac{1}{128}(V_1-V_2){\cal{E}}^3-\frac{3}{64}(V_1+3V_2)E{\cal{E}}^2-\frac{1}{16}(V_1+V_2)E^2{\cal{E}}-\frac{1}{4}V_1(B_1+N_1)\right. \nonumber \\
&&\left.+\frac{1}{2}W_1+\frac{1}{64}(W_1+7W_2){\cal{E}}^2+\frac{5}{32}(W_1+W_2)E{\cal{E}}-\frac{1}{8}W_1E^2\right]\epsilon_0\nonumber \\
&&+\left[-\frac{1}{8}(2V_2-2A_1V_1-A_2V_1-2A_1V_2){\cal{E}}+\frac{1}{4}W_2\right]\epsilon_0^2\nonumber \\
&&+{\cal{C}}_0\left\{\frac{1}{2}M+\frac{1}{4}(L_1+M_1){\cal{E}}+\frac{1}{64}(M_1+3M_2){\cal{E}}^2+\frac{5}{32}M_1E{\cal{E}}-\frac{1}{8}ME^2\right. \nonumber \\
&&\left.+\left[\frac{1}{2}M_1+\frac{1}{4}(L_1+L_2+M_1+M_2)\right]\epsilon_0+\frac{1}{4}M_2\epsilon_0^2\right\},
\label{eq:V_020}
\end{eqnarray}
\begin{eqnarray}
V_{040}&=&\frac{1}{64}(V_1+V_2+A_1V_1){\cal{E}}^2-\frac{1}{32}V_1E{\cal{E}}-\frac{1}{1536}(V_1+V_2){\cal{E}}^4+\frac{1}{256}(V_1+2V_2)E{\cal{E}}^3+\frac{1}{384}(V_1+V_2)E^2{\cal{E}}^2-\frac{1}{64}V_1E^3{\cal{E}}\nonumber \\
&&+\frac{1}{64}V_1(B_1+N_1){\cal{E}}-\frac{1}{16}(W_1+A_1W){\cal{E}}+\frac{1}{8}WE+\frac{1}{768}(W_1-3W_2){\cal{E}}^3-\frac{1}{128}(2W_1+W_2)E{\cal{E}}^2+\frac{5}{192}W_1E^2{\cal{E}} \nonumber \\
&&+\left[\frac{1}{64}(V_1+3V_2+A_2V_1-A_1V_2){\cal{E}}^2-\frac{1}{32}(V_1+V_2-2A_1V_1)E{\cal{E}}-\frac{1}{16}(W_1+W_2+A_1W+A_2W){\cal{E}}+\frac{1}{8}(W_1-A_1W)E\right]\epsilon_0 \nonumber \\
&&+\left[\frac{1}{32}V_2{\cal{E}}^2-\frac{1}{32}V_2E{\cal{E}}-\frac{1}{16}W_2{\cal{E}}+\frac{1}{16}W_2E\right]\epsilon_0^2\nonumber \\
&&+{\cal{C}}_0\left\{-\frac{1}{16}(M_1+A_1M){\cal{E}}+\frac{1}{8}ME-\frac{1}{64}(L_1+L_2+M_1+M_2){\cal{E}}^2+\frac{1}{32}(L_1+M_1)E{\cal{E}}+\left[-\frac{1}{16}(M_1+M_2){\cal{E}}+\frac{1}{8}M_1E\right]\epsilon_0\right\},\nonumber \\
\end{eqnarray}
\begin{eqnarray}
V_{060}&=&-\frac{1}{2304}(2V_1+6V_2+7A_1V_1+1A_2V_1+6A_1V_2){\cal{E}}^3+\frac{1}{1152}(6V_1+6V_2+7A_1V_1)E{\cal{E}}^2-\frac{1}{144}V_1E^2{\cal{E}} \nonumber \\
&&+\frac{1}{384}(2W_1+2W_2+A_1W+A_2W+6A_1W_1){\cal{E}}^2-\frac{1}{192}(6W_1+7A_1W)E{\cal{E}}+\frac{1}{24}E^2W\nonumber \\
&&+\left[-\frac{1}{1152}(V_1+7V_2){\cal{E}}^3+\frac{1}{192}(V_1+3V_2)E{\cal{E}}^2-\frac{1}{144}(V_1+V_2)E^2{\cal{E}}\right.\nonumber \\
&&\left.+\frac{1}{192}(W_1+3W_2){\cal{E}}^2-\frac{1}{32}(W_1+W_2)E{\cal{E}}+\frac{1}{24}W_1E^2\right]\epsilon_0\nonumber \\
&&+{\cal{C}}_0\left\{\frac{1}{192}(M_1+M_2){\cal{E}}^2-\frac{1}{32}M_1E{\cal{E}}+\frac{1}{24}ME^2\right\},
\end{eqnarray}
\begin{eqnarray}
V_{080}&=&\frac{1}{24576}(V_1+7V_2){\cal{E}}^4-\frac{1}{2048}(V_1+3V_2)E{\cal{E}}^3+\frac{11}{6144}(V_1+V_2)E^2{\cal{E}}^2-\frac{1}{512}V_1E^3{\cal{E}}\nonumber \\
&&-\frac{1}{3072}(W_1+3W_2){\cal{E}}^3+\frac{1}{256}(W_1+W_2)E{\cal{E}}^2-\frac{11}{768}W_1E^2{\cal{E}}+\frac{1}{64}WE^3.
\end{eqnarray}
\begin{eqnarray}
V_{001}&=&-{\cal{C}}_1, \\
V_{0-21}&=&-{\cal{C}}_1\left[\frac{1}{4}K+\frac{1}{8}(H_1+K_1){\cal{E}}+\frac{1}{4}K_1\epsilon_0\right], \\
V_{021}&=&{\cal{C}}_1\left[\frac{1}{4}K+\frac{1}{8}(H_1+K_1){\cal{E}}+\frac{1}{4}K_1\epsilon_0\right], \\
V_{0-41}&=&-{\cal{C}}_1\left(-\frac{1}{32}K_1{\cal{E}}+\frac{1}{16}KE\right), \\
V_{041}&=&{\cal{C}}_1\left(-\frac{1}{32}K_1{\cal{E}}+\frac{1}{16}KE\right), \\
V_{002}&=&-\frac{1}{2}{\cal{C}}_2, \\
V_{0-22}&=&-\frac{1}{4}K{\cal{C}}_2, \\
V_{022}&=&\frac{1}{4}K{\cal{C}}_2, \\
V_{003}&=&-\frac{1}{3}{\cal{C}}_3, \\
V_{004}&=&-\frac{1}{4}{\cal{C}}_4, \\
V_{005}&=&-\frac{1}{5}{\cal{C}}_5.
\label{eq:V_005}
\end{eqnarray}
\end{appendix}


\begin{thebibliography}{}

  \bibitem[Borkovits~et~al.(2002)]{borkoetal02} Borkovits, T., Csizmadia, Sz., Heged\"us, T. et al., 2002,
      \aap, 392, 895

  \bibitem[Borkovits(2003)]{borko03} Borkovits, T., 2003, Proc. of the ``3rd Austrian-Hungarian Workshop on 
      Trojans and Related Topics'', held at Vienna, 2002 May 13-15th, (Freistetter, F., Dvorak, R., \'Erdi, B. eds.), p.~161

  \bibitem[Borkovits~et~al.(2003)]{borkoetal03} Borkovits, T., \'Erdi, B., Forg\'acs-Dajka, E., \&
      Kov\'acs, T., 2003, \aap, 398, 1091

  \bibitem[Borkovits~et~al.(2004)]{borkoetal04} Borkovits, T., Forg\'acs-Dajka, E., \& Reg\'aly, Zs., 2004,
      \aap, 2004, \aap, 426, 951

  \bibitem[Claret(1998)]{claret98} Claret, A., 1998, \aap, 330, 533

  \bibitem[Claret \& Gim\`enez(1991)]{claretgimenez91} Claret, A., \& 
     Gim\`enez, A., 1991, \aaps, 91, 217

  \bibitem[Company~et~al.(1988)]{companyetal88} Company, R., Portilla, M., \& 
     Gim\`enez, A., 1988, \apj, 335, 962  

  \bibitem[Eggleton~et~al.(1998)]{eggletonetal98} Eggleton, P. P., Kiseleva, L. G., \&
     Hut, P., 1998, \apj, 499, 853

   \bibitem[Gim\`enez \& Garcia-Pelayo(1983)]{gimenezgarcia-pelayo83} Gim\`enez, A., \&
      Garcia-Pelayo, J.M., 1983, \apss, 92, 203

   \bibitem[Heged\"us \& Nuspl(1986)]{hegenuspl86} Heged\"us, T., \& Nuspl, J.,
      1986, \actaa, 36, 381 
      
   \bibitem[Hilditch(1972a)]{hilditch72a} Hilditch, R.W., 1972a, MemRAS, 76, 141

   \bibitem[Hilditch(1972b)]{hilditch72b} Hilditch, R.W., 1972b, \pasp, 84, 519

   \bibitem[Khaliullin \& Kozyreva(1983)]{khaliullinkozyreva83} Khaliullin, Kh., F., \&
      Kozyreva, V.S., 1983, Ap\&SS, 94, 115

   \bibitem[Khaliullin~et~al.(1991)]{khaliullinetal91} Khaliullin, Kh.F., Khodykin, S.A., \&
      Zakharov, A.I., 1991, \apj, 375, 314

   \bibitem[Khodykin(1989)]{khodykin89} Khodykin, S.A., 1989, Astron. Tsirk. 1536

   \bibitem[Khodykin \& Vedeneyev(1997)]{khodykinvedeneyev97} Khodykin, S.A., \& 
      Vedeneyev, V.G., 1997, \apj, 475, 798

   \bibitem[Khodykin~et~al.(2004)]{khodykinetal04} Khodykin, S.A., Zakharov, A.I., \&
      Andersen, W.L., 2004, \apj, 615, 506

   \bibitem[Kozai(1962)]{kozai62} Kozai, Y., 1962, \aj, 67, 591

   \bibitem[Kozyreva et al.(1999)]{ibvs4690} Kozyreva, V.S., Zakharov, A.I., \& 
      Khaliullin, Kh.F., 1999, IBVS, No. 4690

   \bibitem[Laskar \& Robutel(1993)]{laskarrobutel93} Laskar, J., \& Robutel, P., 1993,
      Nature, 361, 608

   \bibitem[Maloney et al.(1989)]{maloneyetal89} Maloney, F.P., Guinan, E.F., \& 
      Boyd, P.T., 1989, \aj, 98, 1800

   \bibitem[Maloney et al.(1991)]{maloneyetal91} Maloney, F.P., Guinan, E.F., \&
      Mukherjee, J., 1991, \aj, 102, 256

   \bibitem[Padalia \& Srivastava(1975)]{padaliasrivastava75} Padalia, T.D., \&
      Srivastava, R.K., 1975, Ap\&SS, 38, 87

   \bibitem[Pribulla \& Rucinski(2006)]{pribullarucinski06} Pribulla, T., \& 
      Rucinski, S.M., 2006, \aj, 131, 2986

   \bibitem[Semeniuk(1968)]{semeniuk68} Semeniuk, T., 1968, Acta Astr., 18, 1

   \bibitem[Shakura(1985)]{shakura85} Shakura, N.I., 1985, Sov. Astr. Lett., 11, 224
      
   \bibitem[S\"oderhjelm(1984)]{soderhjelm84} S\"oderhjelm, S., 1984, \aap, 141, 232

   \bibitem[Zahn(1977)]{zahn77}Zahn, J.-P., 1977, \aap, 57, 383

\end{thebibliography}
\end{document}